\documentclass[preprint,12pt]{elsarticle}
\usepackage{amsmath,amsfonts}
\usepackage{caption}
\usepackage{siunitx}
\usepackage{amssymb}
\usepackage{booktabs}
\usepackage{multirow}
\usepackage{comment}
\usepackage{microtype}
\usepackage[pdftex,hypertexnames=true,linktocpage=true]{hyperref}
\hypersetup{colorlinks=true,linkcolor=blue,anchorcolor=blue,citecolor=blue,filecolor=blue,urlcolor=blue,bookmarksnumbered=true,pdfview=FitB}
\usepackage{rotating}
\biboptions{sort&compress}
\usepackage[dvipsnames]{xcolor}
\usepackage{bbding}
\usepackage{lineno}
\usepackage[normalem]{ulem}
\definecolor{mat-1}{rgb}{0, 0.4470, 0.7410}
\definecolor{mat-2}{rgb}{0.8500, 0.3250, 0.0980}
\definecolor{mat-3}{rgb}{0.47, 0.67, 0.19}
\definecolor{mat-4}{rgb}{0.4940, 0.1840, 0.5560}
\definecolor{mat-5}{rgb}{0.4660, 0.6740, 0.1880}
\definecolor{mat-6}{rgb}{0.6350, 0.0780, 0.1840}
\definecolor{amber}{rgb}{1.0, 0.75, 0.0}
\definecolor{brose}{rgb}{1.0, 0.33, 0.64}
\definecolor{afb}{rgb}{0.36, 0.54, 0.66}
\definecolor{ao}{rgb}{0.0, 0.5, 0.0}
\usepackage{tikz}
\usetikzlibrary{shapes.geometric}
\usetikzlibrary{calc}

\usepackage{tikz}
\usetikzlibrary{shapes.geometric}
\usetikzlibrary{calc}
\usepackage{pgfplots}
\pgfplotsset{compat=1.18}
\newcommand{\tikzcircle}[2][red,fill=red]{\tikz[baseline=-0.6ex]\node[draw,line width=0.5mm,#1,scale=#2,circle]{} ;}

\newcommand{\tikztriangle}[2][red,fill=red]{\tikz[baseline=-0.6ex]\node[draw,line width=0.5mm,#1,scale=#2,regular polygon, regular polygon sides=3]{};}

\newcommand{\tikzline}[2][red,solid]{\tikz[baseline=-0.6ex]\draw[line width=0.6mm,#1,#2] (0,0)--(0.5,0);}

\newcommand{\tikzsquare}[2][red,fill=red]{\tikz[baseline=-0.6ex]\node[draw,line width=0.5mm,#1,scale=#2,regular polygon, regular polygon sides=4]{};}

\tikzset{cross/.style={cross out, draw=black, minimum size=2*(#1-\pgflinewidth), inner sep=0pt, outer sep=0pt},
cross/.default={1pt}}
\newcommand{\tikzcross}[2][mat-2,fill=mat-2]{\tikz[baseline=-0.6ex]\draw[line width=0.5mm] (0,0.1)--(0.2,-0.1);\hspace{-2.1ex}\tikz[baseline=-0.6ex]\draw[line width=0.5mm] (0,-0.1)--(0.2,0.1);}

\newcommand{\triangleriblet}[2]{%
\begin{tikzpicture}[baseline=0.2ex, scale=0.8]
    \pgfmathsetmacro{\s}{#1}
    \pgfmathsetmacro{\tipAngle}{#2}
    \pgfmathsetmacro{\h}{\s/(2*tan(\tipAngle/2))}
    \pgfmathsetmacro{\n}{3} % number of periods
    \pgfmathsetmacro{\H}{0.8}
    
    \fill[gray!20]
    (0,0)
    \foreach \i in {1,...,\n} {
        -- ++(\s/2,\h)
        -- ++(\s/2,-\h)
    }
    -- (\n*\s,\H)
    -- (0,\H)
    -- cycle;
    \draw[line width=0.6pt]
    (0,0)
    \foreach \i in {1,...,\n} {
        -- ++(\s/2,\h)
        -- ++(\s/2,-\h)
    };
    \draw[line width=0.6pt]
    (0,0) -- (\n*\s,0);
    \node[font=\footnotesize] 
    at (\n*\s/1.7, 0.9*\H) {\scalebox{0.8}{$\pgfmathprintnumber{#2}^\circ$}};
\end{tikzpicture}%
}

\newcommand{\scallopriblet}[2]{%
\begin{tikzpicture}[baseline=-1.8ex, scale=0.8]

    \pgfmathsetmacro{\s}{#1}   % spacing
    \pgfmathsetmacro{\h}{#2}   % height
    \pgfmathsetmacro{\n}{3}    % number of periods
    \pgfmathsetmacro{\H}{0.425}

    \fill[gray!20]
    plot[domain=0:\n*\s, samples=150]
        (\x, {-\h*((sin((180 + 180*\x/\s))^2)})
    -- (\n*\s,\H)
    -- (0,\H)
    -- cycle;

    \draw[line width=0.6pt]
    plot[domain=0:\n*\s, samples=150]
        (\x, {-\h*((sin((180 + 180*\x/\s))^2)});

\end{tikzpicture}%
}

\newcommand{\twotrapezoidriblet}[3]{%
\begin{tikzpicture}[baseline=0.2ex, scale=0.8]
    \pgfmathsetmacro{\s}{#1}      % base width
    \pgfmathsetmacro{\alpha}{#2}  % tip angle
    \pgfmathsetmacro{\g}{#3}      % valley width
    
    \pgfmathsetmacro{\h}{\s/(2*tan(\alpha/2))}
    \pgfmathsetmacro{\rs}{0.25*\s}
    \pgfmathsetmacro{\st}{\s-2*\rs}
    \pgfmathsetmacro{\rh}{\rs/(tan(\alpha/2))}
    
    \fill[gray!20]
    (0,0)
    -- (\rs,\rh)
    -- (\rs+\st,\rh)
    -- (\s,0)
    -- (\s+\g,0)
    -- (\s+\g+\rs,\rh)
    -- (\s+\g+\rs+\st,\rh)
    -- (2*\s+\g,0)
    -- (2*\s+\g,0.8)
    -- (0,0.8)
    -- cycle;
    
    \draw[line width=0.6pt]
    (0,0)
    -- (\rs,\rh)
    -- (\rs+\st,\rh)
    -- (\s,0)
    -- (\s+\g,0)
    -- (\s+\g+\rs,\rh)
    -- (\s+\g+\rs+\st,\rh)
    -- (2*\s+\g,0);
    
    \draw[line width=0.6pt] (0,0) -- (2*\s+\g,0);
\end{tikzpicture}%
}

\newcommand{\twotriangleriblet}[3]{%
\begin{tikzpicture}[baseline=0.2ex, scale=0.8]
    \pgfmathsetmacro{\s}{#1}      % base width
    \pgfmathsetmacro{\alpha}{#2}  % tip angle
    \pgfmathsetmacro{\g}{#3}      % valley width
    
    \pgfmathsetmacro{\h}{\s/(2*tan(\alpha/2))}
    
    \fill[gray!20]
    (0,0)
    -- (\s/2,\h)
    -- (\s,0)
    -- (\s+\g,0)
    -- (\s+\g+\s/2,\h)
    -- (2*\s+\g,0)
    -- (2*\s+\g,0.8)
    -- (0,0.8)
    -- cycle;
    
    \draw[line width=0.6pt]
    (0,0)
    -- (\s/2,\h)
    -- (\s,0)
    -- (\s+\g,0)
    -- (\s+\g+\s/2,\h)
    -- (2*\s+\g,0);
    
    \draw[line width=0.6pt] (0,0) -- (2*\s+\g,0);
\end{tikzpicture}%
}

\newcommand \blacktri {\tikztriangle[black, fill=none]{0.4pt}}

\newcommand \blacksqr {\tikzsquare[black,fill=none]{0.5pt}}

\newcommand \blackcirc{\tikzcircle[black,fill=none]{0.5pt}}

\newcommand \bluecirc {\tikzcircle[blue,fill=none]{0.5pt}}
\newcommand \greycirc {\tikzcircle[black!50,fill=none]{0.5pt}}

\definecolor{darkgreen}{RGB}{0,128,0}

\begin{document}
% \linenumbers
% \maketitle
% \thispagestyle{empty}

% %%%%%%%%%%%%%%%%%%%%
% % Paper text
% %%%%%%%%%%%%%%%%%%%%

%%% Insert here the abstract %%%

\begin{frontmatter}

%% Title, authors and addresses

%% use the tnoteref command within \title for footnotes;
%% use the tnotetext command for theassociated footnote;
%% use the fnref command within \author or \affiliation for footnotes;
%% use the fntext command for theassociated footnote;
%% use the corref command within \author for corresponding author footnotes;
%% use the cortext command for theassociated footnote;
%% use the ead command for the email address,
%% and the form \ead[url] for the home page:
%% \title{Title\tnoteref{label1}}
%% \tnotetext[label1]{}
%% \author{Name\corref{cor1}\fnref{label2}}
%% \ead{email address}
%% \ead[url]{home page}
%% \fntext[label2]{}
%% \cortext[cor1]{}
%% \affiliation{organization={},
%%             addressline={},
%%             city={},
%%             postcode={},
%%             state={},
%%             country={}}
%% \fntext[label3]{}

% \title{Simulations of zero pressure-gradient boundary layers \ar{over} riblets}

% \title{Direct numerical simulations of zero pressure-gradient boundary layers \ar{over} riblets with step change in surface texture}

% \title{Direct numerical simulations of zero pressure-gradient boundary layers over riblets: analyzing non-equilibrium effects associated with step change in surface texture}

\title{Non-equilibrium effects in turbulent boundary layers over riblets: DNS of step changes in surface texture}

%% use optional labels to link authors explicitly to addresses:
\author[bsc]{Vishal Kumar}
\author[unimel]{Melissa Kozul}
\author[olemiss]{Wen Wu}
\author[bsc]{Oriol Lehmkuhl}
\author[ntu]{Amirreza Rouhi}
\affiliation[bsc]{organization={Barcelona Supercomputing Center},
            addressline={Plaça d'Eusebi G\"uell, 1-3},
            city={Barcelona},
            postcode={08034},
            % state={Catalonia},
            country={Spain}}

\affiliation[unimel]{organization={The University of Melbourne},
            addressline={Grattan Street},
            city={Parkville},
            postcode={3010},
            state={Victoria},
            country={Australia}}            

\affiliation[olemiss]{organization={The University of Mississippi},
            addressline={1764 University Circle},
            city={University},
            postcode={38677},
            state={MS},
            country={USA}}  

\affiliation[ntu]{organization={Nottingham Trent University},
            addressline={Clifton Ln},
            city={Nottingham},
            postcode={NG11 8NS},
            % state={Nottinghamshire},
            country={UK}}            

% \author{} %% Author name

% %% Author affiliation
% \affiliation{organization={},%Department and Organization
%             addressline={}, 
%             city={},
%             postcode={}, 
%             state={},
%             country={}}

\begin{comment}
\end{comment}

\begin{abstract}
  We computationally study the response of zero-pressure-gradient (ZPG) turbulent boundary layers (TBLs) to streamwise step changes from a smooth wall to riblets (SM\_RI), and vice versa (RI\_SM). To quantify the departure from equilibrium due to the step changes, we conduct reference calculations of ZPG TBLs over an entirely smooth wall, and an entirely riblet-covered surface.
  %In addition to the step change textures, we also simulate underlying smooth and fully riblet surface textures for comparison purposes. 
  % We consider three variations of triangular riblets for R--S: tip angle $\alpha=90^\circ$ and viscous spacing, $s^+_0=50$ (T950),  $\alpha=60^\circ, \, s^+_0=15$ (T615) and $\alpha=60^\circ, \, s^+_0=50$ (T650); $s_0^+$ is the viscous spacing at a reference location. For S--R, we consider T950 riblet. 
  %We use recently proposed $\eta-$grid meshing approach of \textbf{Rouhi et. al.} to generate optimal meshes
 %  We generate a ZPG TBL upstream of the step change (with thickness $\delta_0$) with momentum thickness Reynolds number $Re_{\theta_0} \simeq 680$.
 % near wall recovery of small scale fluctuations and is complete by $x \simeq 100k$ ($k$ is the riblet height). IEL growth rate in this stage scales as $\sim (x/k)^{0.6}$ and is substantially higher than reported rough-to-smooth growth rates in the literature.
  % persistent history effects within the frozen wake region. Pre-multiplied energy spectra reveals that this phenomenon of frozen wake can be attributed to the slow adjustment of the intermediate scales of turbulence.
  To save the computational cost, we generate an optimal grid for an unstructured spectral-element code, consistent with the size of turbulent scales across the TBL. By the step change, the momentum thickness Reynolds number reaches $Re_{\theta_0} \simeq 680$ (friction Reynolds number $Re_{\tau_0} \simeq 283$), and by the domain outlet downstream of the step change, $Re_{\theta} \simeq 1000$ ($Re_{\tau} \simeq 400$).
   The TBL departure from equilibrium due to the step change, and its subsequent relaxation, recall previous studies on step changes in surface roughness. Downstream of the step change, growth of the internal equilibrium layer thickness $\delta_\text{IEL}$, hence recovery to equilibrium, follows two stages. Stage I corresponds to the recovery up to the buffer region ($y^+ \simeq 10$), which is slower during the RI\_SM step change than the SM\_RI counterpart. For the RI\_SM cases during Stage I, $\delta_\text{IEL} \propto (x/k)^{0.6}$, and this stage is completed by $x \simeq 100k \simeq (5\delta_0 - 20\delta_0)$ downstream of the step change, where $k$ is the riblet height. Stage II recovery i.e.\ recovery of the outer region, is quite slow. Therefore, for drag-increasing riblets with $k^+ \ge 25$, $\delta_\text{IEL}$ does not reach the boundary layer thickness, even up to $50\delta_0$ downstream of the step change, owing to the advected frozen wake from upstream. As a result, skin-friction coefficient reaches to more than $90\%$ of its equilibrium counterpart, but does not reach its $100\%$.
\end{abstract}

%%Graphical abstract
\begin{graphicalabstract}
\includegraphics[width=1\linewidth]{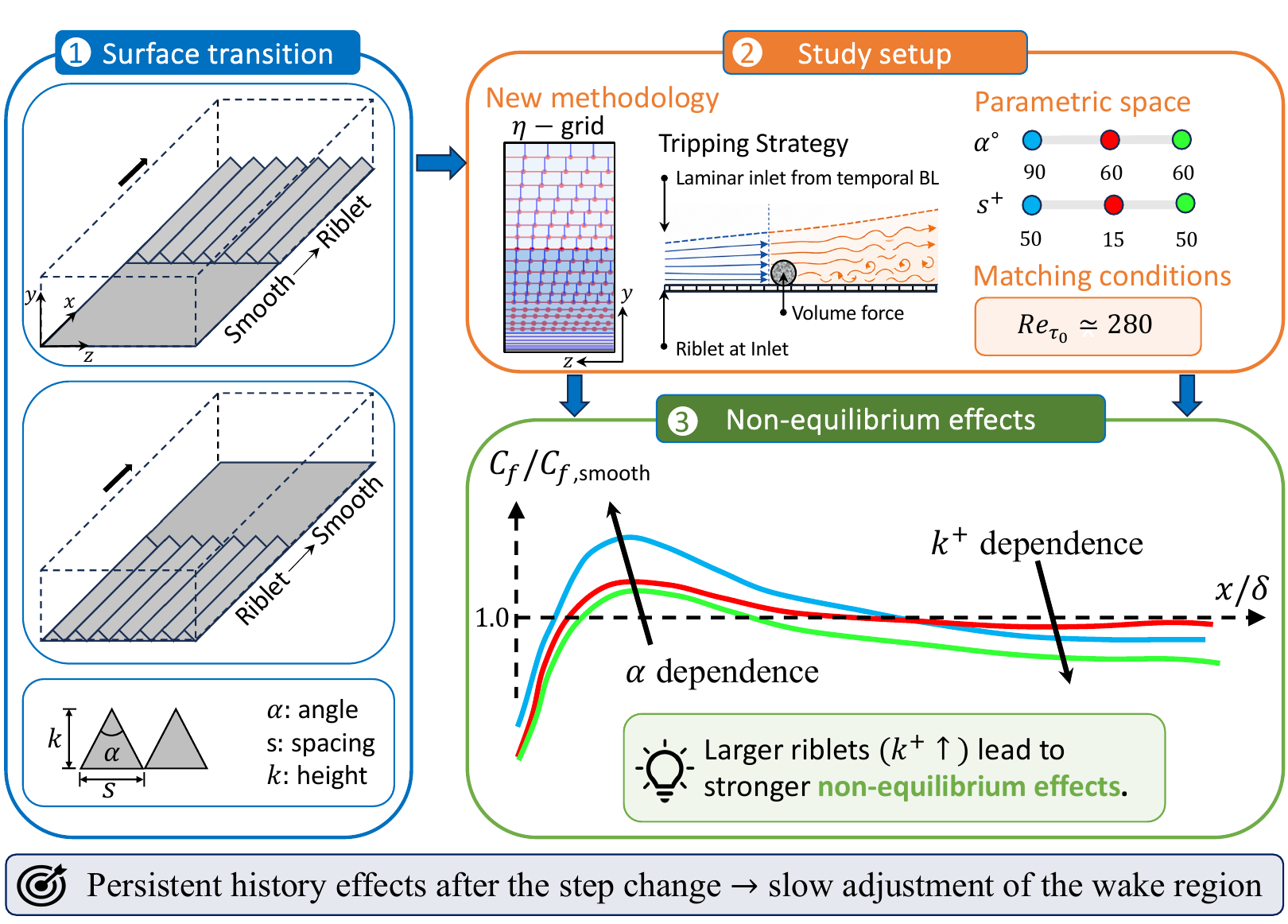}
\end{graphicalabstract}

%%Research highlights
\begin{highlights}
\item Direct numerical simulation of turbulent boundary layers up to $Re_\tau \simeq 400$
\item Comparative study of smooth, riblet, and heterogeneous surface transitions
\item Physics-informed, $\eta-$grid generation strategy for \textit{optimal} mesh design 
\item Balanced inflow generation strategy for smooth and riblet configurations
\item Quantification of non-equilibrium effects induced by surface transitions
\end{highlights}

%% Keywords
\begin{keyword}
%% keywords here, in the form: keyword \sep keyword

%% PACS codes here, in the form: \PACS code \sep code

%% MSC codes here, in the form: \MSC code \sep code
%% or \MSC[2008] code \sep code (2000 is the default)

    Direct numerical simulations    \sep 
    turbulent boundary layer        \sep
    triangular riblets              \sep
    spectral element method         \sep
    $\eta-$grid unstructured mesh   \sep
    surface texture step change     \sep
    % BL tripping                     \sep
    % Rough-to-Smooth step            \sep
    % Smooth-to-Rough step            \sep
    internal equilibrium layer      
    
\end{keyword}

\end{frontmatter}
%%%  Insert here the actual article text %%% 
\begin{comment}
\end{comment}

\section{Introduction} %%%%%%%%%%%%%%%%%%

The pioneering work of Walsh and co-workers at the NASA Langley research center firmly established the drag reducing properties of streamwise aligned surface grooves, termed riblets \citep{Walsh1979DragSurfaces, WALSH1982TurbulentRiblets, Walsh1990ViscousLayers}. Various geometries, including triangular, sinusoidal, and U-shaped riblets, were examined and it was shown that the size of the riblets should be of the order of the viscous-sublayer thickness ($\lesssim$ 15 wall units) to produce a net drag reduction. During the past four decades, numerous studies have investigated the physical mechanisms underlying riblet-induced drag modification in wall-bounded flows. Several review articles summarize the effects of riblet geometry on skin-friction drag and associated flow physics \cite{Tardu1995CoherentRiblets, Garcia-Mayoral2011}. 

%---------------------------------------------%
% \begin{sidewaystable}
\begin{table}[!h]%
  \centering
  % \begin{tabular}{ m{2cm} m{2cm} m{2cm} }
  \begin{tabular}{
  @{}
  @{\extracolsep{\fill}}
  >{\centering\arraybackslash}p{0.15\textwidth} 
  >{\centering\arraybackslash}p{0.05\textwidth} 
  >{\centering\arraybackslash}p{0.5\textwidth} 
  >{\centering\arraybackslash}p{0.2\textwidth} 
  @{}
  }
    \toprule
    Study  & Shape & Setup details & \\
    \midrule
    & \multicolumn{3}{c}{\textbf{Experimental studies ($\spadesuit$ HWA; $\heartsuit$ PIV)}} \\
    \cmidrule{2-4}
    &  &  & Inflow \\
    $^\spadesuit$\cite{Choi1989Near-wallRiblets} &
    \twotrapezoidriblet{0.4}{40}{0.12} &
    $Re_\tau \approx 1.2 \times 10^3$; 
    $(k^+,s^+) = (13,20)$ &   
    NA
          \\
    $^\spadesuit$\cite{Baron1993SomeLayer}  &
    \triangleriblet{0.3}{53} &
    $Re_\tau \approx  400$ ,
    $(k^+,s^+) = (12,12)$
            &
            Trip
          \\ 
    $^\spadesuit$\cite{Park1994FlowLayer} & 
    \triangleriblet{0.3}{90} &
    $Re_\tau \approx  (833,966)$ ,
    $(k^+,s^+) = (14, 28)$
            &
            Trip               
          \\ 
    $^\spadesuit$\cite{Choi1997TurbulenceTransfer} &
    \triangleriblet{0.3}{53}        &
    $Re_\tau \approx 400$         ,
    $s^+ \in (5, 60)$               &
    Trip                        
          \\
    $^\heartsuit$\cite{Lee2001FlowSurface} & 
    \scallopriblet{0.3}{0.4} &
    $Re_\theta \approx (2340, 4950)$,
    $s^+ \approx (25, 40)$    &
    3D-Trip                 
    \\
    $^\heartsuit$\cite{AbuRowin2025ExperimentalSurfaces}     &
    \triangleriblet{0.3}{30} &
    $Re_\tau \in [850, \, 2500]$, $s^+ \in (27, 113)$      &
    P40-Grit                                             
    \\    
    \cmidrule{2-4}     
    & \multicolumn{3}{c}{\textbf{Computational studies ($\spadesuit$ DNS; $\heartsuit$ WRLES)}} \\
    \cmidrule{2-4} 
     &  &  & SG/IBM/Inflow \\
    $^\spadesuit$\cite{STRAND2011DirectSpots}             & 
    \triangleriblet{0.3}{55} &
    \shortstack{
        $Re_{\delta^*} \approx 380$, $s^{+} \approx 22$
    } &
    \CheckmarkBold/\CheckmarkBold/Lam 
    \\
    $^\spadesuit$\cite{Wang2022OnReduction}                  &
    \scallopriblet{0.3}{0.4} & 
    $Re_{\delta} \approx 180$, $s^{+} \approx (20,60)$
    &
    \CheckmarkBold/\XSolidBrush/Lam
    \\
    $^\heartsuit$\cite{Bannier2015RibletIdentity}        & 
    \twotriangleriblet{0.3}{30}{0.32} &
    \shortstack{
        % $h/s=0.5,\theta=30^\circ$ \\
        $Re_\tau \simeq \left\{ 240,500 \right\}$,      
        $k^+ \simeq \left\{ 8,1,7.5\right\}$      
    } &
    \CheckmarkBold/\XSolidBrush/SyEM 
    \\
    $^\heartsuit$\cite{Boomsma2015}                  &
    \scallopriblet{0.3}{0.4} &
    $h/s=0.5$, $Re_\tau^* \approx 450$, $s^+ \in [10,27] $
    & 
    \CheckmarkBold/\XSolidBrush/R\&R 
    \\     
    $^\heartsuit$\cite{MalathiAnanth2023RibletNumbers}                  &
    \triangleriblet{0.3}{60} & 
    $Re_{\tau}^* \approx 400$, $s^{+} \in (14,18)$
    & 
    \CheckmarkBold/\CheckmarkBold/SyEM 
    \\
    \bottomrule
  \end{tabular}
    \caption{ Zero-pressure gradient turbulent boundary layer studies over riblets; momentum-thickness Reynolds number $(Re_\theta)$; friction Reynolds number $(Re_\tau)$,  following acronyms have been used: HWA: hot-wire anemometry; PIV: particle image velocimetry; DNS: direct numerical simulation; LES: large-eddy simulation; SG: structured grid; IBM: immersed boundary method; SyEM: synthetic-eddy method; R\&R: recycling and rescaling.}
\label{tab:studies_zpg_riblets}
\end{table}
% \end{sidewaystable}
%---------------------------------------------%

Most experimental studies on riblets have predominantly considered zero-pressure-gradient (ZPG) turbulent boundary layers (TBLs) (table~\ref{tab:studies_zpg_riblets}). In contrast, computational investigations focus heavily on the canonical turbulent channel configuration \cite{Choi1993,Garcia-Mayoral2011HydrodynamicRiblets, Endrikat2021, Modesti2021, Rouhi2022Riblet-generatedAnalogy, zhdanov2024influence, zhdanov2024net}; resolved numerical simulations of TBLs over riblets are limited (table~\ref{tab:studies_zpg_riblets}). Some direct numerical simulations (DNSs) of TBLs over riblets have focused on instability growth rates in transitional boundary layers at low Reynolds numbers (\cite{STRAND2011DirectSpots,Wang2022OnReduction} in table~\ref{tab:studies_zpg_riblets}). The reason for the limited DNSs of TBLs over riblets can be traced to their massive computational requirements. In particular, the computational hurdle stems from structured meshes as widely adopted by the previous studies (see table~\ref{tab:studies_zpg_riblets}, last column). Most studies in table~\ref{tab:studies_zpg_riblets} employ Cartesian grids with riblets implements via an immersed boundary method (IBM). Well resolving riblets with viscous-scaled spacing $s^+ \sim \mathcal{O}(10)$ requires spanwise grid sizes $\Delta z^+ \simeq 0.3 - 0.5$ ($20 - 30$ grid points per $s^+$). With Cartesian grids and IBM, such stringent $\Delta z^+$ will be generated across the domain, hence over-resolving the TBL in the outer region. In addition, the use of IBM to represent sharp featured riblets is not ideal for resolving the physics at the riblet peaks and valleys. Historically, the dependence on structured meshes has been driven by the need for high-order numerical schemes, which were mainly implemented in in-house structured-grid solvers. This requirement is no longer an area of concern, as several high-order open-source codes that utilize unstructured meshes are available, and optimal meshes for wall-bounded flows can be designed readily. Some work has been reported in the literature in this regard. \citet{Rai1993DirectLayer} proposed the use of zonal embedded meshes (non-conformal meshes with varying $\Delta x^+, \Delta z^+$ with wall normal distance, $y$) to study wall turbulence. \citet{Kravchenko1996ZonalFlows} reported significant computational and memory savings with the use of zonal meshes for ZPG TBLs.  \citet{Pirozzoli2021NaturalFlows} proposed a `natural' wall-normal grid stretching $\Delta y^+$ proportional to the local Kolmogorov scale $\eta^+$, rather than relying on conventional grid stretching, such as hyperbolic tangent or cosine mapping. Recently \cite{Rouhi2025}, we extended this idea to simultaneous coarsen $\Delta y^+$ and $\Delta z^+$ with the wall-normal distance. This idea was implemented via multi-block unstructured grid generation in $yz$-planes, and was tested with finite-volume method and spectral element method solvers. The grid performed as accurate as a Cartesian grid in its application to turbulent channel flow and TBLs over smooth wall and riblets. However, the number of grid points by the proposed $\eta$-based grid could drop to $10\%$ of the one by a Cartesian grid over a smooth wall, and could drop to $3\%$ over riblets. In the present study, we follow this grid-generation approach to afford for DNSs of TBLs over riblets.   

A challenge associated with TBLs over riblets is generating turbulent inflow. The setups of the numerical and experimental works in table~\ref{tab:studies_zpg_riblets} consist of a smooth wall at the inlet, followed by a surface change to riblets. This allows to apply an inlet condition for smooth surfaces, which is more straightforward than generating an inflow over riblets. Although experiments employ some form of flow tripping, numerical studies typically generate a turbulent inflow using techniques developed for smooth walls, such as recycling-rescaling methods \citep{Lund1998} or synthetic eddy methods \citep{Pamies2009}. The surface change from smooth to riblets, alters the equilibrium state of the TBL, which requires some downstream distance to recover. One way to minimize introducing such non-equilibrium effects is to utilize a smoothed `ramp' to connect the smooth and the riblet patches. This strategy was used by experimental \citep{Park1994FlowLayer} and numerical \citep{MalathiAnanth2023RibletNumbers, SavinoAttachedRiblets} works. \citet{MalathiAnanth2023RibletNumbers} noted losses due to uplifting of the shear layer over the sharp riblet. However, detailed analysis of the non-equilibrium effects of smooth-to-riblet step change and associated impact on boundary-layer statistics were not reported. Another way to introduce turbulent inlet is to trip the laminar boundary layer developing on the riblet section itself. To the best of our knowledge, this strategy has not been attempted before.

Abrupt changes in wall texture, for example smooth-to-roughness and vise-versa, trigger a localized but persistent perturbation of the flow, leading to its departure from equilibrium \cite{Antonia_1971}. The recovery of the flow to its new equilibrium state downstream of the step change depends on various parameters, such as the geometrical characteristics of the surface, the Reynolds number, and the quantity of interest \cite{Antonia_1971, Rouhi2019, Li2019}. The flow recovery to equilibrium is a bottom-up process from the surface and is quantified via either the internal boundary layer (IBL), or the internal equilibrium layer (IEL). The IBL is the region of the flow that `feels' the presence of the new wall condition, but still preserves some history effects from the upstream surface \cite{Garratt1990}. An inner part of the IBL, where the flow has reached complete equilibrium with the new surface underneath is IEL \cite{Savelyev2005}.

%---------------------------------------------%
% \begin{sidewaystable}
\begin{table}[!h]%
  \centering
  % \setlength{\tabcolsep}{3pt}
  % \begin{tabular}{ m{2cm} m{2cm} m{2cm} }
  % \begin{tabular}{ c p{1cm} p{1cm} p{1cm} p{1cm} c c p{4cm} }
    \begin{tabular}{ 
  @{}
  @{\extracolsep{\fill}}
  >{\centering\arraybackslash}p{1.8cm} 
  >{\centering\arraybackslash}p{1cm} 
  >{\centering\arraybackslash}p{1.8cm}  
  >{\centering\arraybackslash}p{1cm}
  >{\centering\arraybackslash}p{1cm}
  >{\centering\arraybackslash}p{0.12\textwidth}
  >{\centering\arraybackslash}p{0.19\textwidth}
  @{}
  }
    \toprule
    Study & Shape & $Re_\tau$ & 
    $k/\delta$ & $k^+$ & R-S/S-R  & 
    \raisebox{-0.25\height}{
    \shortstack{
        $\delta_i (\delta_e) \propto x^{\alpha}$
        }
    }\\
    \midrule
    & \multicolumn{6}{c}{\textbf{Experimental studies ($\spadesuit$ HWA; $\heartsuit$ PIV; $\clubsuit$ LDV)}} \\
    \cmidrule{2-7}
    $^\spadesuit$\cite{Antonia1971TheRoughness} &
    % \cite{Antonia1971TheRoughness, Antonia_1971, Antonia_1972} 
    % \squareriblets{0.2}{0.4}{0.32} 
    \raisebox{-.25\height}{\includegraphics[width=0.1\textwidth]{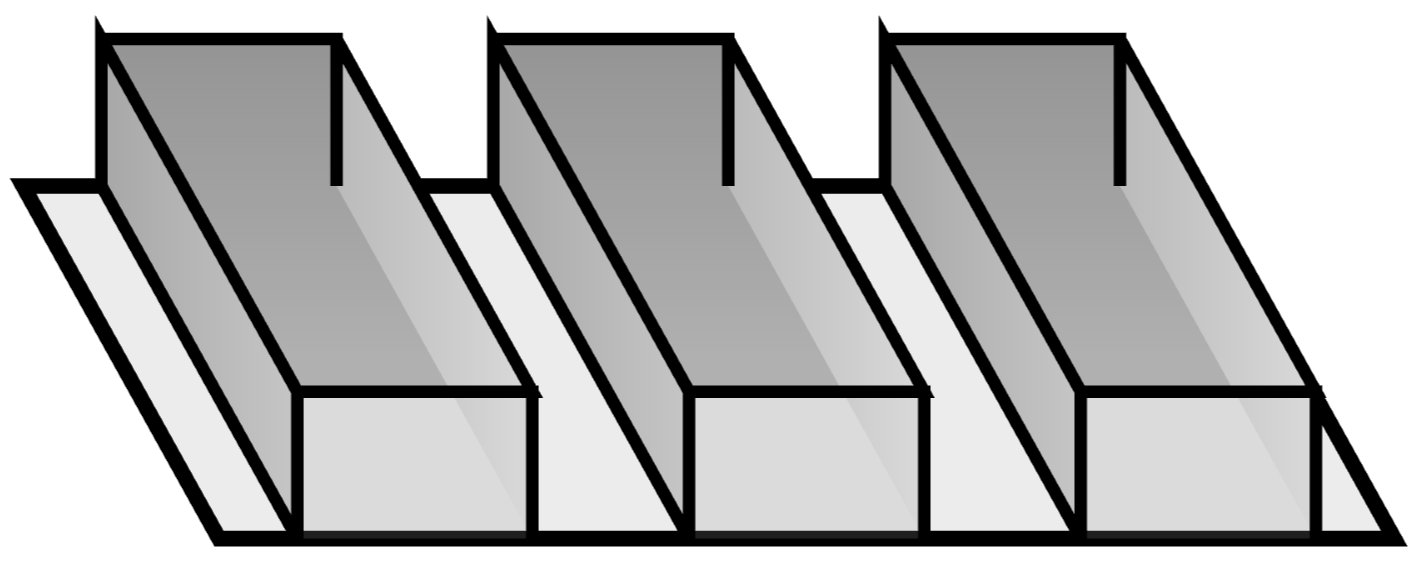}} &
    $790$         &
    $ 6.0\%$      & 
    $ 47$         &   
    \raisebox{-.25\height}{
    \includegraphics[scale=0.5]{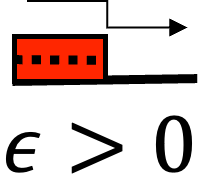}
    } &
    $0.5$
    \\
    \cmidrule{2-7}
    $^\spadesuit$\cite{Antonia_1971, Antonia_1972} &
    % \squareriblets{0.2}{0.4}{0.32} 
    \raisebox{-.25\height}{\includegraphics[width=0.1\textwidth]{etmm_figs/rib_rough.png}} &
    % \staggeredCuboidArray{0.1}{0.2}{0.1}{0.3} 
    $\color{red} 1030, 1800$, $\color{blue} 790, 1210$         & 
    $ 4.5\%$      & 
    $\color{red} 46, 80$, $\color{blue} 35, 53$         &   
    \raisebox{-.25\height}{
    \includegraphics[scale=0.5]{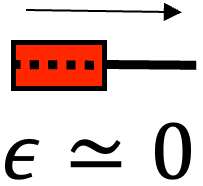}, 
    \includegraphics[scale=0.5]{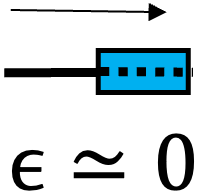}
    } &
    $\color{red}{0.43}$, $\color{blue}{0.72 - 0.79}$
     \\
    \cmidrule{2-7}
    $^\spadesuit$\cite{Cheng_2002} & 
    \raisebox{-.25\height}{\includegraphics[width=0.1\textwidth]{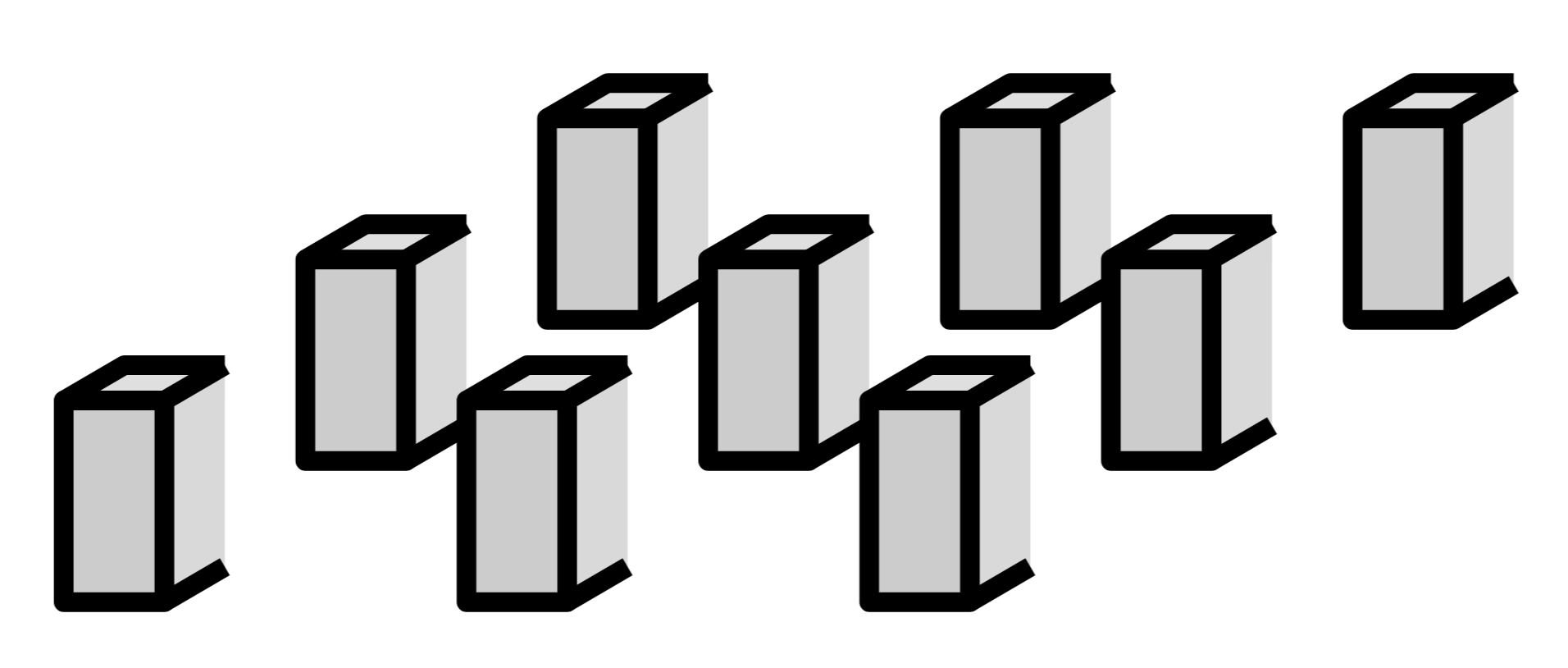}}
    \raisebox{-.25\height}{\includegraphics[width=0.1\textwidth]{etmm_figs/rib_rough.png}} &
    $ 3100$, $4200$                &
    $ 1\%$                &
    $ 31, 42$                   &   
    \raisebox{-.25\height}{\includegraphics[scale=0.5]{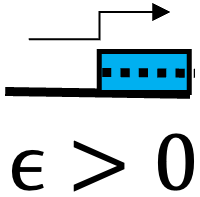}} &
    \shortstack{ 
        $\color{blue}{0.33}$ ($\color{SkyBlue}{0.37}$)
    }
    \\      
    \cmidrule{2-7}
    $^{\spadesuit\heartsuit}$\cite{Carper2007Subfilter-scaleRates} & 
    \raisebox{-.25\height}{\includegraphics[width=0.1\textwidth]{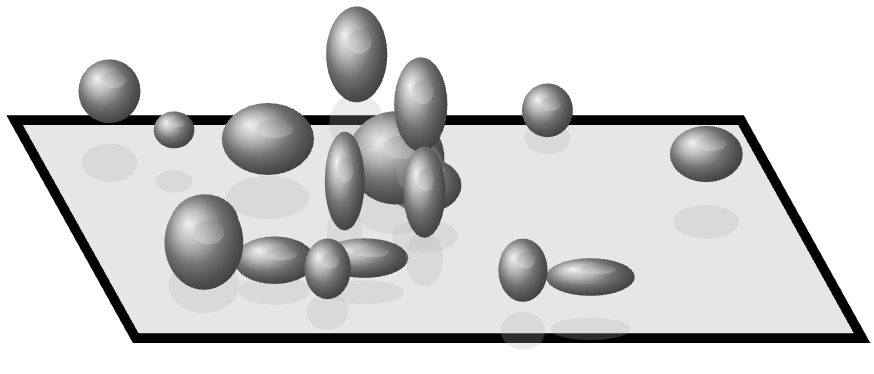}} &
    $ 8800$           &
    $ 0.8\% $        & 
    $ 70$                 &   
    \raisebox{-.25\height}{
    \includegraphics[scale=0.5]{etmm_figs/R-S-2.png}
    } &
    \shortstack{
        $0.6$
    }          \\    
    \cmidrule{2-7}
    $^\clubsuit$\cite{Efros2011DevelopmentSurface} & 
    \raisebox{-.25\height}{\includegraphics[width=0.1\textwidth]{etmm_figs/rib_rough.png}} &
    $ 4200$              & 
    $ 1.3\%$            & 
    $ 120$                &   
    \raisebox{-.25\height}{\includegraphics[scale=0.5]{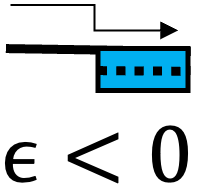}} &
    \shortstack{ 
        $0.73$
    }
    \\    
    \cmidrule{2-7}
    $^{\spadesuit\heartsuit}$\cite{Jacobi2011NewLayer} & 
    \raisebox{-.25\height}{\includegraphics[width=0.1\textwidth]{etmm_figs/rib_rough.png}} &
    $ 910$          &
    $ 5.8\%$            & 
    $ 53$                 &   
    \raisebox{-.25\height}{
    \includegraphics[scale=0.5]{etmm_figs/R-S-2.png}, 
    \includegraphics[scale=0.5]{etmm_figs/S-R-2.png}
    } &
    \shortstack{ 
        $\color{red}{0.2}$, $\color{blue}{0.2}$
    } \\    
    \cmidrule{2-7}
    & \multicolumn{6}{c}{\textbf{Computational studies ($\spadesuit$ DNS; $\heartsuit$ WRLES)}} \\
    \cmidrule{2-7} 
    $^\spadesuit$\cite{Lee2015TurbulentSurface}             & 
    \raisebox{-.25\height}{
        \includegraphics[width=0.1\textwidth]{etmm_figs/rib_rough.png}
    } &             
    $ 710$          &             
    $ 4.5\%$            & 
    $ 32$                 &
    \raisebox{-.25\height}{\includegraphics[scale=0.5]{etmm_figs/S-R-2.png}} &
    \shortstack{ 
        $\color{blue}{0.22}$ ($\color{SkyBlue}{0.38}$)
    } 
    \\
    \cmidrule{2-7}
    $^\spadesuit$\cite{Ismail2018}             & 
    \raisebox{-.25\height}{\includegraphics[width=0.1\textwidth]{etmm_figs/rib_rough.png}} &                                     
    $ 1115$                        &             
    $8.3\%$       &
    $ 185$                         &
    \raisebox{-.25\height}{
    \includegraphics[scale=0.5]{etmm_figs/R-S-2.png}
    } &
    \shortstack{ 
        $0.41$
    }
    \\
    \cmidrule{2-7}
    $^\spadesuit$\cite{Rouhi2019}             & 
    \raisebox{-.25\height}{\includegraphics[width=0.1\textwidth]{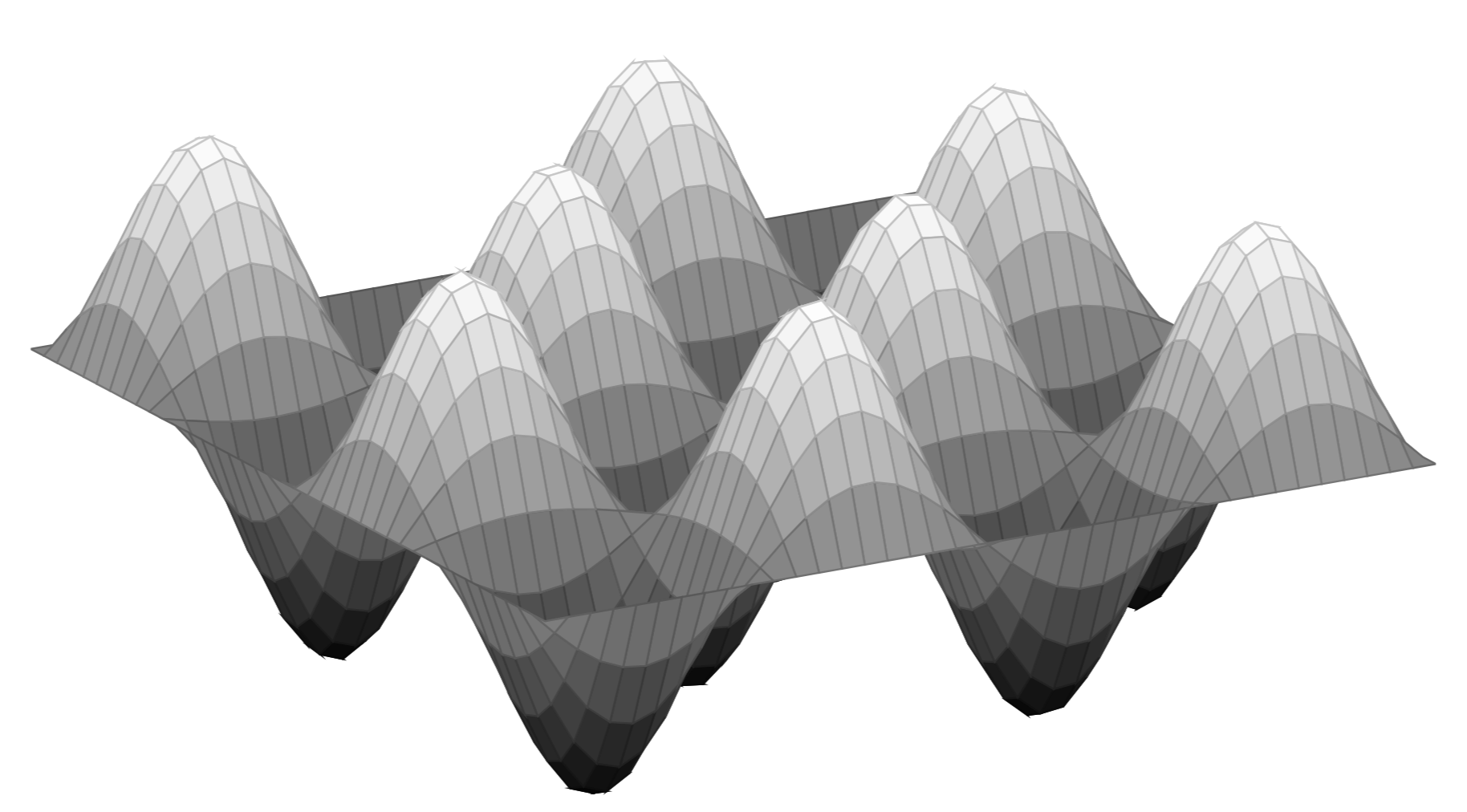}} & 
     $\color{red}690$, $\color{blue}430$          &
    $ 5.6\%$               & 
    $\color{red} 39$,$\color{blue}24$             &
    \raisebox{-.25\height}{
    \includegraphics[scale=0.5]{etmm_figs/R-S-1.png}, 
    \includegraphics[scale=0.5]{etmm_figs/S-R-1.png}
    } &
    \shortstack{
        $\color{red}{0.38}$, $\color{blue}{0.58}$
    }
          \\
    \cmidrule{2-7}
    $^\heartsuit$\cite{Li2022OnRoughness}             &
    \raisebox{-.5\height}{\includegraphics[width=0.1\textwidth]{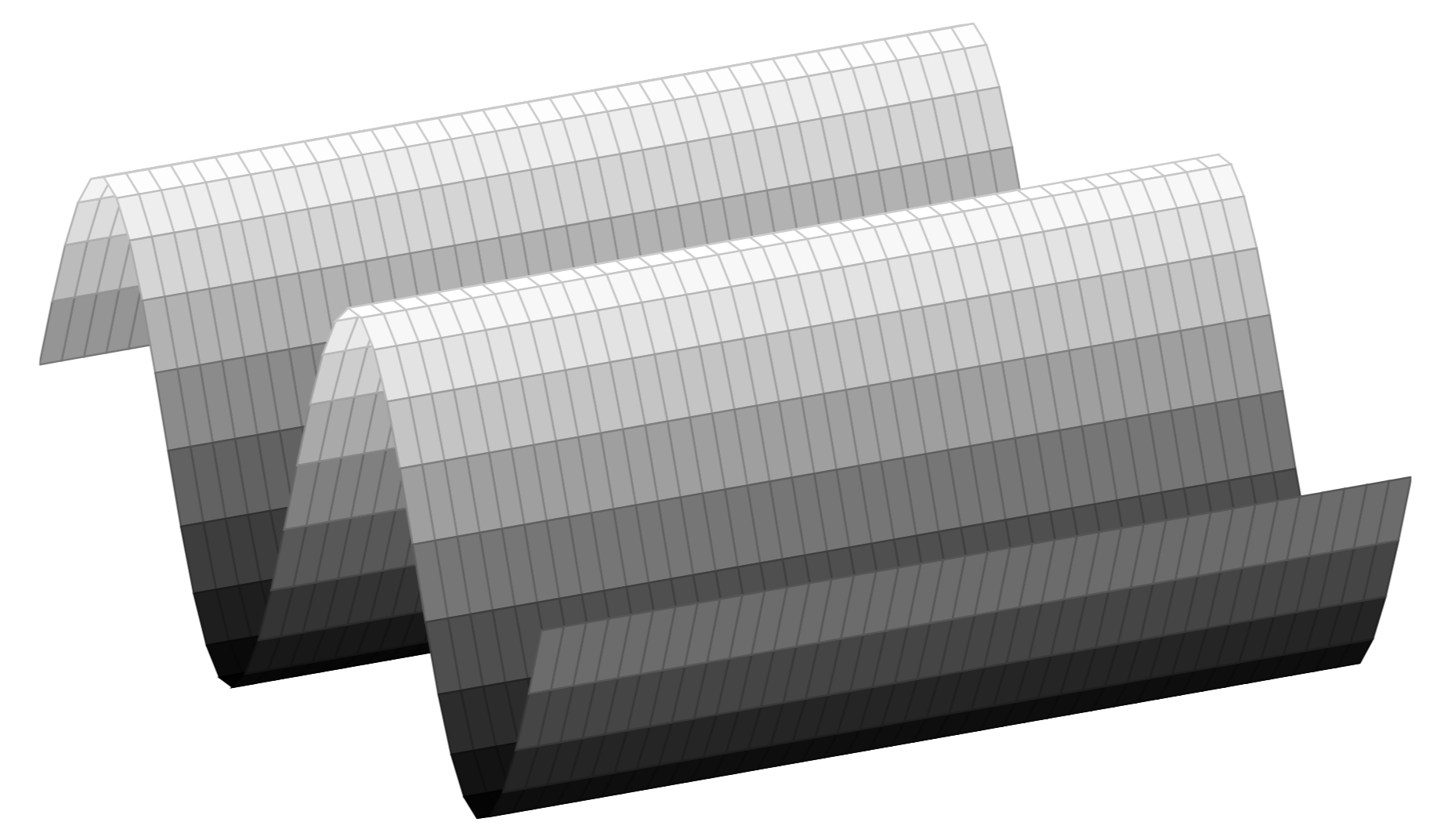}} &     
    $\color{red}680$, $\color{blue}570$               &             
    $ 3.3\%$                  & 
    $\color{red}23$,$\color{blue}18$                  &
    \raisebox{-.25\height}{
    \includegraphics[scale=0.5]{etmm_figs/R-S-2.png}, 
    \includegraphics[scale=0.5]{etmm_figs/S-R-2.png}
    } &
    \shortstack{
        $\color{red}{0.25}$, $\color{blue}{0.25}$
    }
    \\
    \cmidrule{2-7}
    $^\spadesuit$\cite{Cogo2025SurfaceLayers}             &
    \raisebox{-.25\height}{\includegraphics[width=0.1\textwidth]{etmm_figs/bars_rough.png}} &                                
    $ 1500$               &               
    $ 3.6\%$                 & 
    $ 56$                     &
    \raisebox{-.25\height}{
    \includegraphics[scale=0.5]{etmm_figs/S-R-2.png}
    } &
    \shortstack{
        $0.42$ -- $0.62$
    }
    \\     
    \bottomrule
  \end{tabular}
    \caption{\label{tab:studies_step_change} List of studies analyzing non-equilibrium effects associated with step changes in surface texture on zero-pressure gradient turbulent boundary layers; $Re_\tau = u_\tau \delta/\nu$ is reported at the step change; $k$ is the height of roughness; $\delta$ is boundary layer thickness at the step change; R-S (S-R) refers to rough-to-smooth (smooth-to-rough) step change; $\epsilon$ the height of the virtual origin of the rough surface; $\delta_i$ and $\delta_e$ denote IBL and IEL thicknesses, respectively. Entries are color coded: red represents R-to-S values and blue represents S-to-R values. $\delta_i$ growth rate is denote by dark colors and light colors are used for $\delta_e$ (in brackets).}

\end{table}
% \end{sidewaystable}
%---------------------------------------------%

Previous campaigns on quantifying non-equilibrium effects due to step changes have considered both smooth-to-rough and rough-to-smooth step changes (table~\ref{tab:studies_step_change}, first column). Different roughness shapes were investigated in experiments (table~\ref{tab:studies_step_change}, second column): square bars \citep{Antonia_1971, Antonia_1972, Efros2011DevelopmentSurface, Jacobi2011NewLayer}, grit roughness \citep{Hanson_2016} and mesh roughness \citep{Carper2007Subfilter-scaleRates, Chamorro2009VelocityModel, Hanson_2016}. Computational studies have utilized DNS \citep{Lee2015TurbulentSurface, Ismail2018, Rouhi2019}, wall-resolved LES \citep{Li2022OnRoughness} and wall-modeled LES \citep{Miller2013SurfaceTransitions, Lopes2015OnPatches} based strategies.
For resolved studies, two-dimensional geometries, such as square bars \citep{Lee2015TurbulentSurface, Ismail2018}, sine functions \citep{Li2022OnRoughness}, and three-dimensional geometries, such as square blocks \citep{Ismail2018TheFlows, Ismail2023, Cogo2025SurfaceLayers} and egg-carton \citep{Rouhi2019}, have been used as the rough surface.
The computational studies differ from the wind tunnel experiments in some aspects. First, numerical studies typically simulate flow in channels as opposed to boundary layer configurations used in the experiments. Second, wall-resolved simulations are conducted at friction Reynolds numbers $Re_\tau$, that are an order of magnitude lower than those in the experiments (table~\ref{tab:studies_step_change}, third column). Finally, the scale separation between the roughness size $k$ and the channel height $h$, or the TBL thickness $\delta$ is generally larger in the experiments (table~\ref{tab:studies_step_change}, fourth column).

The growth rate of IBL $(\delta_i)$ and IEL $(\delta_e)$ thickness over rough-to-smooth and smooth-to-rough step changes has received considerable attention (table~\ref{tab:studies_step_change}). It is generally accepted that the growth of $\delta_i$ downstream of the rough-to-smooth step change is slower compared to its counterpart downstream of the smooth-to-rough step change. However, when it comes to the growth rates, as quantified through a power $\alpha$ of the downstream distance $x$ from the step change, $\delta_i (\delta_e) \propto x^{\alpha}$, there is a large scatter in the literature. Early studies by Antonia \& Luxton~\cite{Antonia_1972,Antonia_1971} reported $\delta_i \propto x^{0.4-0.5}$ for a rough-to-smooth and $\delta_i \propto x^{0.7-0.8}$ for a smooth-to-rough step change. Later studies proposed lower growth rates for rough-to-smooth ($\delta_i \propto x^{0.2-0.5}$), as well as for smooth-to-rough ($\delta_i \propto x^{0.2-0.7}$) step changes (table~\ref{tab:studies_step_change}). Recently, \citet{Cogo2025SurfaceLayers} observed that the growth rate was not constant along the streamwise direction; they reported a lower growth rate of $\delta_e \propto x^{0.42}$ for $x \lesssim 2.5\delta$ and $\delta_e \propto x^{0.62}$ for $x \gtrsim 2.5\delta$ for the smooth-to-rough step change on a 3-D cubical roughness. The growth rate of IEL ($\delta_e$) is investigated to a lesser extent compared to IBL ($\delta_i$). Of the two studies that reported the growth rates of $\delta_e$ (table~\ref{tab:studies_step_change}, light blue color text), the findings are different: in \citet{Cheng_2002} the growth rates of $\delta_e$ and $\delta_i$ are close to each other, however, in \citet{Lee2015TurbulentSurface}, the growth rate of $\delta_e$ is almost two times of the one for $\delta_i$.

Among the studies compiled in table~\ref{tab:studies_step_change}, we observed noticeable disparity in the growth rates of IBL and IEL (last column). Rouhi et al.~\cite{Rouhi2019,Rouhi2019RoughnessChange} relate such disparity to differences in 1) flow configuration, i.e., channel vs boundary layer; 2) $\delta_i$ and $\delta_e$ definition; 3) Reynolds number; 4) height of the roughness as a fraction of the boundary layer thickness at the step change, $k/\delta$ (or $k/h$ for channels); 5) shape of the roughness; and 6)  virtual origin of the roughness, $\epsilon/\delta$ (or $\epsilon/h$ for channels). \citet{Rouhi2019} focused on (2) by analyzing several definitions of $\delta_i$ proposed in the literature. They concluded that the definition of $\delta_i$ by \citet{Elliott1958TheLayer} better measures the extent of the bottom wall influence on the turbulence characteristics. In a follow-up study, \citet{Rouhi2019RoughnessChange} investigated the other above-mentioned sources of disparities in $\delta_i$ (3,4,5,6); they found that the differences in the roughness origin relative to the smooth surface ($\epsilon/h$) has a considerably larger influence on the growth of $\delta_i$ compared to the roughness size $k/h$ or $Re_\tau$. To our knowledge, investigations of turbulent flows over step changes from smooth to non-smooth surfaces (and vice-versa) have focused on rough surfaces with significant pressure drag. The departure of turbulent flows from equilibrium due to riblet-to-smooth step change (and vice versa) is not investigated yet. Unlike roughness, riblets do not impose pressure drag; also, with spacing $s^+ \lesssim 50$, riblets marginally change the drag  compared to the smooth wall ($\lesssim 10\%$). However, this conclusion is valid for equilibrium turbulent flows over riblets. It is not clear how the drag, hence wall friction, evolves during riblet-to-smooth step change (and vice versa), and how this compares with the rough-to-smooth surface change (and vice versa).

With regard to the discussion above, we pursue two objectives in the present study. Our first objective is to develop an efficient computational methodology for systematic study of TBLs over riblets. This entails: a) utilizing an optimum grid-generation framework for body fitted unstructured solvers, and b) formulating a consistent inflow generation approach to achieve a ZPG TBL over riblets. Our second objective is to study the flow recovery to equilibrium past the step changes from smooth-to-riblets, and vise versa. Both objectives are interlinked: development of a consistent inflow generation enables us to systematically assess the non-equilibrium effects due to a sudden step change, as employed in the past experimental and computational investigations. After introducing our numerical solver ($\S$\ref{sec:method}), we comprehensively validate it ($\S$\ref{sec:code_verify}) for riblet cases in an open channel. Next, we turn to setting up the boundary layer cases. For this, we present a domain-length independence study ($\S$\ref{sec:domain_verify}) and a drag-force-based flow tripping study ($\S$\ref{sec:tripping_setup}). In $\S$\ref{sec:meshing_setup}, we extensively describe the key components of an unstructured meshing strategy that we propose for wall-bounded flows. Then, we present our results and look closely at the non-equilibrium effects of step change ($\S$\ref{sec:results}). Finally, the study is concluded in $\S$\ref{sec:conclusion}.

%%--------------------------------------------------%%

\section{Methodology \label{sec:method}} %%%%%%%%%%%%%%%%%%

\subsection{Governing equations and numerical discretization}
We solve the incompressible conservation of mass and momentum flow equations with constant density $\rho$ and kinematic viscosity $\nu$:
\begin{eqnarray}
    \nabla\boldsymbol{\cdot}\mathbf{u} &=& 0, \label{eq:cont} \\
 \frac{\partial \mathbf{u}}{\partial t} + \nabla \boldsymbol{\cdot}(\mathbf{uu}) &=& -\frac{1}{\rho}\nabla p + \nu \nabla^2 \mathbf{u}. \label{eq:mom}
\end{eqnarray}
Here, $\mathbf{u} = (u,v,w)$ is the velocity vector and $x,y,z$ are the streamwise, wall-normal, and spanwise directions, respectively. Equations~\ref{eq:cont} -- \ref{eq:mom} are solved using a well-validated computational solver (SOD2D) developed at the Barcelona Supercomputing Center. Spatial discretization is based on a spectral formulation of the continuous Galerkin finite element model, coupled with entropy-viscosity stabilization \citep{Guermond2011a}. Skew-symmetric splitting is used for the convective term to counter the aliasing effects \citep{Kennedy2008ReducedFluid}. We utilize a third-order hexahedral element based mesh for all our calculations. For temporal discretization, a BDF-EXT3 high-order operator splitting approach is used to solve velocity–pressure coupling, allowing equal order interpolation of velocity and pressure \citep{Karniadakis1991}. The conjugate gradient method is used to solve the Poisson equation for the residual pressure. The numerical details for SOD2D are reported in \citet{Gasparino2024} and \citet{Blanco-Casares2025Projection-basedSolvers}.

%%--------------------------------------------------%%
\begin{figure}[!h]
    \begin{center}
    \includegraphics*[width=1\linewidth]{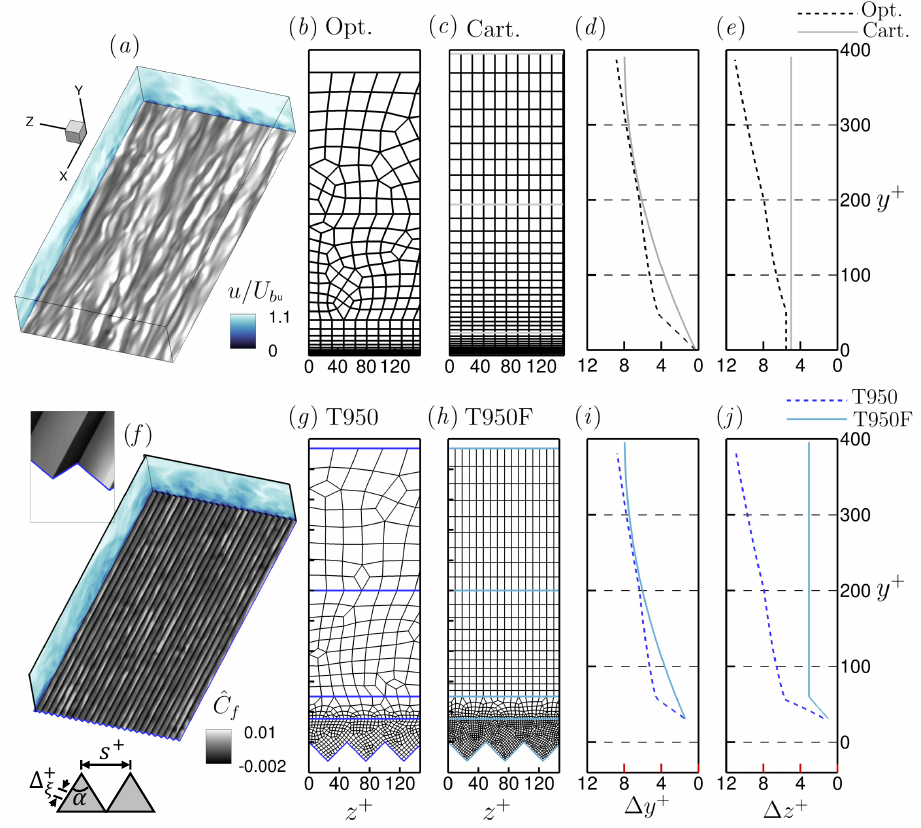}
	\caption{\label{fig:channel_setup} Computational setups and grids for turbulent channel flow over (\textit{a-e}) smooth wall, and (\textit{f-j}) riblets. Plots (\textit{a,f}) show the computational domains as periodic half-channel flow; on the bottom-left corner of (\textit{f}) we draw the riblet dimensions: riblet spacing $s^+$, tip angle $\alpha$. Plots (\textit{b,c}) show Optimal and Cartesian mesh (only elements are shown) for smooth channel, respectively; plots (\textit{g,h}) show Optimal mesh (only elements are shown) for T950 and 950F riblets, respectively. Plots (\textit{d,i}) show the average wall normal grid size $\Delta y^+$ and plots (\textit{e,j}) show the azimuthal grid size $\Delta_z^+$ for smooth and T950 cases.}
    \end{center}
\end{figure}
%%--------------------------------------------------------%%

\begin{table}[!h]
    \centering
    \setlength{\tabcolsep}{4.5pt}
    \begin{tabular}{ c c c c c c c c}
    \toprule
    & Case & $N_{dof}$ & $ \Delta^+_x$ & $\Delta^+_\xi$ & $\Delta^+_z$ & $\Delta^+_y$ & Leg.  \\ 
    \midrule
    \multirow{2}{*}{\fbox{\rotatebox{0}{\parbox{1.5cm}{\centering Smooth}}}}
    & Optimal       & 4.3 M & 10.0 & - & 5.4 -- 10.6 & 0.3 -- 9.4 & \tikzline[dashed]{black} \\
    & Cartesian  & 7.9 M & 10.0 & - & 4.9 & 0.3 -- 8.4 &  \tikzline{gray} \\
    \cmidrule{2-7}
    \multirow{2}{*}{\includegraphics[scale=0.3]{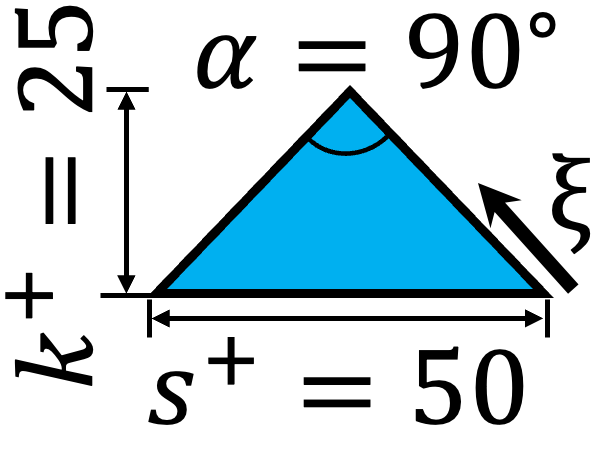}}    
    & T950  & 11.6 M &  $10.0$ & $1.47$ & $1.47 - 11.68$ & $1.47 - 8.91$  & \tikzline[dashed]{blue}  \\
    & T950F & 28.0 M &  $10.0$ & $0.98$ & $0.98 - 4.48$ & $0.98 - 7.86$ & \tikzline{Cyan} \\ 
    \multirow{2}{*}{\includegraphics[scale=0.3]{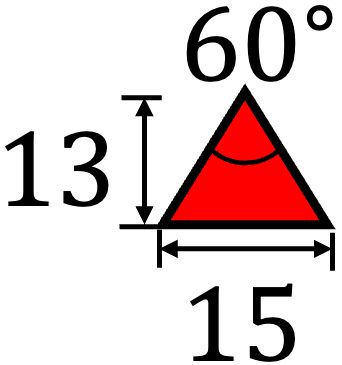}}    
    & T615  & 29.2 M &  $10.0$ & $0.54$ & $0.54 - 11.07$ & $0.54 - 8.36$  & \tikzline[dashed]{red}\\
    & T615F & 87.8 M &  $7.0$ & $0.38$ & $0.38 - 8.09$ & $0.38 - 5.16$ & \tikzline{orange}\\
    \bottomrule
    \end{tabular}
    \caption{\label{tab:channel} Simulation cases for turbulent half-channel flow over smooth and riblet surfaces; riblet shapes and sizes are shown for the current setup. $N_{dof}$ is total degrees of freedom; $\Delta x^+, \Delta y^+, \Delta z^+$ report average streamwise, wall-normal and spanwise spacings in viscous units; $\Delta^+_\xi$ is the viscous-scaled azimuthal grid size over the riblets. All calculation use $L_x \simeq 6h, L_z\simeq 3h$ (domain is adjusted to fit integer number of riblet wavelengths), with $h$ as the half-channel height. Line legends (Leg.) are included that will be used in figures~\ref{fig:channel_setup} and \ref{fig:channel}, where we present the setups and results.} 
\end{table}

\begin{figure}[!h]
	\begin{center}
    \includegraphics*[width=0.9\linewidth]{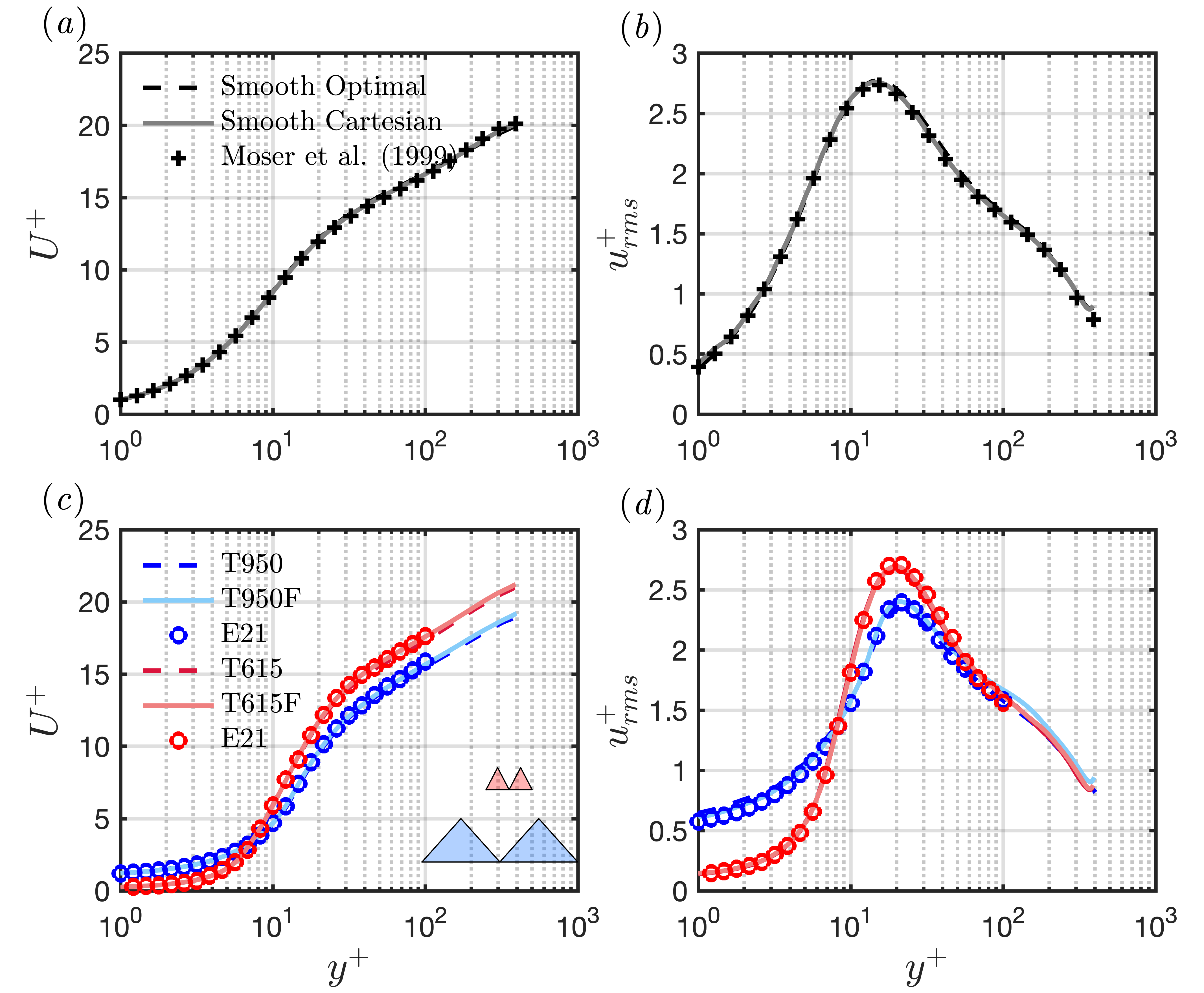}
	\caption{\label{fig:channel} Profiles of (a,c) mean velocity $U^+$, and (b,d) streamwise turbulent stress $\overline{u'^2}^+$ for turbulent channel flow over (a,b) smooth wall and (c,d) riblet cases from table~\ref{tab:channel}. For (a,b) Optimal (\tikzline[dashed]{black}) and Cartesian (\tikzline{gray}) meshes are shown. Reference data of \citet{Moser1999DirectRe=590} ($\boldsymbol{+}$) and \citet{Endrikat2021} (label E21; \blackcirc{}) are shown. For \citet{Endrikat2021}, data is included for the inner region ($y^+ < 100$), since they conducted simulations in minimal channels.}
	\end{center}
\end{figure}

%%%%%%%%%%%%%%%%%%%%%%%%%%%%%%%%%%%%%%%%%%%%%%%%%%%%%%
\subsection{Code verification \label{sec:code_verify}}
We report on comprehensive verification and validation tests of SOD2D for turbulent half-channel flow at $Re_\tau = u_\tau h/\nu \simeq 400$ over a smooth wall and over riblets. For turbulent flow over riblets, we replicate two DNS cases by \citet{Endrikat2021}.

\subsubsection{Turbulent half-channel flow (smooth wall) \label{sec:channel_smooth}}
We employ a Cartesian grid (figure~\ref{fig:channel_setup}c, table~\ref{tab:channel} second row), as well as an optimal grid (figure~\ref{fig:channel_setup}b, table~\ref{tab:channel} first row). The Cartesian grid is the conventional grid for smooth-wall channel flow; its major constraint is the uniform $\Delta z^+$ that must be maintained from the bottom up to the top. This leads to unnecessarily fine resolution in the outer region ($\sim$80\% of the domain height), hence unnecessary increase in the computational cost. In fact, the turbulent scales in the outer region could be well resolved by a several times coarser grid. With SOD2D, as an unstructured solver, we have the advantage of local grid refinement or coarsening. We exploit this advantage for the optimal grid and locally adjust the spanwise and wall-normal grid sizes following the formulation by \citet{Rouhi2025}; the formulation is based on progressive coarsening of $\Delta y^+$ and $\Delta z^+$ proportional to the local Kolmogorov scale. With this approach, we save the number of grid points by 45\% compared to a Cartesian grid (compare $N_\text{dof}$ between Optimal and Cartesian in table~\ref{tab:channel}). Such grid saving becomes significantly larger for turbulent boundary layer simulations, as we discuss shortly. We obtain excellent agreement in the mean velocity $U^+$ and r.m.s of streamwise velocity fluctuations $u^+_{rms}$ between the optimal grid, the Cartesian grid, as well as the reference DNS data of \citet{Moser1999DirectRe=590} (figure~\ref{fig:channel}a,b).

%%---------------------------------------------------------------%%
\subsubsection{Turbulent half-channel flow (riblets) \label{sec:channel_riblets}}
We assess the accuracy of SOD2D and its resolution requirements for riblets by replicating the turbulent channel flow cases of \citet{Endrikat2021} at friction Reynolds number $Re_\tau = 400$ over two triangular riblet geometries; one with tip angle $\alpha = \ang{90}$ and viscous-scaled spacing $s^+ = 50$ (termed T950), and another one with $\alpha = \ang{60}$ and $s^+ = 15$ (termed T615). The riblet geometry T950 increases drag, while T615 reduces drag. Similar to the smooth case, we generated the optimal grids for riblets (figure~\ref{fig:channel_setup}g,l). On a $yz$-plane, we fill the space from the riblet valley up to the riblet sublayer $y^+ \in [-k^+, 0.6s^+]$ with square cells with size $\Delta \xi^+ = 1.47$ (T950) and $0.54$ (T615). Then, we expand $\Delta y^+$ and $\Delta z^+$ following the formulation by \citet{Rouhi2025}.
Table~\ref{tab:channel} lists the grid size details. To ensure grid convergence with the optimal grids, we conducted finer grid cases T950F (figure~\ref{fig:channel_setup}h) and T615F. In T950F, $\Delta y^+$ and $\Delta z^+$ are refined by at least $1.5$ times, and in T615, $\Delta x^+, \Delta y^+$ and $\Delta z^+$ are refined by $1.4$ times. We obtain very good agreement in the profiles of $U^+$ and $u^+_{rms}$ between the optimal grid and the fine grid, as well as the reference DNSs by \citet{Endrikat2021} (figure~\ref{fig:channel}c,d).
%%----------------------------------------------------------%%
\begin{table}[!h]
    \centering
    \begin{tabular}{c c c c c}
    \toprule
    Case & Configuration& $L_\mathrm{SM}/\delta_0$ & $L_\mathrm{RI}/\delta_0$  & Riblet geometry \\ 
    \midrule
    SM                          &
    \raisebox{.25\height}{\includegraphics[scale=0.1]{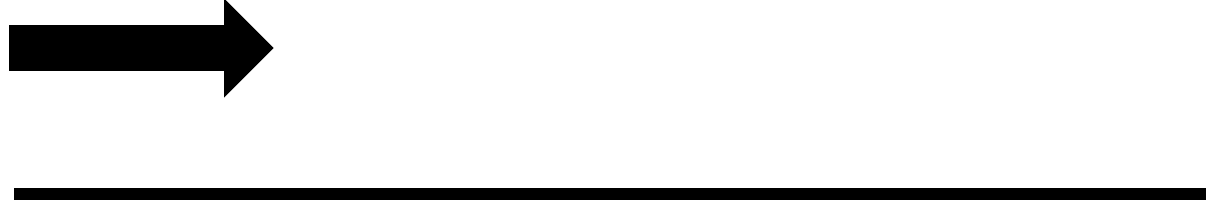}} &    
    $120 $                      & 
    $-$                         &  
    $-$ \\
    RI                          & 
    \raisebox{.4\height}{
    \includegraphics[scale=0.1]{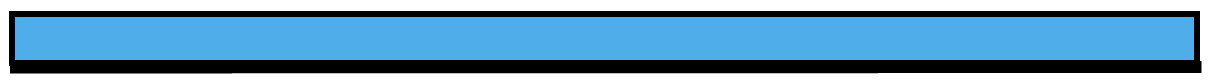}
    } &
    $-$    & 
    $120 $ &  
    \raisebox{-.4\height}{
    \includegraphics[scale=0.3]{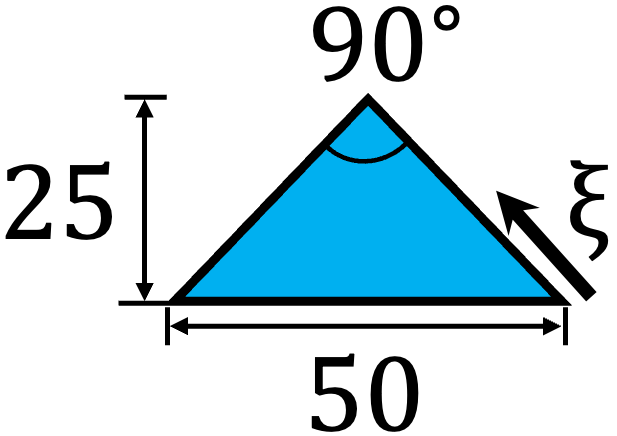}
    } \\
    RI\_SM                      &
    \raisebox{.4\height}{\includegraphics[scale=0.1]{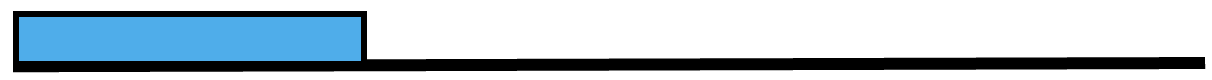}} &    
    $80 $                       & 
    $40 $                       & 
    \raisebox{-.4\height}{
    \includegraphics[scale=0.3]{etmm_figs/T950.png}
    \includegraphics[scale=0.3]{etmm_figs/T615.png}
    \includegraphics[scale=0.3]{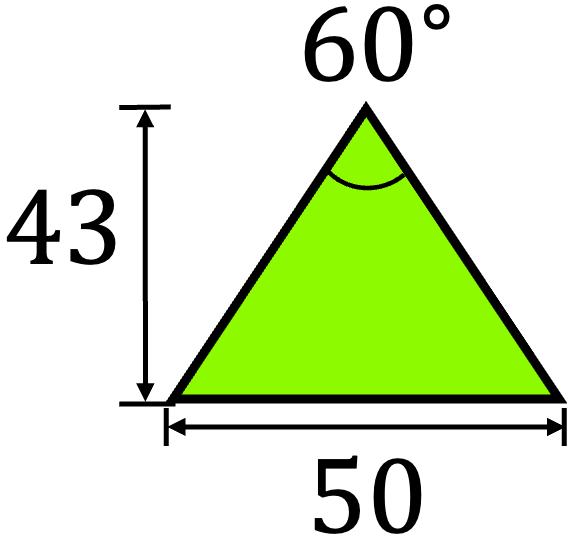}
    }\\
    SM\_RI                      &
    \raisebox{.4\height}{\includegraphics[scale=0.1]{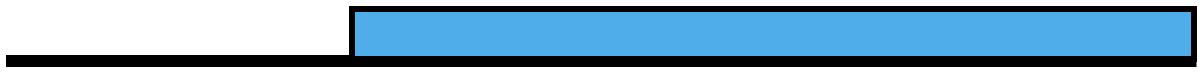}} &    
    $40 $                       & 
    $80 $                       & 
    \raisebox{-.4\height}{\includegraphics[scale=0.3]{etmm_figs/T950.png}
    }\\ 
    \bottomrule
    \end{tabular}
    \caption{\label{tab:zpg} Simulation cases for the ZPG TBLs; SM: smooth; RI: riblet; RI\_SM: riblet-to-smooth; SM\_RI: smooth-to-riblet. Black arrow in second column shows flow direction. $L_\mathrm{SM}$ and $L_\mathrm{RI}$ indicate the length of the smooth and riblet patches; $\delta_0$ is the boundary layer thickness at the reference location. RI will be substituted with T950 ($\alpha=90^\circ, s_0^+=50$), T615 ($\alpha=60^\circ, s_0^+=15$) or T650 ($\alpha=60^\circ, s_0^+=50$) in the main text for clarity. Riblet dimensions are mentioned at the reference location (where $Re_\theta\simeq 670, Re_\tau \simeq 283$).} 
\end{table}
%%----------------------------------------------------------%%

\subsection{Calculation cases for ZPG TBLs} 
We perform direct numerical simulations of ZPG TBLs in four different configurations (table~\ref{tab:zpg}): a fully smooth configuration (SM), a fully riblet configuration (RI), a riblet-to-smooth configuration (RI\_SM), and finally, a smooth-to-riblet configuration (SM\_RI).  For the configurations with riblets, we chose three triangular riblet shapes, with tip angles $\alpha = \ang{90}, s_0^+ \simeq 50$ (legend will use T950), $\alpha = \ang{60}, s_0^+ \simeq 15$ (legend will use T615) and $\alpha = \ang{60}, s_0^+ \simeq 50$ (legend will use T650); $s^+_0$ is scaled based on the friction velocity $u_{\tau_0}$ at the step change, where we place the origin ($x=0$). At the origin, $Re_{\theta_0} \simeq 670$ $(Re_{\tau_0} \simeq 283)$. The patch length upstream of the step change is $40\delta_0$, and the one downstream is $80\delta_0$ (table~\ref{tab:zpg}).
 
%%----------------------------------------------------------%%
\begin{figure}[!h]
	\begin{center}
	\includegraphics*[width=1.0\linewidth,trim={{0.0\linewidth} {0.0\linewidth} {0.0\linewidth} {0.0\linewidth}},clip]{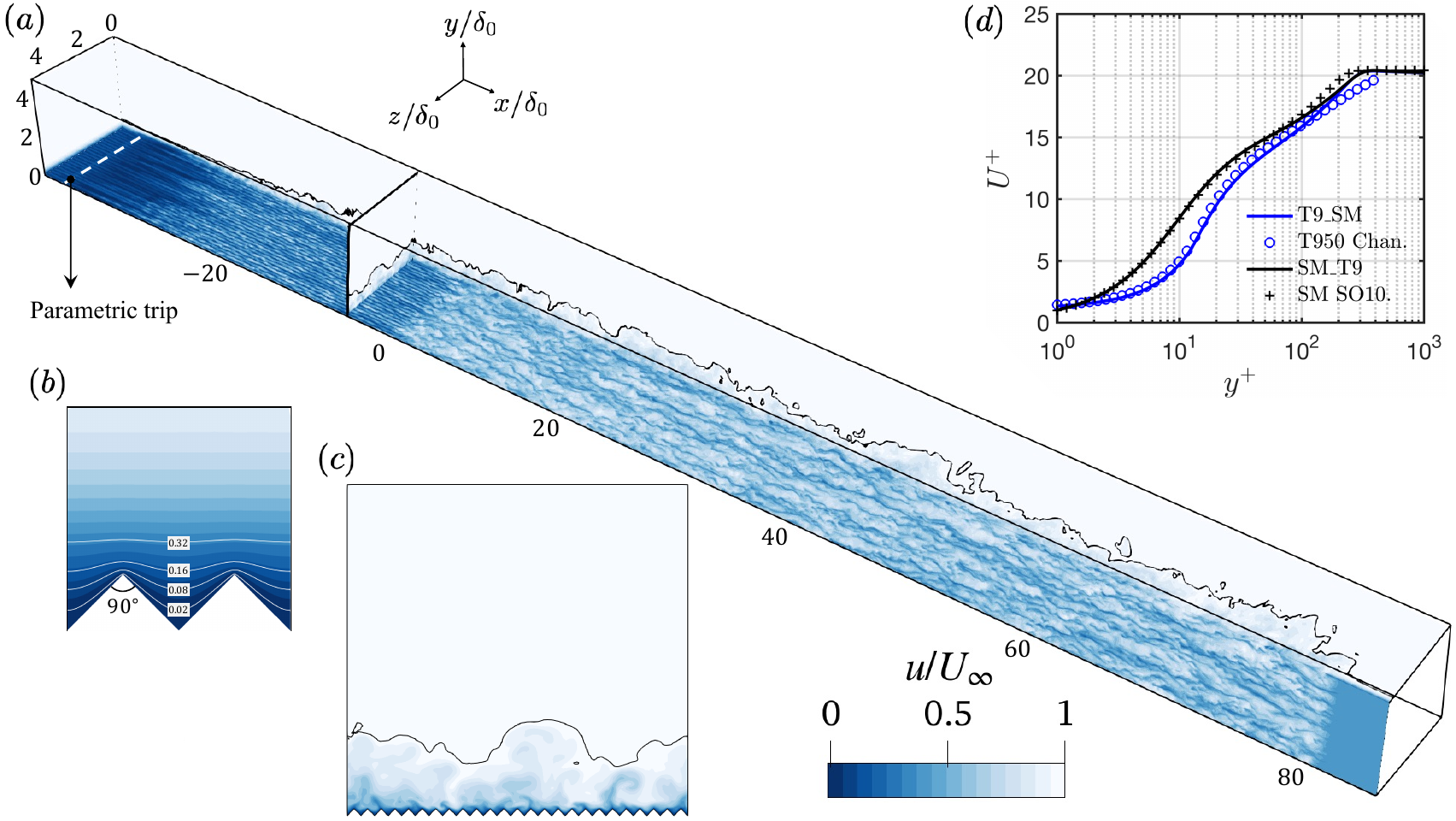}%
	\caption{\label{fig:zpg} (\textit{a}) Flow setup and visualization of streamwise velocity for T950\_SM. (\textit{b}) laminar inlet from precursor calculation; contour lines of $u/U_\infty$ are shown in white. (\textit{c}) Flow field at reference location. (\textit{d}) Mean velocity profiles upstream of the step change ($x = -4 \delta_0$) for the SM\_T950 (black line) and T950\_SM (blue line); the profiles are compared with ones from the smooth-wall ZPG TBL of \citet{Schlatter2010} ($\mathbf{+}$), and T950\_fine case (\bluecirc{}) from table~\ref{tab:channel}.}
	\end{center}
\end{figure}
%%----------------------------------------------------------%%

\subsection{Boundary layer setup \label{sec:bl_setup}}
Figure~\ref{fig:zpg} presents our setup for ZPG TBL simulations. We set the domain width and height $(L_z,  L_y)$ equal to three times the maximum boundary layer thickness near the outlet, where the momentum thickness Reynolds number reaches $Re_\theta \simeq 1000$ (friction Reynolds number $Re_\tau \simeq 400$). These dimensions are larger than the ZPG TBL setup by \citet{Schlatter2009}. A laminar boundary layer with displacement thickness Reynolds number $Re_{\delta^*_\mathrm{in}} = 775$ is applied at the inlet (figure~\ref{fig:zpg}\textit{a}). For the setups with the smooth wall at the inlet, we apply a laminar Blasius profile. However, for the cases with riblets at the inlet, we generate the inflow from a precursor calculation of a laminar temporal boundary layer (figure~\ref{fig:zpg}\textit{b}; procedure is discussed in $\S$\ref{sec:temporal_bl_setup}). The laminar boundary is numerically tripped (details to be discussed next) a short distance away from the inlet section. We apply a no-slip condition at the bottom boundary, zero spanwise-vorticity condition at the top boundary, and periodic boundary condition in the spanwise direction. At the outlet, we apply a buffer region which compels the flow to re-transition to laminar, following similar approaches in the literature \citep{Schlatter2009, Munters2016TurbulentFarms}. The effect of the buffer region is discussed in $\S$\ref{sec:domain_verify}.

In figure~\ref{fig:zpg}(\textit{d}), we assess the equilibrium state of the TBL upstream of the step change at $x = -4\delta_0$. For the smooth-to-T950 step change (SM\_T950), the mean velocity profile agrees well with the reference profile
of \citet{Schlatter2010}, with matched $Re_\theta \simeq 670$. For the T950-to-smooth step change (T950\_SM), the mean velocity profile over the riblets patch at $x=-4\delta_0$ agrees well with the mean profile of turbulent half-channel flow over T950 (table~\ref{tab:channel}, third row) up to $y^+ \simeq 100$. Beyond $y^+ \simeq 100$, the profiles are not comparable, due to the different wake profiles of TBL and half-channel flow.

%%----------------------------------------------------------%%

\begin{table}[!h]
\centering
\begin{tabular}{ c c c c c c c c c c }
\toprule
 
       \multicolumn{10}{c}{\textbf{Trip study} (figure~\ref{fig:trip_tests}, left and middle columns)} \\
       \multicolumn{10}{c}{$L_x = 279\delta_0^*, L_y = 22\delta_0^*, L_z = 22\delta_0^*$} \\      
%  \cmidrule{2-9} 
\midrule
 Case &  & $l_{\text{buff}}$ & $\Delta x^+$ & $C_D$ & $x_\text{trip}$ & $l_x/\delta_0^*$ & $l_y/\delta_0^*$ & $Re_\text{trip}$ & Leg.\\     
 T1   &   &  &  &   1 &  & 2.0 & 2.0 & 1550 & \tikzline{blue} \\
 T2   &   &  &  &   3 &  & 1.5 & 1.5 & 1160 & \tikzline{red,dashed} \\
 T3   &   &  &  &   1 &  & 1.0 & 1.0 & 775 & lam. \\
 T4   &   & $29\delta_0^*$ & 20 & 3 & $12\delta_0^*$ & 2.0 & 1.0 & 775 & \tikzline{green,dotted} \\ 
 T5   &   &  &  & 3 &  & 2.0 & 0.5 & 390 & lam. \\
 T6   &   &  &  & 2 &  & 2.0 & 1.0 & 775 & \tikzline{magenta,dashdotted} \\
 T2F  &   &  & 12 & 3 &  & 1.5 & 1.5 & 1160 & \tikzline{black, dashed} \\
 \midrule
       \multicolumn{10}{c}{\textbf{Domain length and buffer study} (figure~\ref{fig:trip_tests}, right column) } \\
       \multicolumn{10}{c}{$L_y = 36\delta_0^*, L_z = 36\delta_0^*$} \\
\midrule
 Case & $L_x$ & $l_{\text{buff}}$ & $\Delta x^+$ & $C_D$ & $x_\text{trip}$ & $l_x/\delta_0^*$ & $l_y/\delta_0^*$ & $Re_\text{trip}$ & Leg.\\
 D1   &  $840\delta_0^*$ & $40\delta_0^*$ & & & & & 
  &  
 & \tikzline{black} \\
 D2   &  $880\delta_0^*$ & $80\delta_0^*$ & 30 & 3 & $5.8\delta_0^*$ & 1.5 & 1.5 & 1160 & \tikzline{blue,dashed} \\
 D3   &  $940\delta_0^*$ & $40\delta_0^*$ & & & & &  &  & \tikzline{red,dotted} \\ 
 \bottomrule
  \end{tabular}
\caption{\label{tab:trip_tests} Setup for tripping and domain tests for ZPG TBL over a smooth wall. The top seven cases (T1--T6, T2F) test the tripping parameters $C_D,l_x,l_y$ (setup shown in figure~\ref{fig:zpg}a). The bottom three cases (D1--T3) test the effect of buffer thickness $l_\text{buff}$. Line legends (Leg.; lam. refers to a case staying laminar) are mentioned for each case, which will used for results in figure~\ref{fig:trip_tests}.} 
\end{table}

\begin{figure}[!h]
	\begin{center}
	\includegraphics*[width=0.33\linewidth]{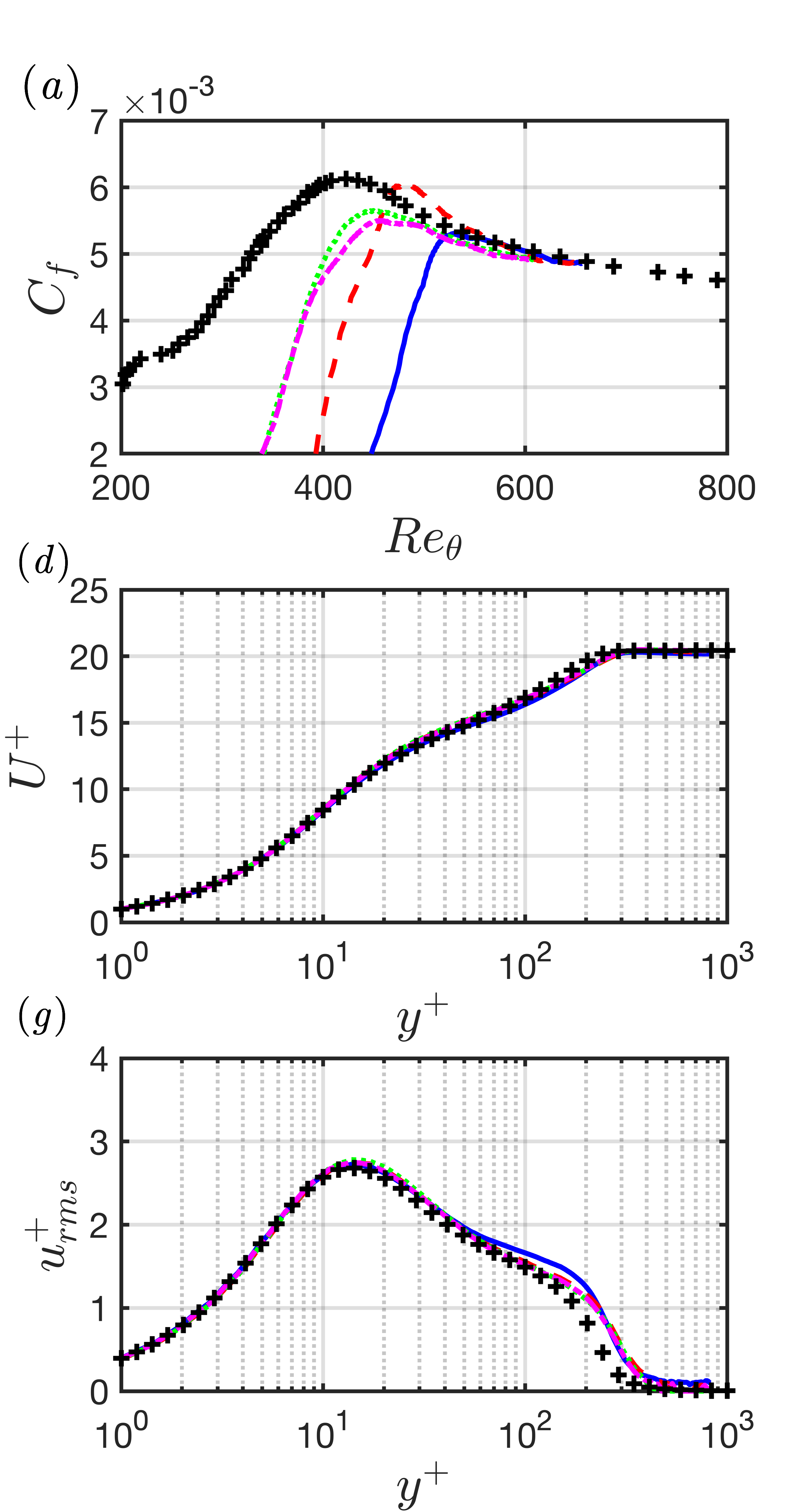}%
    \includegraphics*[width=0.33\linewidth]{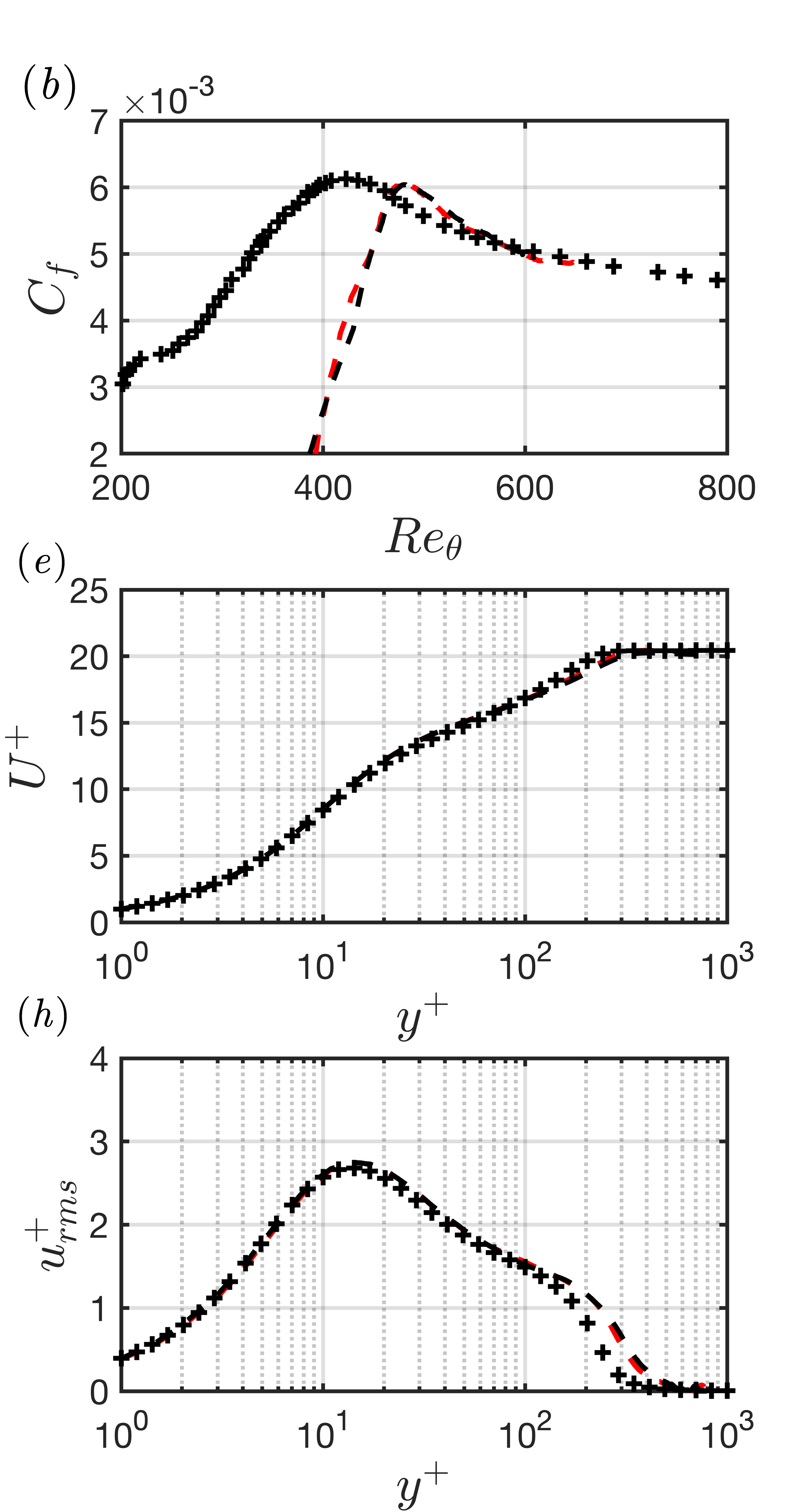}
    \includegraphics*[width=0.33\linewidth]{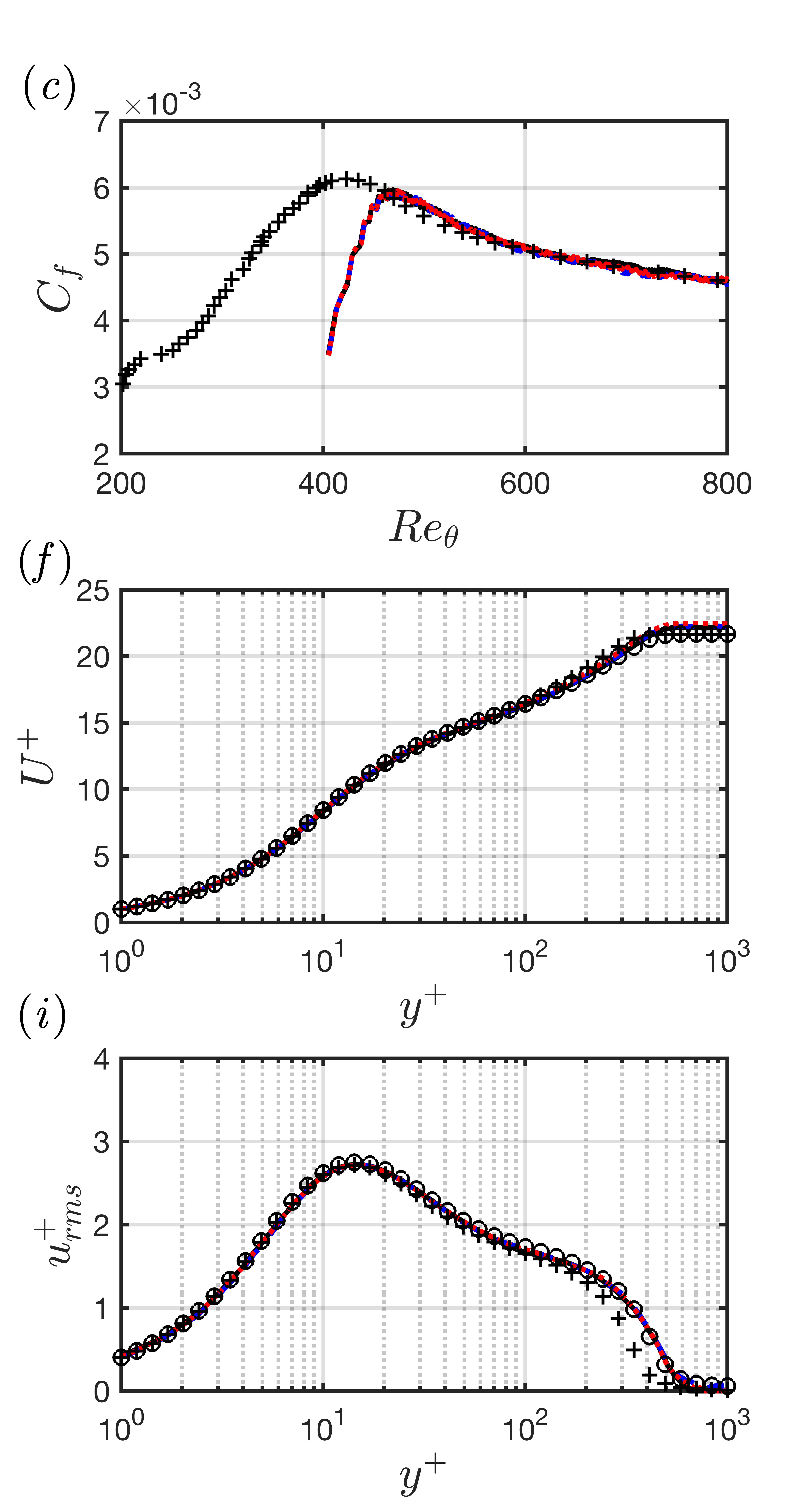}
	\caption{\label{fig:trip_tests} Results for the trip study and the domain study cases from table~\ref{tab:trip_tests}. (\textit{a,b,c}) Skin-friction coefficient $C_f$; (\textit{d,e,f}) mean velocity $U^+$, and (\textit{g,h,i}) r.m.s. of streamwise fluctuating velocity $u^+_\text{rms}$. (\textit{a,d,g}) plot the cases T1 (\tikzline{blue}), T2 (\tikzline{red,dashed}), T4 (\tikzline{green,dotted}) and T6 (\tikzline{magenta,dashdotted}); (\textit{b,e,h}) plot T2 (\tikzline{red, dashed}) and T2F (\tikzline{black, dashed}) from table~\ref{tab:trip_tests}. Reference data of  \citet{Schlatter2010,Schlatter2010SimulationsTo} ($\boldsymbol{+}$), where $U^+$ and $u^+_\text{rms}$ are at $Re_\theta \simeq 670$. In (\textit{c,f,i}) the profiles from cases D1 (\tikzline{black}), D2 (\tikzline{blue,dashed}) and D3 (\tikzline{red,dotted}) are plotted at $Re_\theta \simeq 1000$, and are compared with the reference data of \citet{Schlatter2010SimulationsTo} at $Re_\theta = 1000$ ($\boldsymbol{+}$), and \citet{Simens2009ALayers} at $Re_\theta = 1100$ (\blackcirc{}).}
	\end{center}
\end{figure}

%%----------------------------------------------------%%
\subsubsection{Study of tripping parameters \label{sec:tripping_setup}}

Slightly downstream of the inlet, we trip the boundary layer via volumetric forcing added to equation~\ref{eq:mom}. This forcing mimics the drag force by a trip wire with length and height $l_x,l_y$, respectively. The equations read,

\begin{eqnarray}
 f_x = -\frac{1}{2}\rho C_D u |u|/l_x, \quad  f_y = 0, \quad f_z = 0. \label{eq:force}
\end{eqnarray}
Here, $C_D$ is a constant factor. The forcing is applied for $x_{\text{trip}} \le x \le x_{\text{trip}}+l_x$, $0 \le y \le l_y$. Some past computations of ZPG TBL employ tripping technique, albeit with a different parameterization; \citet{Chen2023Two-scaleLayer} explicitly place a trip wire in the computational domain, \citet{MalathiAnanth2023RibletNumbers} use isolated roughness element and \citet{Kozul2016} apply an inflectional velocity profile similar to the downstream profile of a trip wire. 

We conducted a series of calculations to systematically study the tripping parameters $C_D,l_x,l_y$ (table~\ref{tab:trip_tests}; trip study). For all cases, we used a fixed domain size $279\delta_0^* \times 22\delta_0^* \times 22\delta_0^*$, a fixed buffer layer thickness $l_\text{buff} = 29\delta_0^*$, as well as a fixed tripping location at $12\delta^*_0$ from the inlet. Our choice of $L_x$ resolves the boundary layer up to $Re_\theta \simeq 600, \, Re_\tau \simeq 250$. Such domain length is sufficient to study the effect of tripping on the state of the TBL. We use same mesh resolution across T1-T6 cases ($\Delta x^+=20, \, \Delta y^+=0.3-35, \, \Delta z^+ = 5-26$); for T2F case we decrease the streamwise grid size to $\Delta x^+ = 12$ for assessing the resolution effect of tripping. Note that the grid sizes are scaled based on $u_\tau$ of the highest local $Re_\tau \simeq 250$. 
% For domain tests (table~\ref{tab:trip_tests}, D1--D3), only $L_x$ is reported; wall-normal and spanwise domain sizes remain the same, $L_y = L_z = 36\delta_0^*$.
Our aim is to find the proper tripping parameters for fast transition to turbulence. The numerical trip (\ref{eq:force}) acts as an obstacle that leads to the local flow reversal, and perturbs the boundary layer to hasten transition to turbulent. 
Details on such transition mechanisms can be found in \citep{narasimha1985laminar, Rist1995}.
We define a trip based Reynolds number, $Re_{\text{trip}} = l_y U_{\text{top}}/\nu$ (where $U_{\text{top}}$ is the streamwise velocity at $y=l_y$ in the unperturbed flow) and compare the values in table~\ref{tab:trip_tests}. It is important to note that transition to a turbulent regime will not take place if $Re_{\text{trip}}$ is lower than a critical value. In fact, we find that when $Re_{\text{trip}}\lesssim 800$, the perturbations are not strong (with $C_D \sim 1$) to trigger transition, and the boundary layer stays laminar (case T3 and T5). In this regard, the numerical trip displays a behavior that is similar to a physical trip in wind tunnel experiments. In scenarios where transition is initiated, the case with smallest $Re_\text{trip}$ (least perturbed) leads to a natural development from the smallest possible $Re_\theta$, similar to the observations of \citet{Kozul2016}. This can be observed in figure~\ref{fig:trip_tests}\textit{a} where T2 ($Re_\text{trip}\simeq 1160$) and T1 ($Re_\text{trip}\simeq 1550$) follow the T4 ($Re_\text{trip}\simeq 775$) curve. Further, once transition is initiated, change in $C_D$ for the same $Re_\text{trip}$ (case T4 and T6) shows marginal influence on $C_f$ (figure~\ref{fig:trip_tests}\textit{a}; green and magenta curves). 

In figures~\ref{fig:trip_tests}(\textit{a,d,g}), we compare the cases that become turbulent (T1, T2, T4, T6). For these cases, the skin-friction coefficient $C_f$ profiles approach the reference data of \citet{Schlatter2010} by $Re_\theta \simeq 550$ (figure~\ref{fig:trip_tests}\textit{a}). Beyond this $Re_\theta$, the tripped cases T1, T2, T4, T6 agree well with each other, as seen in the profiles of $U^+$ and $u^+_{rms}$ at $Re_\theta \simeq 607$ (figures~\ref{fig:trip_tests}\textit{d,g}); these profiles also agree well with the reference data of \citet{Schlatter2010} at $Re_\theta \simeq 670$. We observe some difference between our cases and the data of \citet{Schlatter2010} for $y^+\ge 100$. Such difference is partly due to the different $Re_\theta$, and partly due to the different tripping techniques \citep{Schlatter2012}. \citet{Schlatter2012} concluded that the boundary layer needs to be resolved up to $Re_\theta \simeq 2000$ for the upstream tripping effects to completely disappear. Based on our tripping study, for the rest of our study we set the tripping parameters $C_D = 3.0$, $l_x = l_y = 1.5\delta_0^*$ (case T2). We conducted case T2 with a finer grid resolution (T2F in table~\ref{tab:trip_tests}). Figures~~\ref{fig:trip_tests}(\textit{b,e,h}) show that the resulting statistics are almost identical between T2 and T2F, indicating the insensitivity of our chosen tripping parameters to the grid resolution. 
% \ar{\textit{Lastly, it should be emphasized that the optimal tripping parameters used for this study are not universal; a change in underlying numerical scheme can result in differences in tripping behavior. Therefore, studies utilizing such tripping mechanisms should conduct an independent study on the effect of these tripping parameters on transition to turbulence.} \textbf{I feel this statement is shooting ourselves in the foot.}}

%%----------------------------------------------------------%%
\subsubsection{Domain length and buffer study \label{sec:domain_verify}}
We also performed a set of calculations to obtain the suitable domain length $L_x$ and buffer thickness $l_\mathrm{buff}$ to well resolve the TBL up to $Re_\theta \simeq 1000$ $(Re_\tau \simeq 400)$ (the three bottom rows in table~\ref{tab:trip_tests}). We changed the buffer thickness from $40\delta_0^*$ to $80\delta_0^*$, and extended the effective domain length from $800\delta_0^*$ to $900\delta_0^*$. We compare our results with the reference DNS data of \citet{Schlatter2010SimulationsTo} and \citet{Simens2009ALayers} (figure~\ref{fig:trip_tests}\textit{c,f,i}). All our tested cases, yield identical $C_f$, and they agree with the data of \citet{Schlatter2010SimulationsTo} for $600 \le Re_\theta \le 1100$ (figure~\ref{fig:trip_tests}c). Also, the profiles of $U^+$ and $u^+_\text{rms}$ at $Re_\theta = 1000$ ($x = 600\delta_0^*$) are identical between our tested cases, and show good agreement with the DNS of \citet{Simens2009ALayers} at close $Re_\theta \simeq 1100$. Similar to the results for lower $Re_\theta$ (figure\ref{fig:trip_tests}\textit{d,e,g,h}), we observe some discrepancy between our r.m.s. profiles and the ones by \citet{Schlatter2010SimulationsTo} for $y^+ \simeq 100$; similar discrepancy exists between the two reference datasets~\cite{Simens2009ALayers,Schlatter2010SimulationsTo}, owing to the inevitable history effect of inflow for $Re_\theta \lesssim 2000$. From our domain independent study, we conclude that doubling the buffer thickness from $40\delta_0^* \simeq 3 \delta_\text{max}$ to $80\delta_0^* \simeq 6 \delta_\text{max}$, does not modify the solution up to our region of interest. Furthermore, with an effective domain length of $800\delta^*_0$, we resolve the TBL up to $Re_\theta \simeq 1000$ (Case D1, table~\ref{tab:trip_tests}). Therefore, for the rest of our study we use $L_x = 840\delta^*_0$, with a buffer thickness of $40\delta_0^*$.
 
%%--------------------------------------------------------------%%
\begin{figure}[!h]
	\begin{center}
	\includegraphics*[width=1.0\linewidth]{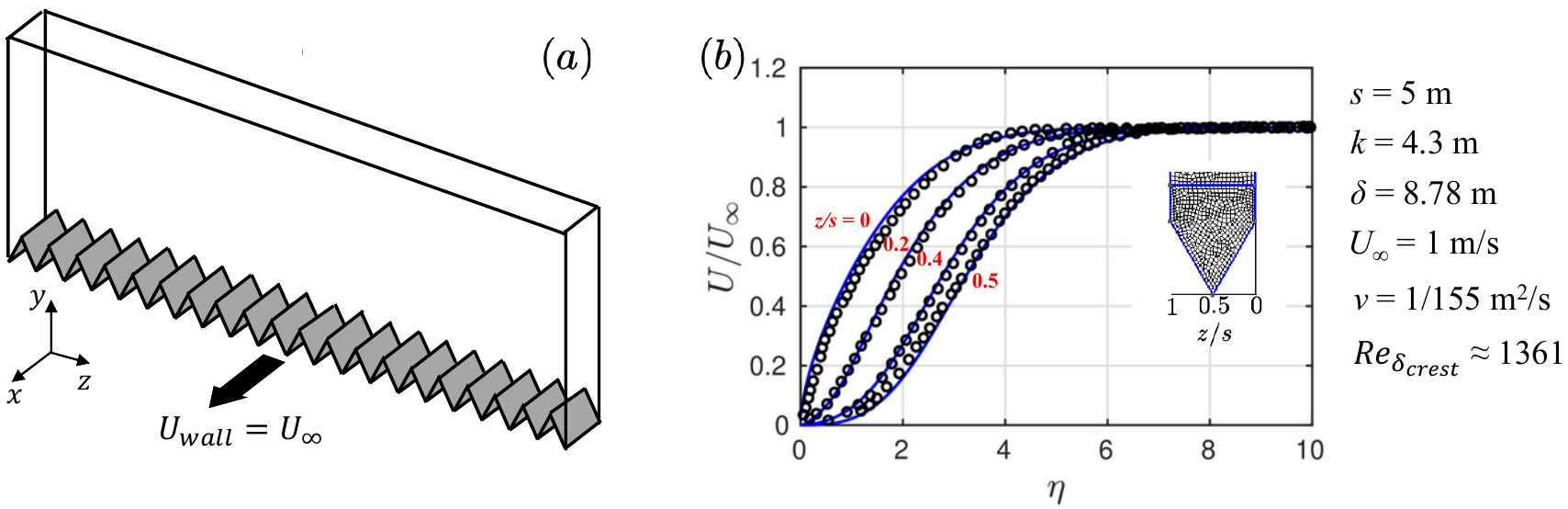}
	\caption{\label{fig:laminar_inflow} Setup and validation of the precursor temporal laminar boundary layer simulation. (a) Simulation setup with the bottom wall moving with the free-stream velocity $U_\infty$. (b) Comparison of the streamwise velocity profiles at several spanwise locations over the riblet between our temporal boundary layer simulation (\tikzline{blue}) with the reference experimental and computational study of \citet{Djenidi1994LaminarRiblets} (\blackcirc{}); $\eta$ is the non-dimensionalized wall-normal coordinate, $\eta = y/\sqrt{U_\infty x/\nu}$; flow conditions and riblet dimensions are reported on the right.}
	\end{center}
\end{figure}

%%------------------------------------------------------------------%%
\subsubsection{Inflow generation via temporal boundary layer \label{sec:temporal_bl_setup}} 
The setup for smooth-to-riblet cases (SM\_RI in table~\ref{tab:zpg}) is similar to the past computational studies of TBL over riblets (\citet{Boomsma2015}; \citet{Bannier2015RibletIdentity}); these studies applied inflow generation techniques that are developed for a smooth wall, namely recycling-rescaling technique \citep{Lund1998} or synthetic-eddy method \citep{Pamies2009}. As a result, the computational domain must have a smooth patch at the entrance. However, our setups with riblets placed from the inlet (RI and RI\_SM in table~\ref{tab:zpg}) are never simulated before; they consist of a laminar boundary layer over riblets at the inlet, that is tripped with parametric forcing. To generate the laminar boundary layer, we conduct a precursor calculation of a temporal boundary layer over riblets (figure~\ref{fig:laminar_inflow}). Temporal boundary layer is previously applied to a smooth wall, for studying the evolution of ZPG TBL \citep{Spalart1988Direct1410, Kozul2016}. It is a computationally efficient approach to obtain boundary layer at a target Reynolds number. The computational domain is a box with periodic boundary conditions in the streamwise and spanwise directions (figure~\ref{fig:laminar_inflow}\textit{a}). The bottom wall is moving with a target velocity $u = U_\text{wall} = U_\infty, v= w = 0$, while the top wall is stationary $u = v = w = 0$. As the simulation starts, the setup appears as a camera that is traveling along a spatial boundary layer with the free-stream velocity $U_\infty$. As the simulation is progressing, the boundary layer grows, hence the Reynolds number increases. The simulation is stopped once the boundary layer reaches a target thickness (i.e. target Reynolds number). We validated the accuracy of temporal boundary layer by replicating a computational and experimental study of spatial laminar boundary layer over riblets (\citet{Djenidi1994LaminarRiblets}, figure~\ref{fig:laminar_inflow}\textit{b}); the local velocity profiles at different spanwise locations over the riblet are in excellent agreement between our temporal setup and the reference data. For our production calculations, we run the temporal boundary layer simulation, on a given riblet shape, up to $Re_{\delta^*} = 775$. Then we apply the resulting velocity field as the inlet condition for RI and RI\_SM cases (with the same riblet dimensions as the precursor run) from table~\ref{tab:zpg}. 

%%----------------------------------------------------------%%
\begin{figure}[!h]
	\begin{center}
	% \includegraphics*[width=0.5\linewidth,trim={{0.0\linewidth} {0.0\linewidth} {0.0\linewidth} {0.0\linewidth}},clip]{etmm_figs/fig3}
    % \includegraphics*[width=0.9\linewidth]{etmm_figs/fig7.pdf}
    % \caption{\label{fig:fig3} (\textit{a}) Visualization of the optimal mesh (only spectral elements are shown) for the T950\_SM case on an $yz$-plane at the reference location. Plot (\textit{b}) gives a zoomed-in view over the riblet and smooth surface near the step change. The viscous-scaled spanwise grid size $\Delta^+_z$ is reported at different viscous-scaled wall-normal distances $y^+$; $\delta^+_{400}$ is the viscous-scaled boundary layer thickness at $Re_\tau = 400$ (i.e.\ $\delta^+_{400} = 400$).}
    
    \includegraphics*[width=1\linewidth]{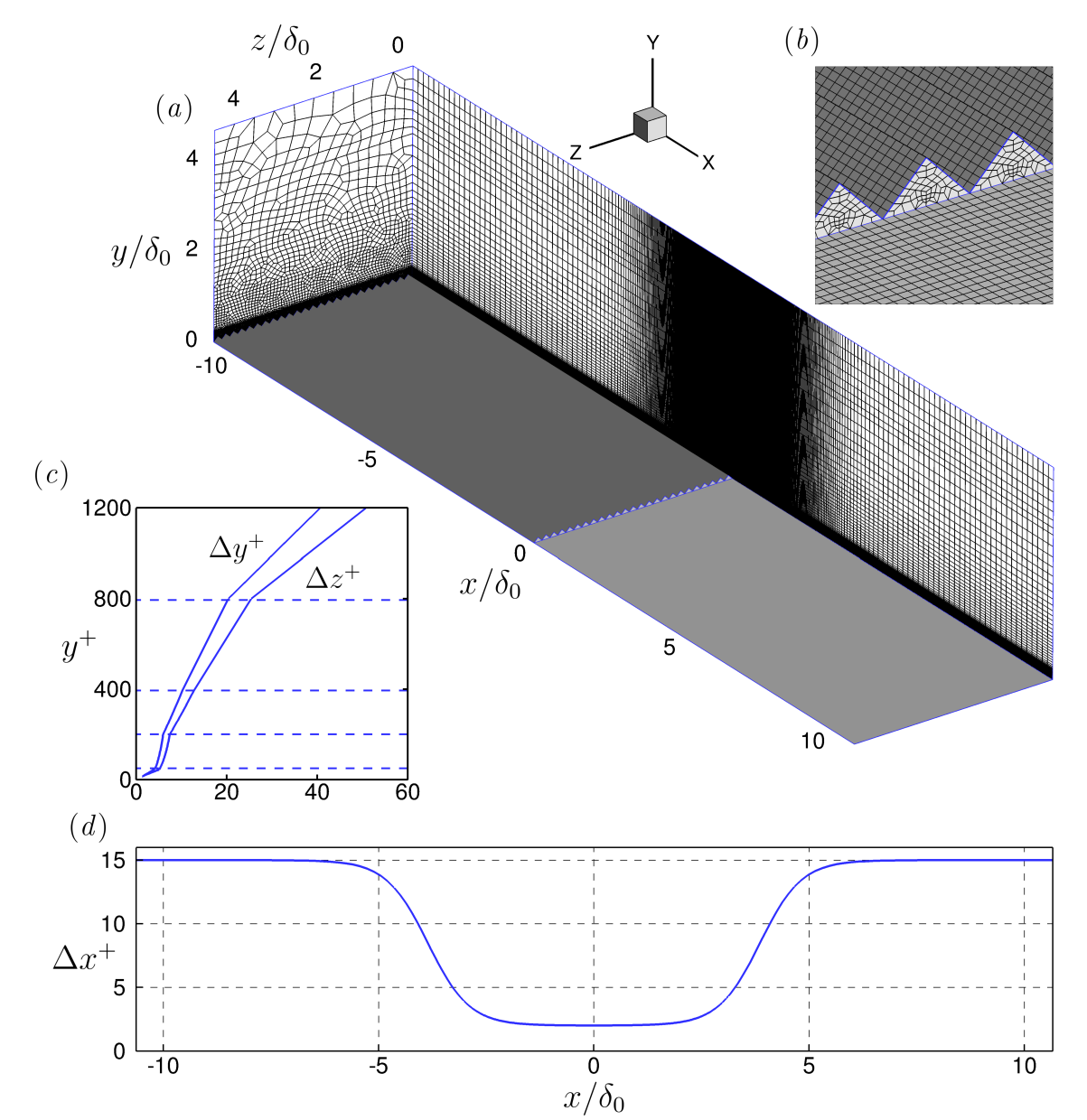}
    \caption{\label{fig:fig3} (\textit{a}) Visualization of the mesh (only spectral elements are shown) for the T950\_SM case. Plot (\textit{b}) gives a zoomed-in view near the riblet-to-smooth step change. Plot (\textit{c}) shows $\Delta y^+$ and $\Delta z^+$ distribution as a function of wall-normal distance. Plot (\textit{d}) shows $\Delta x^+$ distribution in the streamwise direction, near the step change.}
	
	\end{center}
\end{figure}

%%----------------------------------------------------------%%
\begin{table}[!h]
\centering
    \begin{tabular}{c c c c c c c }
    \toprule
    Case     & $\Delta_x^+$ & $\Delta^+_\xi$ & $\Delta_{z_w}^+, \Delta z^+_{\delta_\mathrm{max}}$ & $\Delta_{y_w}^+, \Delta y^+_{\delta_\mathrm{max}}$ & $N_\text{dof}$ & Leg. \\ 
    \midrule         
    SM       & $13$ & - & $5.8-14.6$ & $0.3-11.6$ &  44 M  & \tikzline{black} \\
    T950     & $15$ & $1.5$ & $1.5-12.4$ & $1.5-10.3$ & 99 M & \tikzline{black,dashed} \\
    T950\_SM & $2-15$ & $1.5$ & $1.5-12.4$ & $1.5-10.3$ & 150 M & \tikzline{blue} \\
    T615\_SM & $1-15$ & $0.7$ & $0.7 - 12.8$ & $0.7 - 10.2$ & 268 M & \tikzline{red, dashed} \\
    T650\_SM & $3-15$ & $1.2$ & $1.2 - 12.6$ & $1.2 - 10.2$ & 242 M & \tikzline{green, dotted} \\
    SM\_T950 & $2-15$ & $1.5$ & $1.5-12.4$ & $1.5-10.3$ & 141 M & \tikzline{blue,dashdotted} \\
    \bottomrule
    \end{tabular}
    \caption{\label{tab:simulation_cases} Mesh resolution for our production cases (configurations are mentioned in table~\ref{tab:zpg}; first two columns). All calculations use a domain size of $L_x \times L_y \times L_z \simeq 840\delta_0^* \times 36 \delta_0^* \times 36 \delta_0^*$; for riblet cases the domain size is adjusted slightly to fit an integer number of riblet wavelengths. The grid sizes in viscous units are reported based on $u_\tau$ of the highest resolved $Re_\tau \simeq 400$ ($Re_\theta \simeq 1000$). $w$ and $\delta_\text{max}$ refer to resolution near the wall and near the boundary layer edge (at maximum thickness), respectively.} 
\end{table}

\subsection{Production cases and grid details \label{sec:meshing_setup}}
In table~\ref{tab:simulation_cases}, we report the grid size details for our production cases from table~\ref{tab:zpg}; in figure~\ref{fig:fig3} we visualise the grid and plot the grid sizes for the case T950\_SM. To save the number of grid points, we follow the grid-generation framework by \citet{Rouhi2025}, where $\Delta y^+$ and $\Delta z^+$ are progressively coarsened away from the wall (figure~\ref{fig:fig3}c), as explained in Section \ref{sec:channel_riblets}. We also refine $\Delta x^+$ to well-resolve the flow near the step change (figure~\ref{fig:fig3}d). Near the step change ($-40k \lesssim x \lesssim 40k$), $\Delta x^+ \simeq k^+/12$ (e.g.\ $\Delta x^+ = 2$ for T950\_SM with $k^+ = 12.5$), and away from the step change, $\Delta x^+$ is coarsened to $15$. Note that the grid sizes in viscous units are scaled based on $u_\tau$ of the highest resolved $Re_\tau \simeq 400$ ($Re_\theta \simeq 1000$). 

\begin{comment}
\end{comment}

%%%%%%%%%%%%%%%%%%%%%%%%%%%%%%%%%%%%%%%%%%%%%%%%%%%%%

\begin{figure}[!h]
	\begin{center}
	\includegraphics*[width=0.5\linewidth,trim={{0.0\linewidth} {0.0\linewidth} {0.0\linewidth} {0.0\linewidth}},clip]{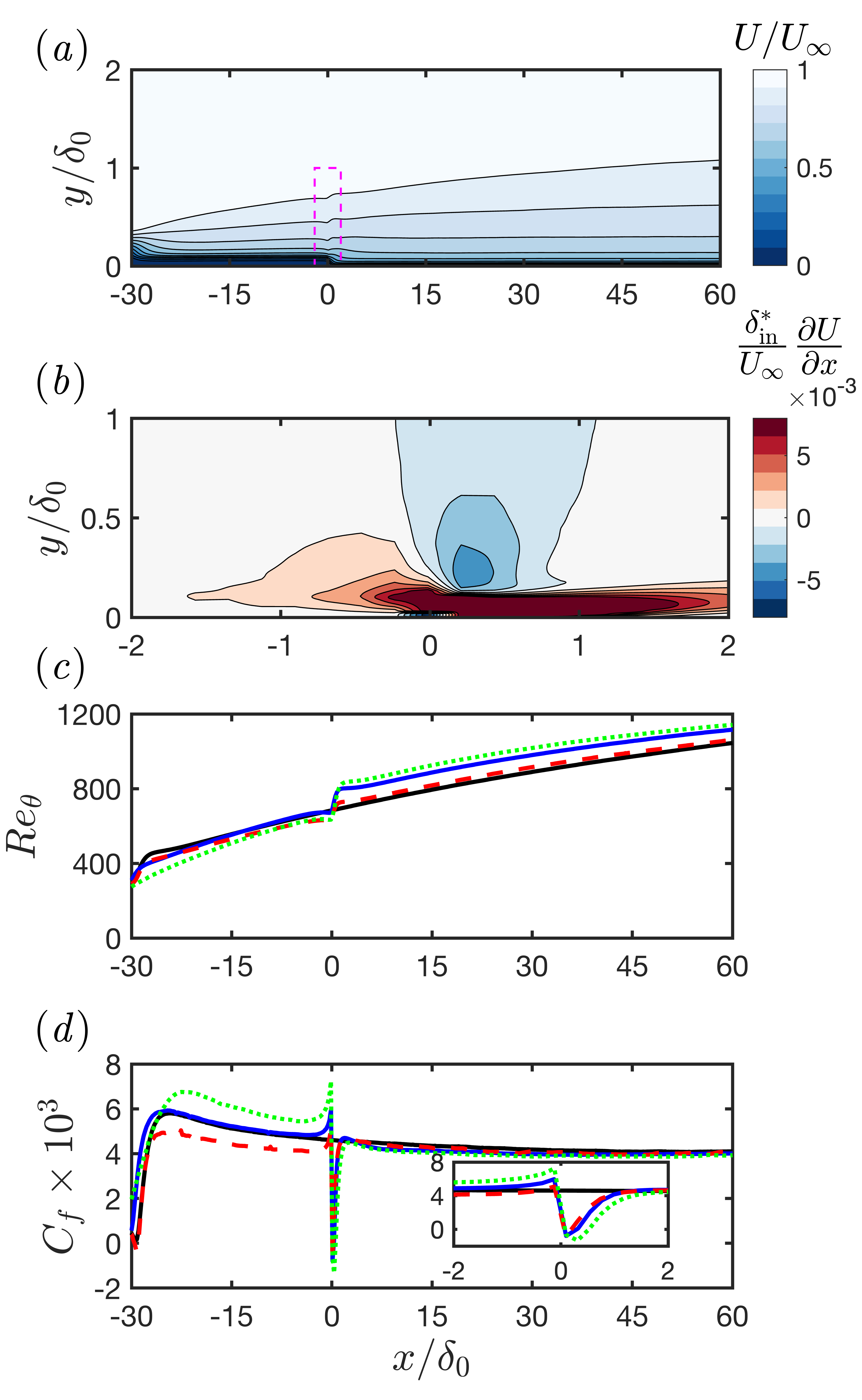}%
    \includegraphics*[width=0.5\linewidth,trim={{0.0\linewidth} {0.0\linewidth} {0.0\linewidth} {0.0\linewidth}},clip]{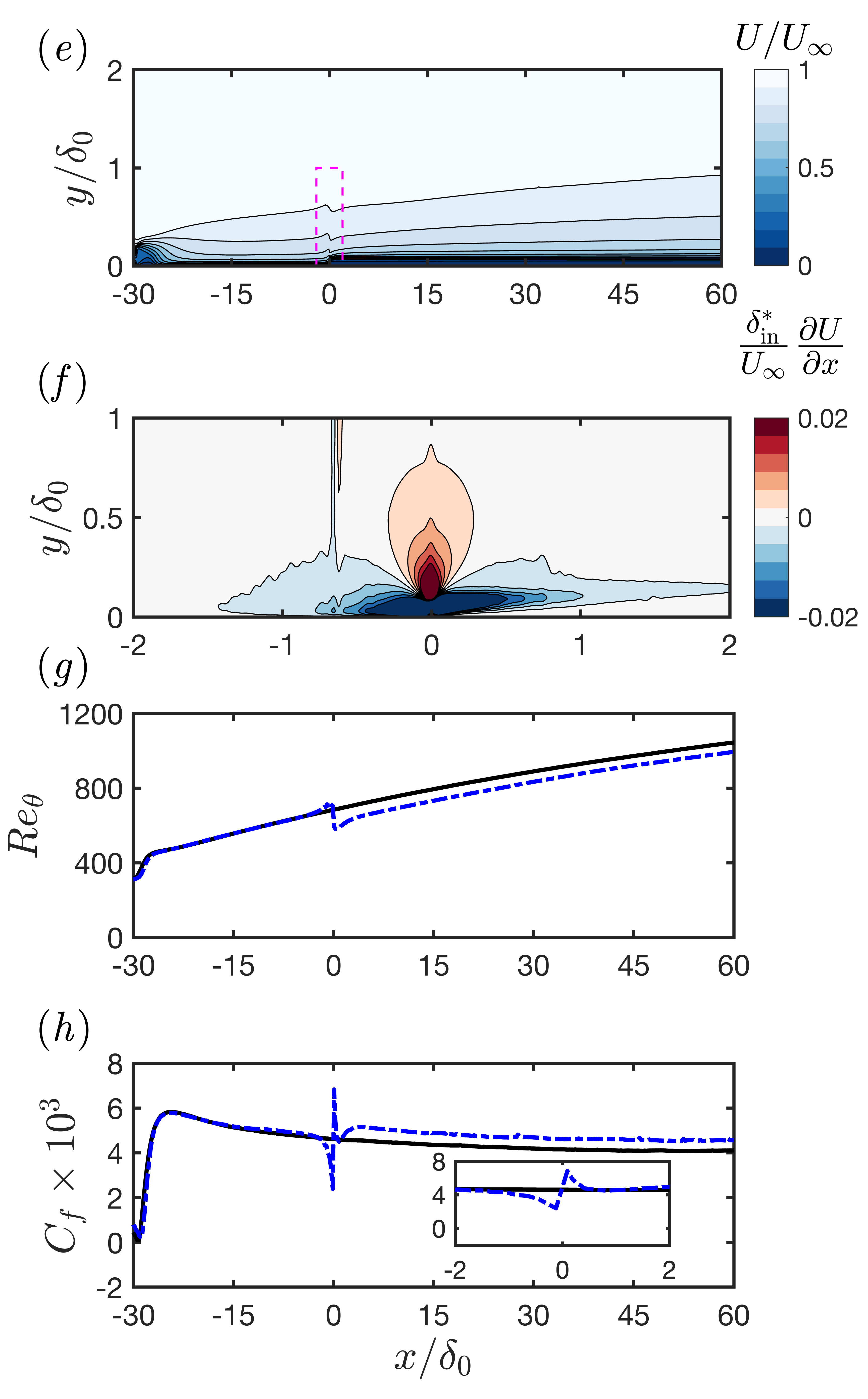}
	\caption{\label{fig:fig4_0} Contours of  (\textit{a,e}) streamwise velocity, $U/U_\infty$, and (\textit{b,f}) pressure gradient parameter, $\delta_{\text{in}}^*/U_\infty (\partial U/\partial x)$. Plots (\textit{b,f}) show zoom in on the regions marked by magenta box in (\textit{a,e}). Results in (\textit{a,b}) are for T950\_SM and in (\textit{e,f}) for SM\_T950 case. Streamwise evolution of (\textit{c,g}) $Re_\theta$ and (\textit{d,h}) $C_f$.  In (\textit{c,d}) results are plotted for all riblet-to-smooth cases and in (\textit{g,h}) for SM\_T950 case; SM case (solid black line) is shown in \textit{(c,d,g,h)} for comparison purposes. Line legends: T950\_SM (blue), T615\_SM (red), T650\_SM (green).}
	\end{center}
\end{figure}

\section{Results \label{sec:results}} %%%%%%%%%%%%%%%%%%
%The pressure gradient parameter, $\delta_{\text{in}}^*/U_\infty (\partial U/\partial x)$, gives a measure of near wall flow acceleration during riblet-to-smooth step change (figure~\ref{fig:fig4_0}\textit{b}, red region). To conserve mass, the flow away from the wall decelerates (figure~\ref{fig:fig4_0}\textit{b}, blue region). The mechanism for smooth-to-riblet step change (figure~\ref{fig:fig4_0}\textit{f}) is reversed; near wall flow decelerates and flow away from the wall accelerates.
%%-------------------------------------------------%%
\subsection{General flow features with step change}
Flow features during the step change are described through the streamwise evolution of time and spanwise-averaged velocity $U$, as well as its normalised streamwise gradient $\delta_{\text{in}}^*/U_\infty (\partial U/\partial x)$ for T950\_SM and SM\_T950 (figures~\ref{fig:fig4_0}\textit{a,b,e,f}). The parameter $\delta_{\text{in}}^*/U_\infty (\partial U/\partial x)$ highlights a significant near-wall acceleration ($\partial U/\partial x > 0$) during riblet-to-smooth step change (figure~\ref{fig:fig4_0}\textit{b}, red region), with a mild flow deceleration ($\partial U/\partial x < 0$) away from the wall (figure~\ref{fig:fig4_0}\textit{b}, blue region). For smooth-to-riblet step change (figure~\ref{fig:fig4_0}\textit{f}), this process is reversed; there is a significant near-wall deceleration and a mild flow acceleration away from the wall.
This impact of step change in riblets is similar to the step change in roughness (see figure 7b in \citet{Rouhi2019}). The streamwise extent of flow acceleration in the T950\_SM case (the extent of $\partial U/\partial x > 0$ in figure~\ref{fig:fig4_0}\textit{b}) is larger than the streamwise extent of flow deceleration in the SM\_T950 case (the extent of $\partial U/\partial x < 0$ in figure~\ref{fig:fig4_0}\textit{f}).

% the growth of $\theta$ depends on the sign on $\partial U/\partial x$; the contribution from the term inside brackets is marginal close the wall, since $U\ll U_\infty$ . Consequently, for riblet-to-smooth step change $Re_\theta$ suddenly rises ($\partial U/\partial x >0$ close to the wall). 

We study the variations of $Re_\theta$ and $C_f$ for the riblet-to-smooth cases (figure~\ref{fig:fig4_0}\textit{c,d}) and for SM\_T950 case (figure~\ref{fig:fig4_0}\textit{g,h}). We also add the respective baseline cases (SM for riblet-to-smooth cases and T950 for SM\_T950 case) to assess the flow recovery downstream of the step change. During the step change, there is a sudden jump in $Re_\theta$ (figure~\ref{fig:fig4_0}\textit{c,g}). By definition, $\frac{\partial \theta}{\partial x} = \frac{1}{U_\infty} \int \frac{\partial U}{\partial x} \left( 1-{2U}/{U_\infty}\right) \partial y$; that means the sudden jump in $\theta$ during the step change, hence large $\partial \theta/\partial x$, is related to the large $\partial U/\partial x$ near the wall. Near the wall, $(1-2U/U_\infty)>0$, as $U \ll U_\infty$; therefore, the sign of $\partial \theta / \partial x$ depends on the sign of $\partial U/\partial x$ near the wall. During the riblet-to-smooth step change, $\partial U/\partial x > 0$ is dominant near the wall (figure~\ref{fig:fig4_0}\textit{b}), hence $\partial \theta/\partial x > 0$ (figure~\ref{fig:fig4_0}\textit{c}). On the other hand, during the smooth-to-riblet step change, $\partial U/\partial x < 0$ is dominant near the wall (figure~\ref{fig:fig4_0}\textit{f}), hence $\partial \theta/\partial x < 0$ (figure~\ref{fig:fig4_0}\textit{g}).
During the riblet-to-smooth step change, there is an overshoot in $C_f$ followed by an undershoot below zero, demonstrating averaged flow reversal (figure~\ref{fig:fig4_0}\textit{d}).  Then, riblet-to-smooth $C_{f}$ quickly rises towards $C_{f_\mathrm{SM}}$ within a distance of $\sim \delta_0$ (figure~\ref{fig:fig4_0}\textit{d}, inset). During the smooth-to-riblet step change, we observe an undershoot in $C_f$, followed by an overshoot (figure~\ref{fig:fig4_0}\textit{h}), consistent with the past studies on smooth-to-rough step change \citep{Rouhi2019, Cogo2025SurfaceLayers}.

%The variation of $Re_\theta$ and $C_f$ for smooth-to-riblet transition follows reversed trend and is less vigorous than riblet-to-smooth (figure~\ref{fig:fig4_0}\textit{g,h}). For the SM\_T950 case, there is a sudden drop in $Re_\theta$ during the step change (figure~\ref{fig:fig4_0}\textit{g}), as the flow decelerates near the wall ($\partial U/\partial x < 0$). For $C_f$ (figure~\ref{fig:fig4_0}\textit{h}) 
%The intensity of flow disruption due to riblet-to-smooth step change strengthens as $\epsilon/\delta_0$ grows at the step location (figure~\ref{fig:fig4_0}\textit{c,d}).

% In fact, T650\_SM experiences larger flow separation and takes longer streamwise distance to recover to the $C_{f_{SM}}$ levels (figure~\ref{fig:fig4_0}\textit{d}, inset). In comparison, T615

The jump in $Re_\theta$, as well as the impulsive response of $C_f$ during the riblet-to-smooth step change depends on $k^+_0$, hence the step height $\epsilon/\delta_0$ (figure~\ref{fig:fig4_0}\textit{c,d}).
The jump in $Re_\theta$ and the impulse in $C_f$ is highest for the T650\_SM case ($\epsilon/\delta_0 \simeq 0.056, k_0^+ \approx 43$), followed by the T950\_SM case ($\epsilon/\delta_0 \simeq 0.031, k_0^+ = 25$), at matching $s^+_0$. Both cases increase $C_f$ upstream of the step change (see $C_f$ in figure~\ref{fig:fig4_0}\textit{d} for $x<0$). However, downstream of the step change, the T650\_SM experiences larger flow separation and its $C_f$ approaches $C_{f_{SM}}$ at a farther distance downstream (figure~\ref{fig:fig4_0}\textit{d}, inset). The T615\_SM case ($\epsilon/\delta_0 \simeq 0.015,k_0^+ = 15$) observes a modest disruption in $Re_\theta$ and $C_f$ during the step change, and its $Re_\theta$ closely follows the SM case downstream of the step change (figure~\ref{fig:fig4_0}\textit{c}). 

\begin{figure}[!h]
	\begin{center}
	\includegraphics*[width=0.5\linewidth,trim={{0.0\linewidth} {0.0\linewidth} {0.0\linewidth} {0.0\linewidth}},clip]{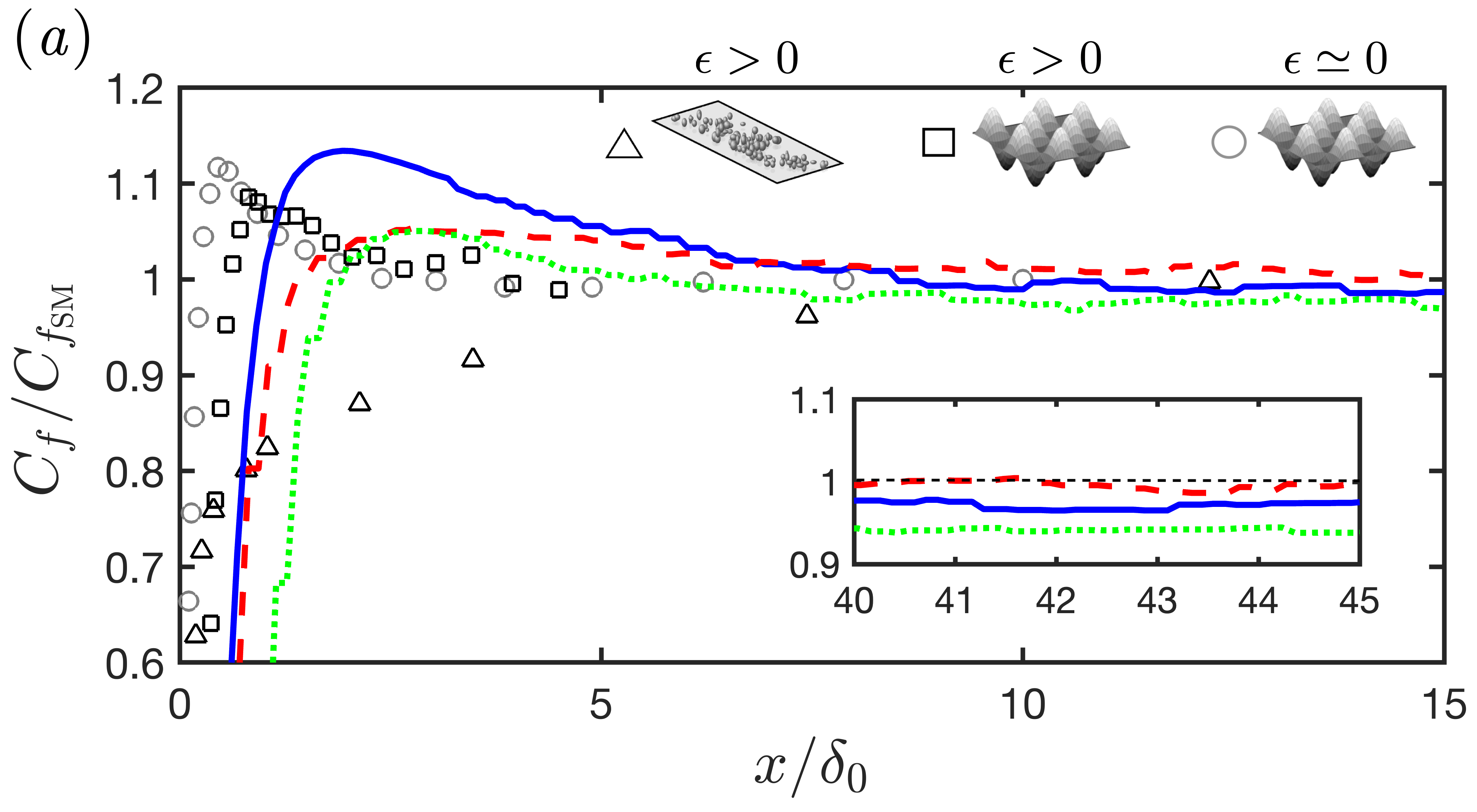}%
    \includegraphics*[width=0.5\linewidth,trim={{0.0\linewidth} {0.0\linewidth} {0.0\linewidth} {0.0\linewidth}},clip]{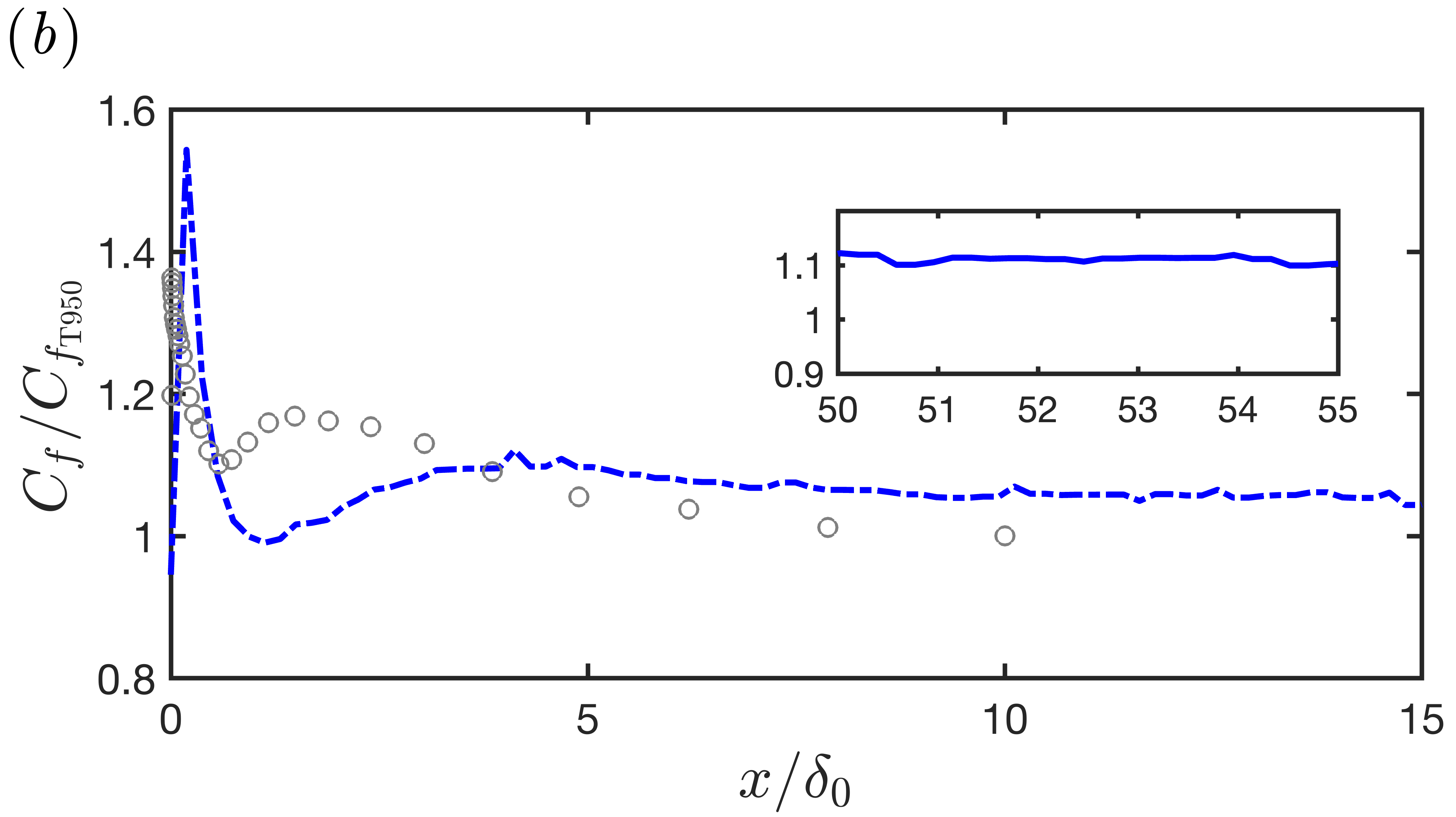}
	\caption{\label{fig:fig4} Streamwise variations of $C_f$ for the (\textit{a}) riblet-to-smooth cases and (\textbf{b}) smooth-to-riblet step change. In (\textit{a}), $C_f$ is normalized by $C_f$ from the smooth case ($C_{f_\mathrm{SM}}$) at matched $Re_\theta$. We include rough-to-smooth step change data of \citet{Rouhi2019}  ($\epsilon \simeq 0$, \greycirc{}), \citet{Rouhi2019RoughnessChange} ($\epsilon > 0$, \blacksqr{}) and \citet{Li2019} ($\epsilon > 0$, \blacktri{}). In (\textit{b}), $C_f$ for SM\_T950 is normalized by $C_f$ from the T950 only case ($C_{f_\mathrm{T950}}$) at matched $Re_\theta$; smooth-to-rough step change data of \citet{Rouhi2019} ($\epsilon \simeq 0$, \greycirc{}) is included. Line legends: T950\_SM (blue), T615\_SM (red), T650\_SM (green).}
	\end{center}
\end{figure}

% We now turn to the question of recovery of $C_f$ in step change with riblets vis-\'a-vis other roughness shapes reported in literature and the effect of $\epsilon/\delta$ in the recovery process (figure~\ref{fig:fig4}). 
%high and low shear regions downstream of the step change, which in turn depend on

%Recovery in $C_f$ is slowest for our T650\_SM case (figure~\ref{fig:fig4}\textit{a}), most likely due to larger separation it experiences. Additionally, the overshoot in $C_f$ weakens as $\epsilon/\delta$ grows from T950\_SM to T650\_SM, at matching $s^+_0$ (figure~\ref{fig:fig4}\textit{a}). This is in-line with the results of \citet{Rouhi2019RoughnessChange} for egg-carton roughness.

%the recovery of $C_f$ towards the smooth value plateaus in about $10 \delta_0$. For comparison, the recovery to $C_{f_\mathrm{SM}}$ is within $2\delta_0$ in \citet{Rouhi2019} and within $12 \delta_0$ in \citet{Li2019}.
%For T650\_SM and T950\_SM cases, $C_f/C_{f_\mathrm{SM}}$ reaches an asymptotic limit ($\sim 0.94$ for T650\_SM and $\sim 0.97$ for T950\_SM) within $10\delta_0$, but does not recover to $1.0$, even after $\sim 50 \delta_0$ downstream (figure~\ref{fig:fig4}\textit{a}, inset). In contrast, the reported $C_f$ values of \citet{Rouhi2019} and \citet{Li2019} in figure~\ref{fig:fig4}(\textit{a}) reach an asymptotic value of 1.0 by the end of the domain.

 %Note that to calculate $C_f/C_{f_\mathrm{SM}}$, $C_f$ and $C_{f_\mathrm{SM}}$ for our cases are at matched $Re_\theta$, following \citet{Li2021ExperimentalNumbers}.

\subsection{Recovery of $C_f$ to equilibrium downstream of the step change}
The variation of $C_f$ downstream of the step change, and its subsequent recovery to equilibrium depends on the geometrical details of the upstream surface, as well as Reynolds number. In figure~\ref{fig:fig4}, we compare our results to the numerical data of \citet{Rouhi2019, Rouhi2019RoughnessChange} (circles) and experimental data of \citet{Li2019} (triangles). \citet{Rouhi2019} conducted DNS of egg-carton roughness-to-smooth step change ($Re_\tau \simeq 590$, $k^+ \simeq 39$, $\epsilon/h \simeq 0$) and vice-versa ($Re_\tau \simeq 430$, $k^+ \simeq 24$, $\epsilon/h \simeq0$) in a turbulent half-channel flow. In a follow-up study~\cite{Rouhi2019RoughnessChange}, they conducted DNS of rough-to-smooth step change with a shifted virtual origin of the egg-carton roughness ($Re_\tau \simeq 590$, $k^+ \simeq 39$, $\epsilon/h = 0.056$). \citet{Li2019} conducted experiments of ZPG TBL over sandpaper rough-to-smooth step change at $Re_\tau \simeq 4100$, $k^+ \simeq 139$ and with $\epsilon/\delta > 0$.
Up to $x \simeq 5 \delta_0$, the variation of $C_f$ is unique to each case. Our riblet cases, as well as the egg-carton roughness cases of \citet{Rouhi2019,Rouhi2019RoughnessChange}, yield an overshoot in $C_f$ ($C_f/C_{f_{SM}}>1$) that differ in amplitude and location, whereas the sand-paper roughness of \citet{Li2019} does not yield an overshoot in $C_f$. Such variation of $C_f$ is resulted from an interplay between the high and low shear regions downstream of the step change, that depend on the complexity of the upstream surface (discussed in section \ref{sec:cf_modulation}). For the riblet-to-smooth cases, $C_f$ reaches a plateau by $x \simeq 10 \delta_0$; however, except T615\_SM (red dashed line), $C_f/C_{f_{SM}}$ does not reach unity, even up to $x = 45 \delta_0$ (figure~\ref{fig:fig4}\textit{a}, inset). The smaller is $k^+_0$, the closer is $C_f/C_{f_{SM}}$ to unity. Similarly, $C_f/C_{f_\mathrm{T950}}$ downstream of the smooth-to-riblet step change approaches unity, but does not reach it (figure~\ref{fig:fig4}\textit{b}, inset). In contrast, $C_f$ data-points from \citet{Rouhi2019,Rouhi2019RoughnessChange} and \citet{Li2019} downstream of the step change reach their equilibrium counterparts by the farthest resolved distance downstream (figure~\ref{fig:fig4}\textit{a,b}), that is by $x \simeq 10 \delta_0$ in \citet{Rouhi2019}, by $x \simeq 12 \delta_0$ in \citet{Li2019}, and by $x \simeq 10 \delta_0$ in \citet{Rouhi2019RoughnessChange}. We conjecture that such discrepancy is related to the different ways of obtaining the equilibrium $C_f$, e.g.\ $C_{f_{SM}}$ in figure~\ref{fig:fig4}(\textit{a}). For our riblet-to-smooth cases, following \citet{Li2021ExperimentalNumbers}, $C_f$ at each location downstream of the step change is normalized by $C_{f_{SM}}$ from our reference calculation at a location with matched $Re_\theta$; we follow the same approach to normalize $C_f$ by the equilibrium $C_{f_\mathrm{T950}}$ downstream of the smooth-to-riblet step change  (figure~\ref{fig:fig4}\textit{b}). \citet{Rouhi2019} and \citet{Li2019}, on the other hand, used the last measured value of $C_f$ at the end of the domain to normalize their $C_f$ values. This approach intrinsically assumes that $C_f$ has recovered by the end of the domain. As discussed later, this is not the case with step change over riblets. Consistent with our results (figure~\ref{fig:fig4}\textit{a}), \citet{Li2021ExperimentalNumbers} report the quick approaching of $C_f/C_{f_{SM}}$ to unity, within $8\delta_0$ downstream of the rough-to-smooth step change. However, similar to our observation, $C_f/C_{f_{SM}}$ deviates from unity by $3\%$, even up to $x = 120\delta_0$. They attribute such deviation to data uncertainty and using an empirical relation to obtain the equilibrium $C_{f_{SM}}$.

%Our observed difference in $C_f/C_{f_{SM}}$ is similar to the difference ($\sim 3\%$) reported by \citet{Li2021ExperimentalNumbers} for rough-to-smooth studies. They used an empirical relation between $C_f$ and $Re_\theta$ to arrive at equilibrium smooth values used for scaling. However, unlike our observations, they attributed their observed differences in $C_f/C_{f_{SM}}$ to experimental uncertainty. 

%The SM\_T950 step change case follows a similar trend. The $C_f/C_{f_\mathrm{T950}}$ reaches an asymptotic value of $1.04$ (\vk{why is it rising? $\sim 10\%$ at $x=50\delta_0$}) within $5\delta_0$ (figure~\ref{fig:fig4}\textit{b}), which is shorter than the asymptotic distance for riblet-to-smooth cases (figure~\ref{fig:fig4}\textit{a}). Nevertheless, $C_f/C_{f_\mathrm{T950}}$ does not converge to $1.0$, even after $55\delta_0$ (figure~\ref{fig:fig4}\textit{b}, inset). Again, to calculate $C_f/C_{f_\mathrm{T950}}$, $C_f$ and $C_{f_\mathrm{T950}}$ are at matched $Re_\theta$.

%%---------------------------------------%%
\begin{figure}[!htbp]
	\begin{center}
	\includegraphics*[width=0.95\linewidth]{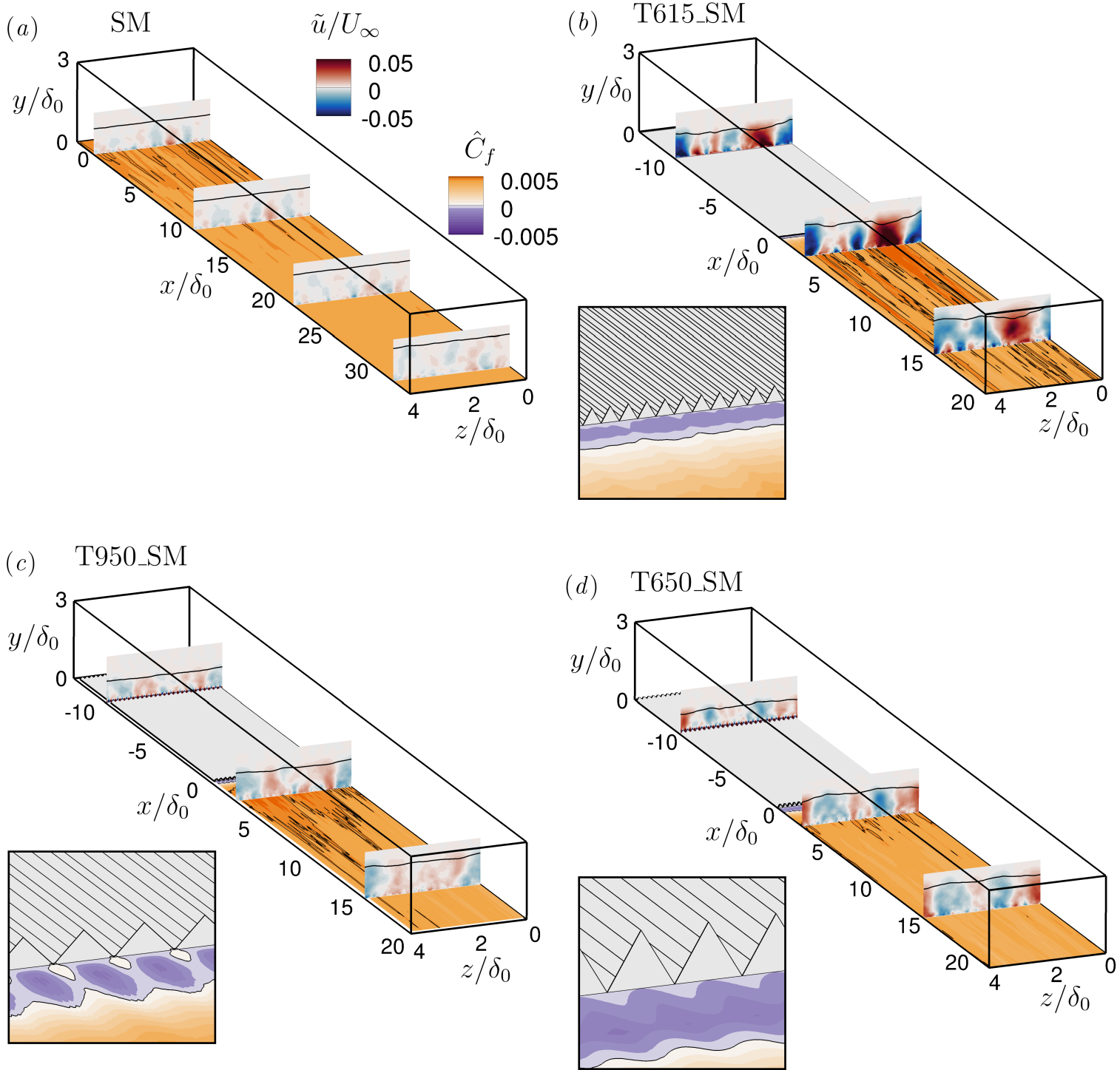}
    \includegraphics*[width=1\linewidth]{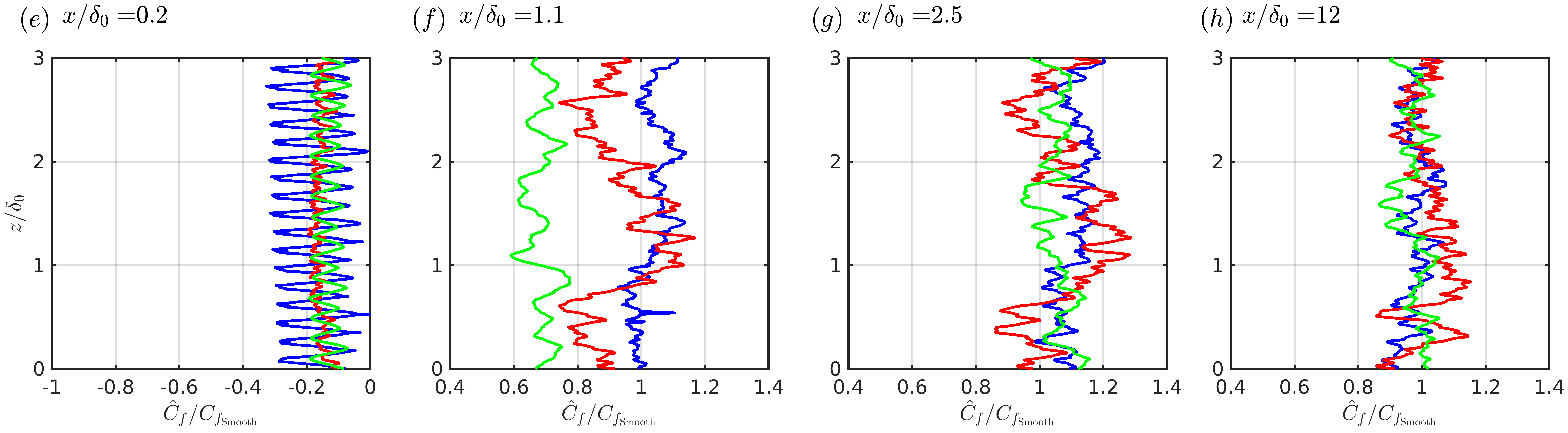}
	\caption{\label{fig:fig5} (\textit{a--d)} Contours of time averaged friction coefficient ($\hat{C}_f$) on the wall, after the step change and contours of $\tilde{u} = \langle u \rangle - U$, shown on $yz-$planes, at three streamwise locations; (\textit{a}) Smooth case (\textit{b}) T615-to-smooth case (\textit{c}) T950-to-smooth case and (\textit{d}) T650-to-smooth case. Solid black lines on $\tilde{u}$ plots represent local boundary layer thickness. Plots \textit{(e-h)} show variation of $\hat{C}_f/C_{f_\text{smooth}}$ in the spanwise direction, for three riblet-to-smooth cases, at four streamwise locations after the step change; $C_{f_\text{smooth}}$ is at matching $Re_\theta$. Line legends: T950\_SM (blue), T615\_SM (red), T650\_SM (green).}
	\end{center}
\end{figure}

%, and displays three-dimensionality
% In fact, the cross flow is strongest for the smallest riblet we tested (T615; figure~\ref{fig:fig5}\textit{a}).
%%---------------------------------------------%%
%\subsection{Secondary flow on riblet surface and modulation of $C_f$}\label{sec:cf_modulation}
\subsection{Local $C_f$ downstream of the riblet-to-smooth step change \label{sec:cf_modulation}}
In figure~\ref{fig:fig5}, we study the local variations of time-averaged skin-friction coefficient $\hat{C}_f$ downstream of the riblet-to-smooth step change. At the immediate downstream distance of the step change (figure~\ref{fig:fig5}\textit{e}), $\hat{C}_f$ is negative. This is due to the local flow reversal with a three-dimensional structure (figure~\ref{fig:fig5}\textit{b--d}, inset), with wider separated region behind the riblet peaks and narrow separation behind riblet valleys. For T950\_SM case, the separated region shows strong modulation (figure~\ref{fig:fig5}\textit{c}, inset); local patches of positive $\hat{C}_f$ are trapped between regions of negative $\hat{C}_f$, and vice versa. Such flow modulation is absent from riblets with $\alpha=60^\circ$ (figure~\ref{fig:fig5}\textit{b,d}; inset). We speculate that such modulation of $\hat{C}_f$ is afforded by the wider angle of the $90^\circ$ riblets; wide angle allows the upstream flow to channel through the valley regions, thus creating attached flow behind the riblet valleys and separated flow behind riblet peaks (figure~\ref{fig:fig5}\textit{c}, inset). Such flow channeling phenomenon have been observed in the works of \citet{Kumar2021, Kumar2023}, with large-scale roughness on airfoils, where significant spanwise modulation of wall-shear was reported (e.g. see figure 5 in \citep{Kumar2023}). The $60^\circ$ riblets, on the other hand, act as flow obstacles by limiting seeping of the flow into valley regions, thus creating relatively uniform separation in spanwise direction (figure~\ref{fig:fig5}\textit{b,d}, inset). Additionally, we observe that the amplitude of $\hat{C}_f$ modulation for T650\_SM  case ($k_0^+\simeq 43$; figure~\ref{fig:fig5}\textit{e}) is larger than the T615\_SM case ($k_0^+\simeq 13$), reflecting the difference in their $k_0^+$ sizes. Further downstream of the step change at $x = 1.1\delta_0$ (figure~\ref{fig:fig5}\textit{f}), the spanwise non-uniformity in $\hat{C}_f$ consists of a short wavelength in the order of the riblet spacing $s$, and a long wavelength in the order of $\delta_0$.
The large-scale spanwise non-uniformity in $\hat{C}_f$ is due to its modulation by the large-scale structures generated by the riblets at upstream of the step change.

To further investigate the modulation of $\hat{C}_f$, in figure~\ref{fig:fig5}(\textit{a--d}) we visualize $\tilde{u} = \langle u \rangle - U$, where $\langle u \rangle$ is the time-averaged $u$, and $U$ is the time- and spanwise-averaged $u$; $\tilde{u}$ highlights the spanwise heterogeneity of the flow. We plot $\tilde{u}$ at several streamwise locations for the riblet-to-smooth cases (figure~\ref{fig:fig5}\textit{b--d}), as well as the reference smooth case (figure~\ref{fig:fig5}\textit{a}). The $\tilde{u}$ fields over the riblets highlight the presence of large-scale secondary flows (with length scale $\sim \delta_0$), leading to alternating high- and low-momentum regions. These high and low-momentum regions are insignificant over the smooth wall (figure~\ref{fig:fig5}\textit{a}), but over riblets, their strength depends on the riblet geometry; they are stronger over T615 (figure~\ref{fig:fig5}\textit{b}), compared to T950 and T650 (figure~\ref{fig:fig5}\textit{c,d}). Downstream of the step change, these high/low-momentum streams modulate local skin-friction (see $\hat{C}_f$ contours on wall in figure~\ref{fig:fig5}\textit{b--d}); high $\hat{C}_f$ values coincide with high-momentum regions, and vice versa. Further downstream on the smooth patch, the modulation of $\hat{C}_f$ fades away (figure~\ref{fig:fig5}\textit{h}), yet the riblet-generated large-scale secondary flows remain persistent even up to $x = 15 \delta_0$ (see $yz$-slices at $x=15\delta_0$ in figure~\ref{fig:fig5}\textit{b--d}).

 To eliminate the possible role of insufficient temporal sample size in the presented spanwise inhomogeneities, we collected statistics of T950\_SM case for more than 15 domain flow-through time ($\approx 1700 \, \delta_0/U_\infty$); the secondary flow structures remained impervious to averaging time. The origin of the large $\delta_0$-scale mean secondary flows could be associated with the tripping near the inlet. However, these persistent large-scale structures are negligibly weak over the smooth wall, despite the tripping (figure~\ref{fig:fig5}\textit{a}). We conjecture that these secondary flows result from the interaction between the TBL and the spanwise heterogeneity of the surface elevation. Our conjecture is supported by the previous studies that have reported these $\delta$-scale secondary flows in turbulent half-channel flow~\cite{willingham2014turbulent,nikora2019friction} and TBL~\cite{Mejia-Alvarez2010, Barros2014, anderson2015numerical, Kaminaris2023} over irregular roughness, streamwise aligned ridges and truncated cones. Similar to our observation, they observe these secondary flows in the time-averaged field; regardless of the surface length-scale, the large-scale mean secondary flows have size $\sim \delta$. They identify these secondary flows either as Prandtl’s first kind \cite{Kaminaris2023}, attributed to vortex stretching and tilting mechanisms, or as Prandtl’s secondary flow of the second kind, that are sustained by the spatial gradients in the Reynolds stresses~\cite{anderson2015numerical}.

\begin{figure}[!htbp]
	\begin{center}
	\includegraphics*[width=1\linewidth,trim={{0.0\linewidth} {0.0\linewidth} {0.0\linewidth} {0.0\linewidth}},clip]{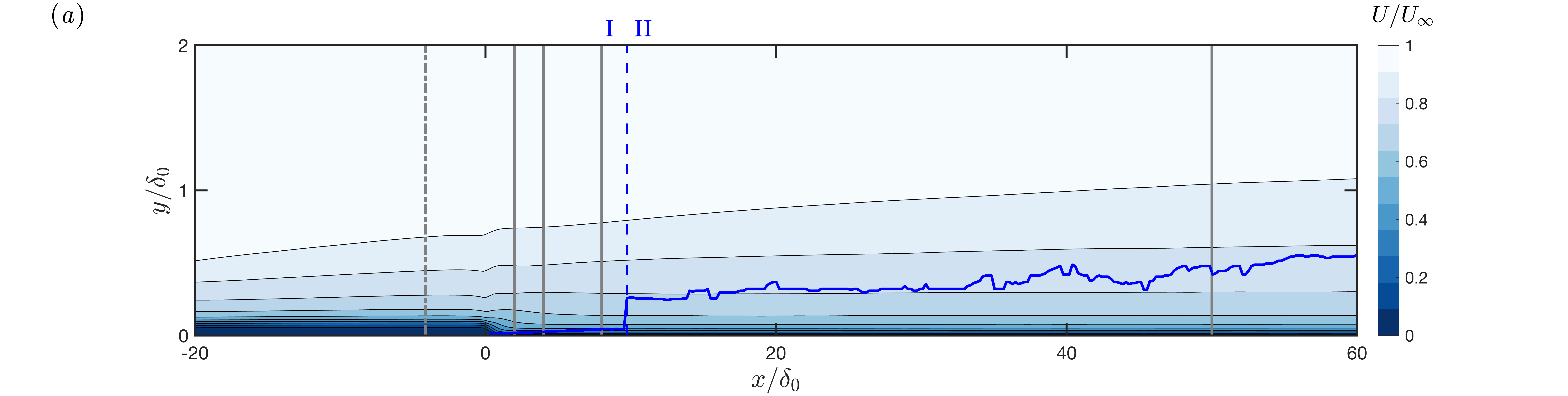}
    \includegraphics*[width=1\linewidth,trim={{0.0\linewidth} {0.0\linewidth} {0.0\linewidth} {0.0\linewidth}},clip]{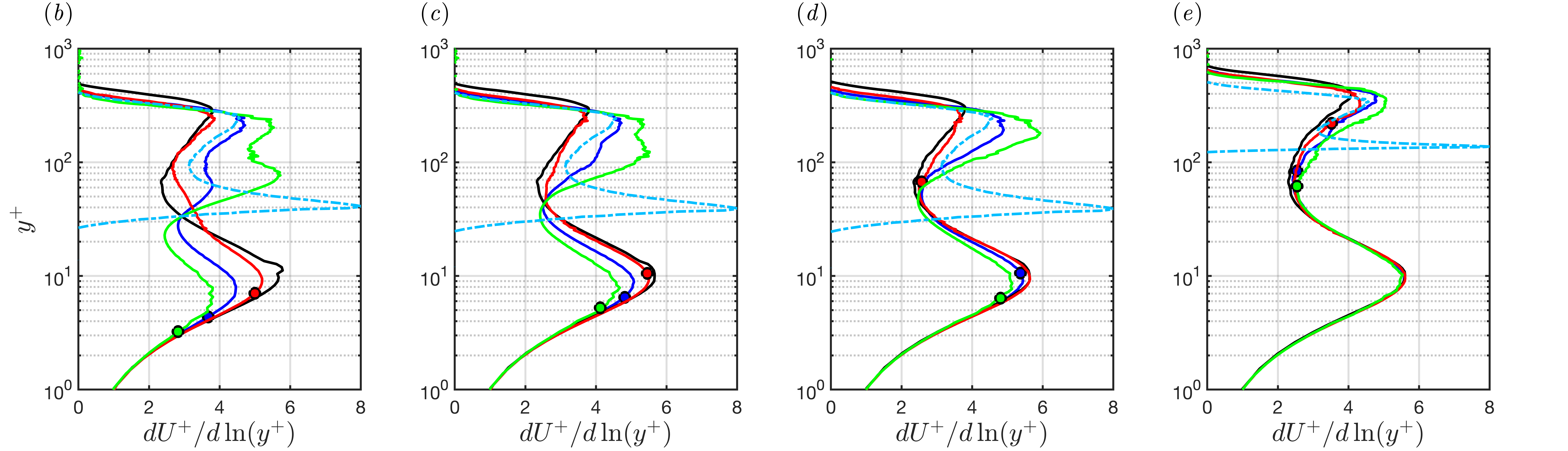}
    \includegraphics*[width=1\linewidth,trim={{0.0\linewidth} {0.0\linewidth} {0.0\linewidth} {0.0\linewidth}},clip]{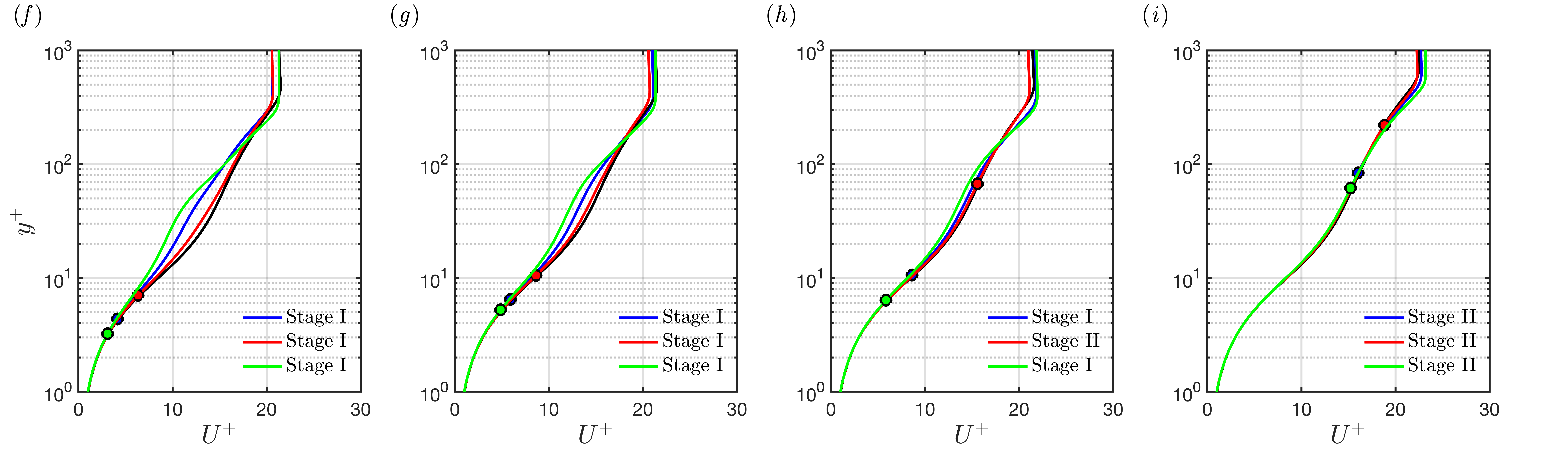}
    \includegraphics*[width=1\linewidth,trim={{0.0\linewidth} {0.0\linewidth} {0.0\linewidth} {0.0\linewidth}},clip]{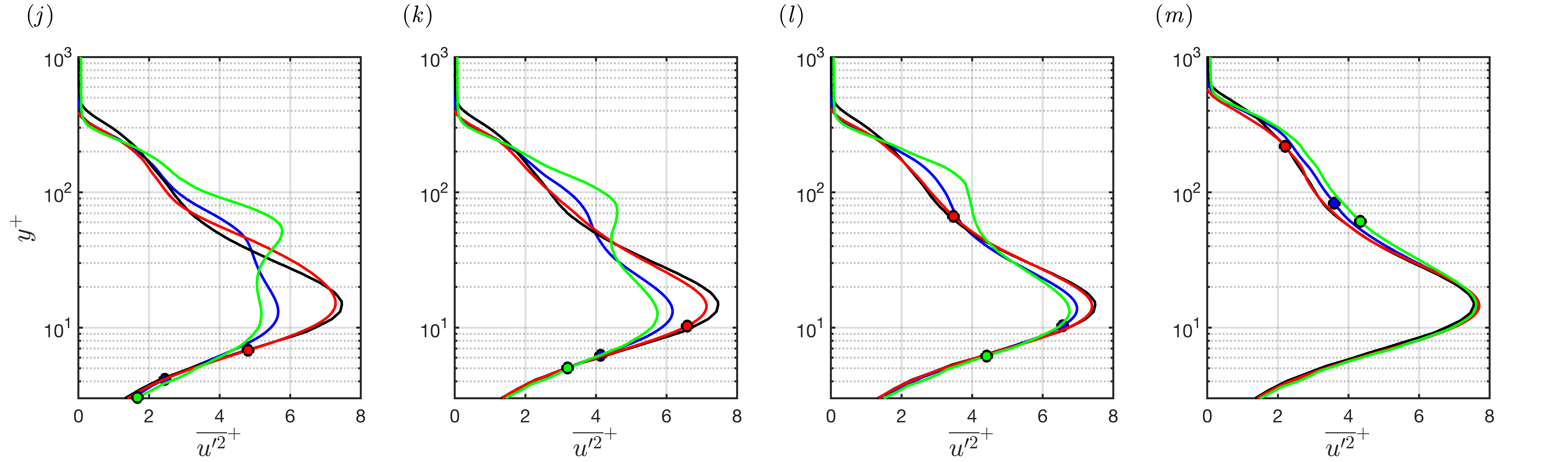}
	\caption{\label{fig:fig6} (\textit{b--e}) Identification of IEL based on $dU^+/d\ln y^+$ for the riblet-to-smooth step change at different locations marked in \textit{a} (gray lines); the profiles are compared with the ones from the smooth case (black lines) at matched $Re_\theta$. Profiles of $dU^+/d\ln y^+$ at $x=-4\delta_0$ for T950\_SM case (light-blue lines) upstream of the step change are also shown in (\textit{b--e}) to highlight frozen wake region. For each streamwise location marked by gray lines in (\textit{a}), profiles of (\textit{f-i}) streamwise velocity, $U^+$ and (\textit{j-m}) streamwise Reynolds stress component, $\overline{u^2}^+$, are shown; the IEL is marked with a colored bullet on each plot. Line legends: T950\_SM (blue), T615\_SM (red), T650\_SM (green).}
	\end{center}
\end{figure}
%%--------------------------------------------------%%

\subsection{Growth of internal equilibrium layer (IEL)}
Downstream of the riblet-to-smooth and smooth-to-riblet step change, $C_f$ does not converge to its equilibrium counterpart $C_{f_{SM}}$. By definition $C_f \equiv 2\tau_w/(\rho U^2_\infty) = 2/{U^+_\infty}^2$; therefore, recovery of $C_f$ depends on the recovery of $U^+$ profiles. In figure~\ref{fig:fig6}(\textit{b-e}), we asses the recovery in the profiles of $dU^+/d\ln(y^+)$ at $2\delta_0, 4\delta_0, 8\delta_0$ and $50\delta_0$ downstream of the riblet-to-smooth step change (lines in color), with their locations shown in figure~\ref{fig:fig6}(\textit{a}) (gray lines). To assess the flow recovery, for each downstream profile, we overlay the profile from the smooth case (SM) at matched $Re_\theta$ (black curves). Since $U^+_\infty$ is the integral of the curve $dU^+/d\ln(y^+)$,
%\begin{align}
%  \ar{U^+_\infty = \int_0^{\ln(\delta^+)} \frac{dU^+}{d\ln(y^+)} d\ln(y^+)},   
%\end{align}
full recovery of $C_f$ to $C_{f_{SM}}$ is reached if the entire profile of $dU^+/d\ln(y^+)$ collapses onto its equilibrium counterpart. Considering figures~\ref{fig:fig6}(\textit{b--e}), the profiles downstream of the step change progress towards the equilibrium profile from the SM case, but they do not fully collapse onto it. By $x = 50\delta_0$ (figure~\ref{fig:fig6}\textit{e}), the profiles are recovered up $y^+ \simeq 50$, but the wake profile ($y^+ \gtrsim 100$) remains unrecovered. Among the riblet-to-smooth cases, the wake profiles of T650\_SM and T950\_SM cases are noticeably stronger than the one from the SM case. However, the wake profile of the T615\_SM case is quite close to the one from the SM case. As a result, $C_f/C_{f_{SM}}$ recovers to unity for the T615\_SM case, but slightly falls below unity for the T950\_SM and T650\_SM cases (figure~\ref{fig:fig4}\textit{a}, inset).

The amplified wake profiles downstream of the riblet-to-smooth step change are carrying the flow history from the riblets patch at upstream of the step change, and remain uninfluenced by the smooth patch downstream of the step change. To support this argument, in figures~\ref{fig:fig6}(\textit{b--e}) we overlay the $dU^+/d\ln(y^+)$ profile from T950\_SM at $x = -4\delta_0$ upstream of the step change (light blue curve). In the wake region ($y^+ \gtrsim 100$), this profile closely agrees with its counterparts at downstream of the step change (dark blue curves). 
% \ar{The persistent wake from upstream of the step change must be linked to the persistent $\delta_0$-scale secondary flows, as discussed in section \ref{sec:cf_modulation}.}

To measure the flow recovery to equilibrium, we calculate the internal equilibrium layer thickness $\delta_\text{IEL}$. We overlay $\delta_\text{IEL}$ onto the mean velocity field for the T950\_SM case (figure~\ref{fig:fig6}\textit{a}), and we mark $\delta_\text{IEL}$ with bullets on the profiles in figure~\ref{fig:fig6}(\textit{b--m}). At each location, we locate $\delta^+_\text{IEL}$ at the distance $y^+$ where the difference between the $dU^+/d\ln(y^+)$ profile from the riblet-to-smooth case (colored profiles in figure~\ref{fig:fig6}\textit{b--e}) and the one from the equilibrium SM case at matched $Re_\theta$ (black profiles in figure~\ref{fig:fig6}\textit{b--e}) is less than a threshold of $0.2$. Our way of quantifying $\delta_\text{IEL}$ based on $dU^+/d\ln(y^+)$ \sout{is} also marks the distance up to which the streamwise Reynolds stress profiles are recovered (figure~\ref{fig:fig6}\textit{j--m}).

%Also,  The internal equilibrium layer thickness ($\delta_\text{IEL}$) is the $y^+$ to which the profiles of $dU^+/d\ln(y^+)$ for riblet-to-smooth curves and SM collapse on each other. $\delta_\text{IEL}$ locations are marked with blue, red and green bullets on figures~\ref{fig:fig6}(\textit{b--e}) for riblet-to-smooth cases with T950, T615 and T650 riblet, respectively. We also identified $\delta_\text{IEL}$ based on the agreement in the streamwise turbulent stress profiles $\overline{u'^2}^+$; the resulting $\delta_\text{IEL}$ is almost identical to the one based on $dU^+/d\ln(y^+)$. We show the profiles of viscous-scaled streamwise velocity, $U^+$, (figure~\ref{fig:fig6}\textit{f--i}) and streamwise turbulent stress, $\overline{u'u'}^+$, (figure~\ref{fig:fig6}\textit{j--m}) to analyze the recovery in different $y^+$ regions.

%Since recovery is a bottom-up process, flow close to the wall ($y^+\lesssim 10$; Stage I) recovers earliest (with streamwise distance) across cases.

%; the riblet-to-smooth profiles agree up to a height $y^+ \simeq 70$ ($\delta^+_\text{IEL} \simeq 70$) with smooth profile, and beyond $y^+ \simeq 70$, the profiles significantly differ from each other, which corresponds to their wake region.

%Recovery of wake region seems dependent on the riblet size.

%T615\_SM recovers earliest (by \ar{$x \sim 4\delta_0$}) while T650\_SM is the slowest (by \ar{$x \sim 20\delta_0$}).

\begin{figure}[!h]
	\begin{center}
	\includegraphics*[width=0.5\linewidth,trim={{0.0\linewidth} {0.0\linewidth} {0.0\linewidth} {0.0\linewidth}},clip]{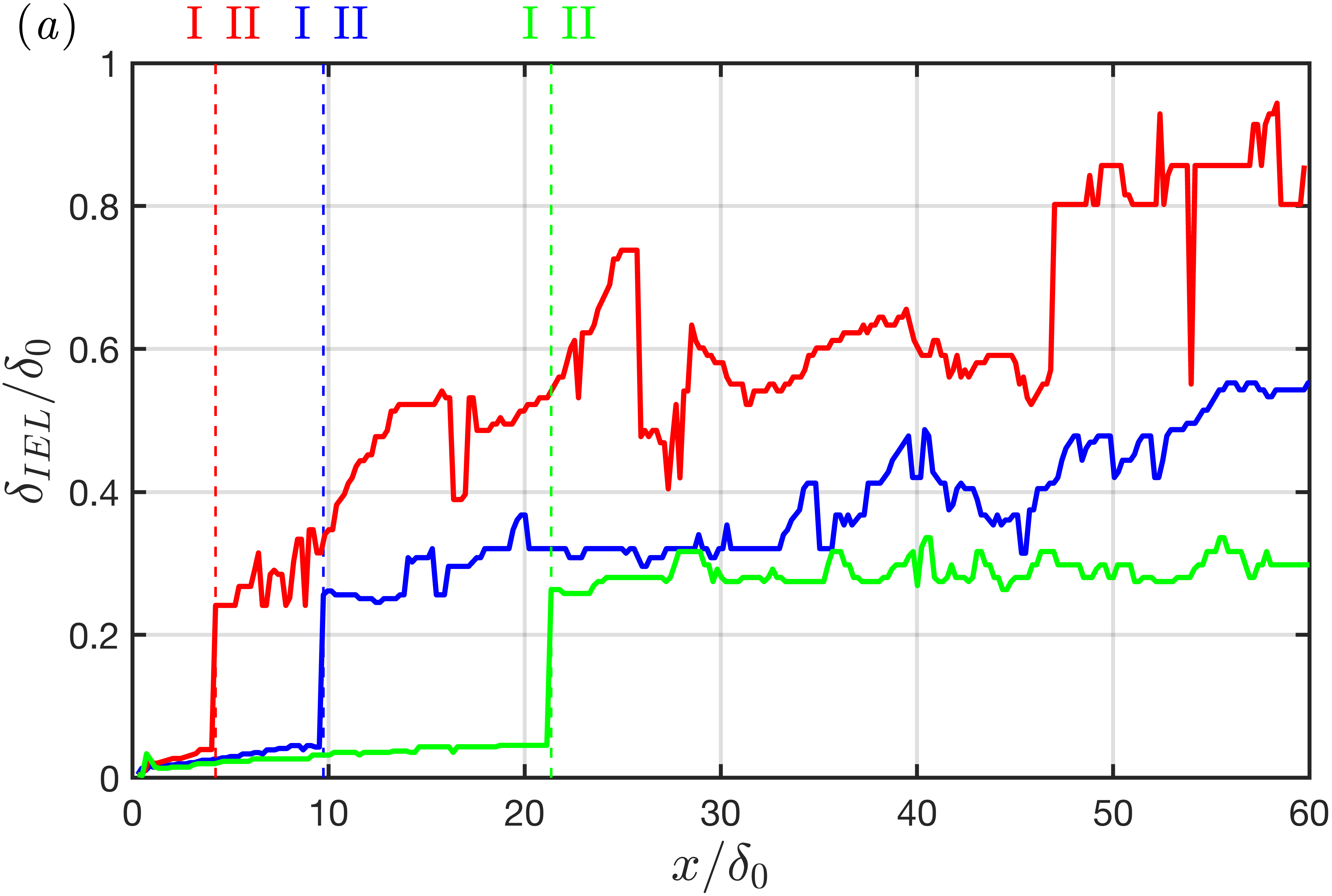}%
    \includegraphics*[width=0.5\linewidth,trim={{0.0\linewidth} {0.0\linewidth} {0.0\linewidth} {0.0\linewidth}},clip]{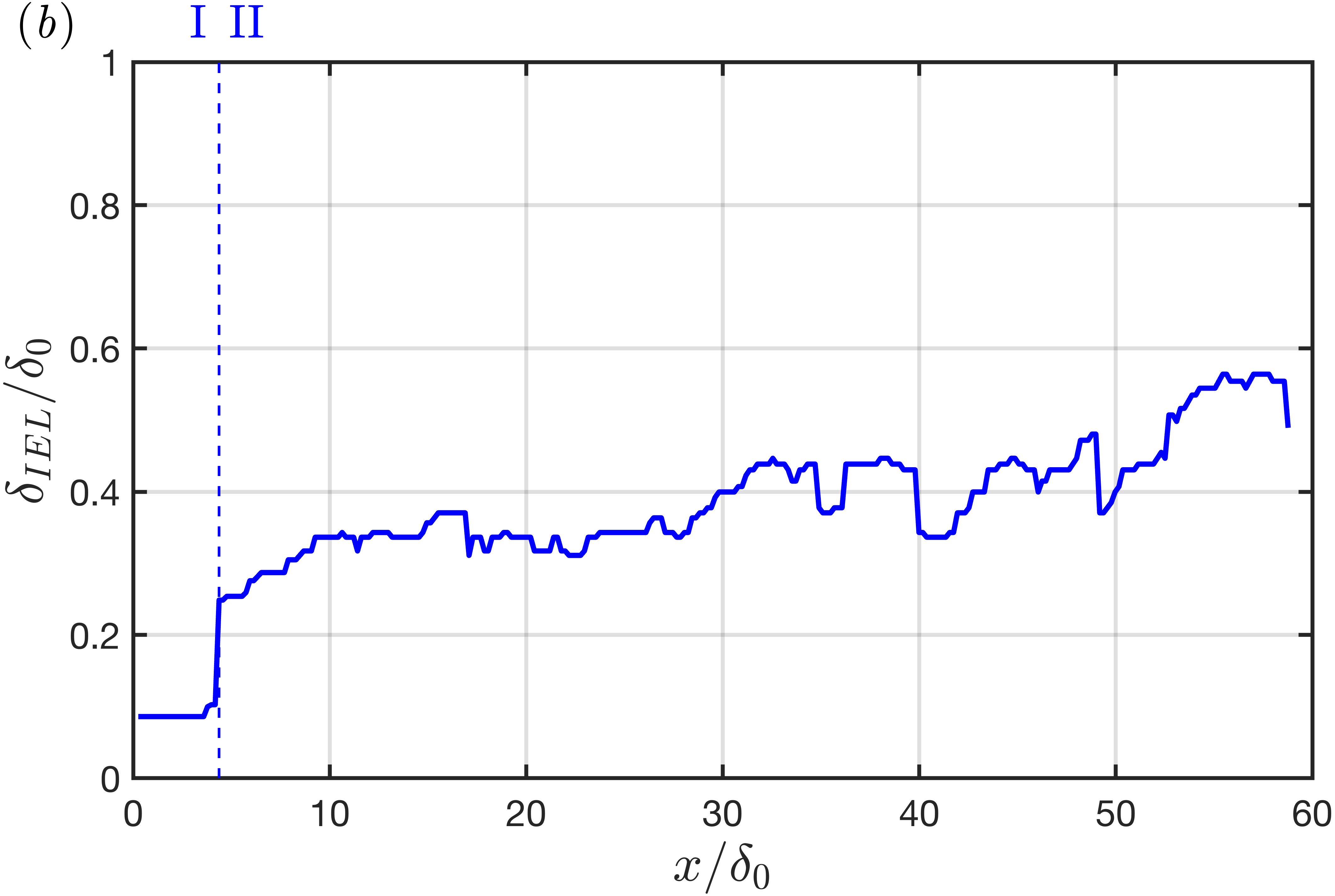}
	\caption{\label{fig:fig7} Streamwise growth of IEL thickness $\delta_\text{IEL}$ for (\textit{a}) riblet-to-smooth step change, and (\textit{b}) smooth-to-T950 step change. Different stages of recovery for each case are distinguished by vertical, dotted lines in plots (\textit{a,b}). Line legends: T950\_SM (blue), T615\_SM (red), T650\_SM (green).}
	\end{center}
\end{figure}

In figure~\ref{fig:fig7}, we plot $\delta_\text{IEL}$ for the riblet-to-smooth cases (figure~\ref{fig:fig7}\textit{a}), as well as the SM\_T950 case (figure~\ref{fig:fig7}\textit{b}). For the latter case, we obtain $\delta_\text{IEL}$ by following the same approach as the one for the riblet-to-smooth cases, except we regard the T950 case as the reference equilibrium case. The growth of $\delta_\text{IEL}$, hence the flow recovery to equilibrium as a bottom-up process, consists of two stages, I and II. The flow recovery during Stage I is limited to the near-wall recovery up to the buffer region ($y^+ \lesssim 10$), while during Stage II the flow in the outer region is recovering and near wall region has already attained equilibrium. Stage II is associated with the slow recovery of the wake region. By $x \sim 20\delta_0$, Stage I is complete for all riblet-to-smooth cases (figure~\ref{fig:fig7}\textit{a}). The growth of $\delta_\text{IEL}$, and the rate of recovery during Stage I across cases, however, is dependent on the riblet geometry, and inversely mirrors the riblet height $k^+$: T650\_SM ($k^+ \simeq 43$) has the slowest growth, followed by T950\_SM ($k^+ \simeq 25$) and T615\_SM ($k^+\simeq 12$). This is also evident in figure~\ref{fig:fig6}(\textit{d,h,l}) at $x/\delta_0=8$ where recovery for T615\_SM is at Stage II, while the inner-regions of T650\_SM and T950\_SM are still recovering (at Stage I). Beyond $x \sim 20\delta_0$, all riblet-to-smooth cases fall at Stage II of flow recovery, where $\delta_\text{IEL} \gtrsim 0.3 \delta_0$ (figure~\ref{fig:fig7}\textit{a}); recovery during this stage depends on the recovery of the riblet-generated wake, advected from upstream of the step change. As discussed in section \ref{sec:cf_modulation} (figure~\ref{fig:fig5}\textit{b,c,d}), the structure of the mean $\delta_0$-scale secondary flows depends on the riblet geometry; this manifests in different wake strengths, with T615\_SM and T650\_SM triggering the weakest and strongest wake regions, respectively (figure~\ref{fig:fig6}\textit{b}). As a result, the wake region of the T615\_SM case recovers to the one from the equilibrium SM case by $x = 50\delta_0$ (red curve in figure~\ref{fig:fig6}\textit{e}), and its $\delta_\text{IEL}$ grows to $\sim 0.9\delta_0$ by $x \sim 60 \delta_0$  (red curve in figure~\ref{fig:fig7}\textit{a}). On the other hand, the strong wake of T650\_SM case, deviates significantly from the equilibrium case, even up to $x = 50 \delta_0$ (green curve in figure~\ref{fig:fig6}\textit{e}). Thus, its $\delta_\text{IEL}$ stays at $0.3\delta_0$ by $x = 60 \delta_0$ (green curve in figure~\ref{fig:fig7}\textit{a}).

%For the smallest riblet (\ar{in T615\_SM}), the wake \ar{nearly} recovers by \ar{$x = 50\delta_0$} (figure~\ref{fig:fig6}\textit{e,i,m}). However, for T950 and T650 the difference in the wake region persists up to the farthest distance that we simulated (figure~\ref{fig:fig6}\textit{e,i,m}). Interestingly, the wake profile of the T950\_SM case downstream of the step change (dark blue curve) is close to its counterpart from upstream of the step change (light blue curve). This indicates a frozen wake region that preserves its history, and remains uninfluenced by the underneath smooth surface. Since the wake profiles of riblet-to-smooth cases does not recover to the one from the SM case, $C_f/C_{f_\mathrm{SM}}$ does not recover to $1.0$ (figure~\ref{fig:fig4}\textit{a}). To elaborate more, $C_f \equiv 2\tau_w/(\rho U^2_\infty) \equiv 2/{U^+_\infty}^2$, and $U^+_\infty$ is the integral of the $dU^+/d\ln(y^+)$ profile. Therefore, agreement in $C_f$ values requires agreement in the entire $dU^+/d\ln(y^+)$ profiles. For the SM\_T950 case, we observe similar frozen wake region with strong history effects, hence $C_f/C_{f_\mathrm{T950}}$ does not converge to $1.0$.

%%--------------------------------------------------%%
%%--------------------------------------------------%%
Comparing $\delta_\text{IEL}$ downstream of the step change between T950\_SM and SM\_T950 cases (blue curves in figure~\ref{fig:fig7}\textit{a,b}), highlights that Stage I is shorter downstream of the smooth-to-riblet step change, than vice-versa.
Thus, the near-wall turbulence has a slower recovery during the T950-to-smooth step change.
This observation is consistent with the previous studies that report faster flow recovery downstream of the smooth-to-rough step change, than vice-versa~\cite{Antonia_1971,Antonia_1972,Rouhi2019}. However, Stage II is similar between T950\_SM and SM\_T950 cases, and $\delta_\text{IEL}$ grows to maximum $0.5\delta_0$ by $x \sim 60 \delta_0$. 

\begin{figure}[!h]
	\begin{center}
	\includegraphics*[width=1\linewidth]{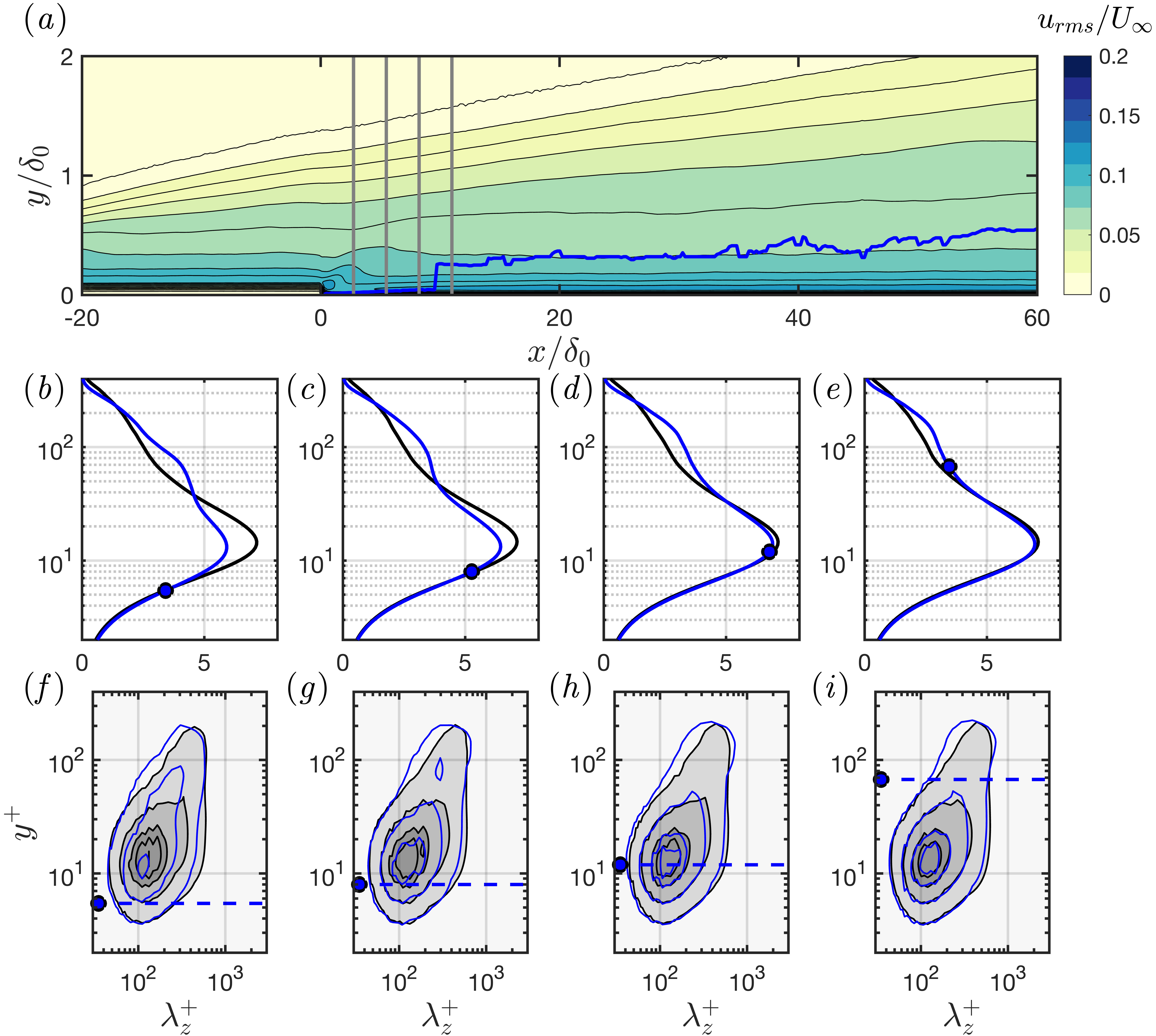}%
	\caption{\label{fig:fig12} (\textit{a}) Contour of $u_{rms}$ for T950\_SM case; growth rate of $\delta_\text{IEL}$ is shown by blue line. Plots (\textit{b--e}) show profiles of $u_{rms}$ at $x/\delta_0 = 2.7, 5.5, 8.2, 11$, respectively (marker by gray lines in \textit{a}); corresponding profiles for the SM case are extracted at matching $Re_\theta$. Plots (\textit{f--i}) show pre-multiplied energy spectra of $u'$ for T950\_SM case (blue contour lines), compared with smooth case (filled contours) at several streamwise locations (marked by gray lines in \textit{a}). Blue bullets in (\textit{b--i}) represent $\delta_\text{IEL}$ location.}
	\end{center}
\end{figure}
%%--------------------------------------------------%%
We study the recovery in the turbulence scales at downstream of the riblet-to-smooth step change, by analyzing the pre-multiplied energy spectrograms of streamwise velocity fluctuations $u'$ for T950\_SM case, at four distances downstream of the step change (figure~\ref{fig:fig12}\textit{a}). As expected, the spectrogram for T950\_SM (line contours in blue) shows significant differences compared to the SM case (gray-scale filled contour) in the vicinity of the step change at $x = 2.7\delta_0$ (figure~\ref{fig:fig12}\textit{f}). These differences are both in the inner ($y^+ \lesssim 30$) and the outer regions. Nevertheless, scales with $\lambda_z^+ \lesssim 60$ seem to recover remarkably well after the step change; the difference between T950\_SM and SM in figure~\ref{fig:fig12}(\textit{f}) for $\lambda_z^+ \lesssim 60 $ is minimal. The faster recovery of the small scales is in line with observations by rough-to-smooth DNS study of \citet{Ismail2018} and experimental study of \citet{Li2019}. Further downstream of the step change (figure~\ref{fig:fig12}\textit{g--i}), recovery progresses to larger $\lambda^+_z$ and higher $y^+$. At $x = 11 \delta_0$ (figure~\ref{fig:fig12}\textit{e,i}), flow recovery falls into Stage II ($\delta_\text{IEL} \simeq 0.3 \delta_0, \delta^+_\text{IEL} \simeq 70$), and only the wake region remains unrecovered. At this stage, the T950\_SM spectrogram agrees well with the equilibrium SM case up to $\delta^+_\text{IEL} \simeq 70$ across all scales (figure~\ref{fig:fig12}\textit{i}). Above $\delta^+_\text{IEL} \simeq 70$, the wake region of T950\_SM has a noticeably different spectral energy distirbution compared to the SM case.

%the large scales incrementally recover the energy equivalent to the smooth values. The intermediate scales, however, retain memory of the non-equilibrium effects, and do not recover to the levels of their smooth counterpart (figure~\ref{fig:fig12}\textit{d}). This suggests slow readjustment of turbulence cascade, as pointed out by \citet{Ismail2018}.

%We explain the source of the observed frozen wake
%The small scale ($\lambda_z^+ \lesssim 100 $) fluctuations 
%by rough-to-smooth DNS study, using transverse square ribs, of \citet{Ismail2018} and experimental investigation, using grit roughness, of \citet{Li2019}.
% On the other hand, the non-equilibrium effects of the step affects intermediate range ($\lambda_z^+ \simeq 100-400 $) and large scale fluctuations ($\lambda_z^+  \gtrsim 400$) significantly.
%%--------------------------------------------------%%
\begin{figure}[!h]
	\begin{center}
	\includegraphics*[width=0.5\linewidth,trim={{1.8cm} {0.5cm} {1cm} {0.cm}},clip]{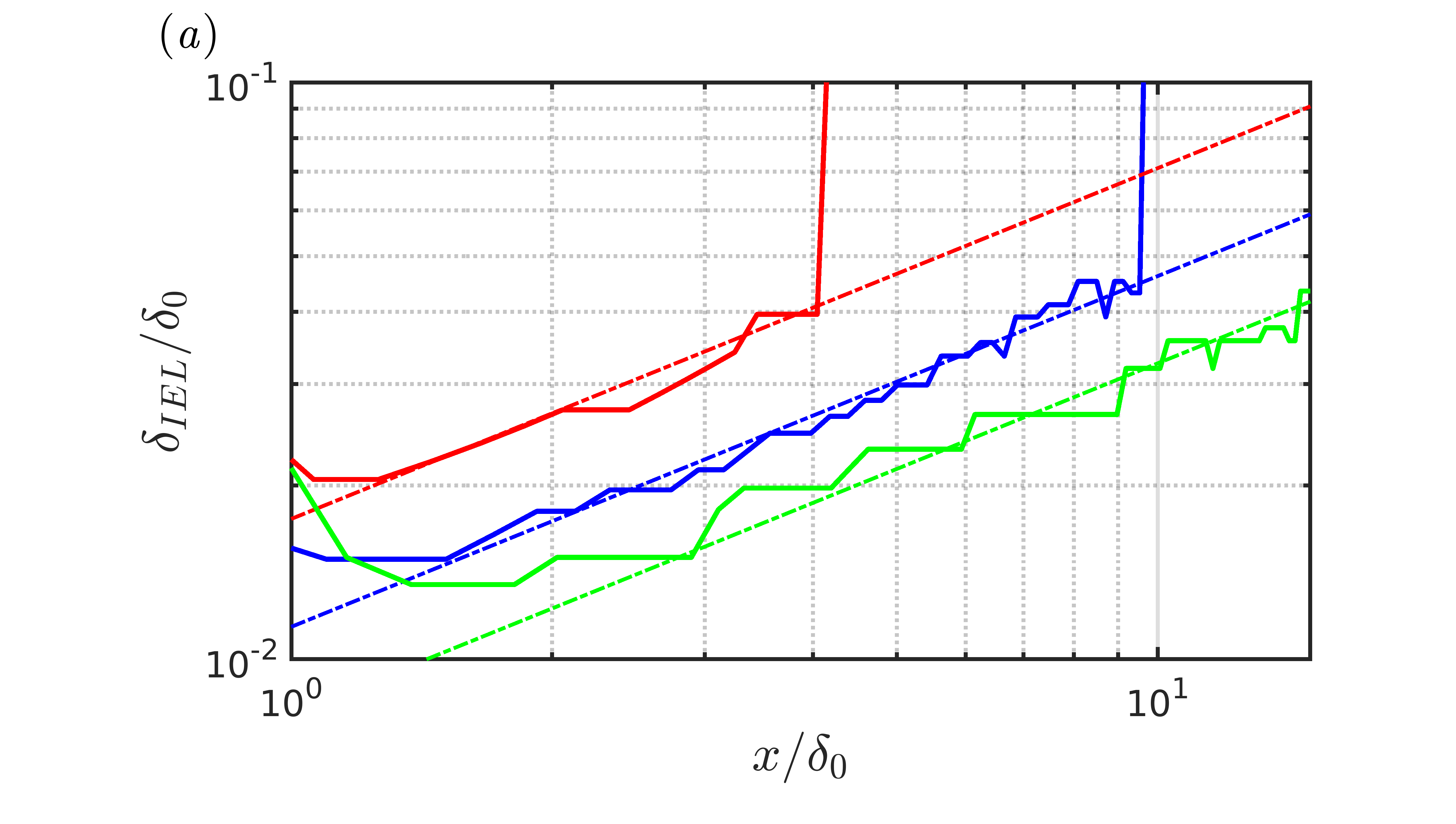}%
    \includegraphics*[width=0.5\linewidth,trim={{1.8cm} {0.5cm} {1cm} {0.cm}},clip]{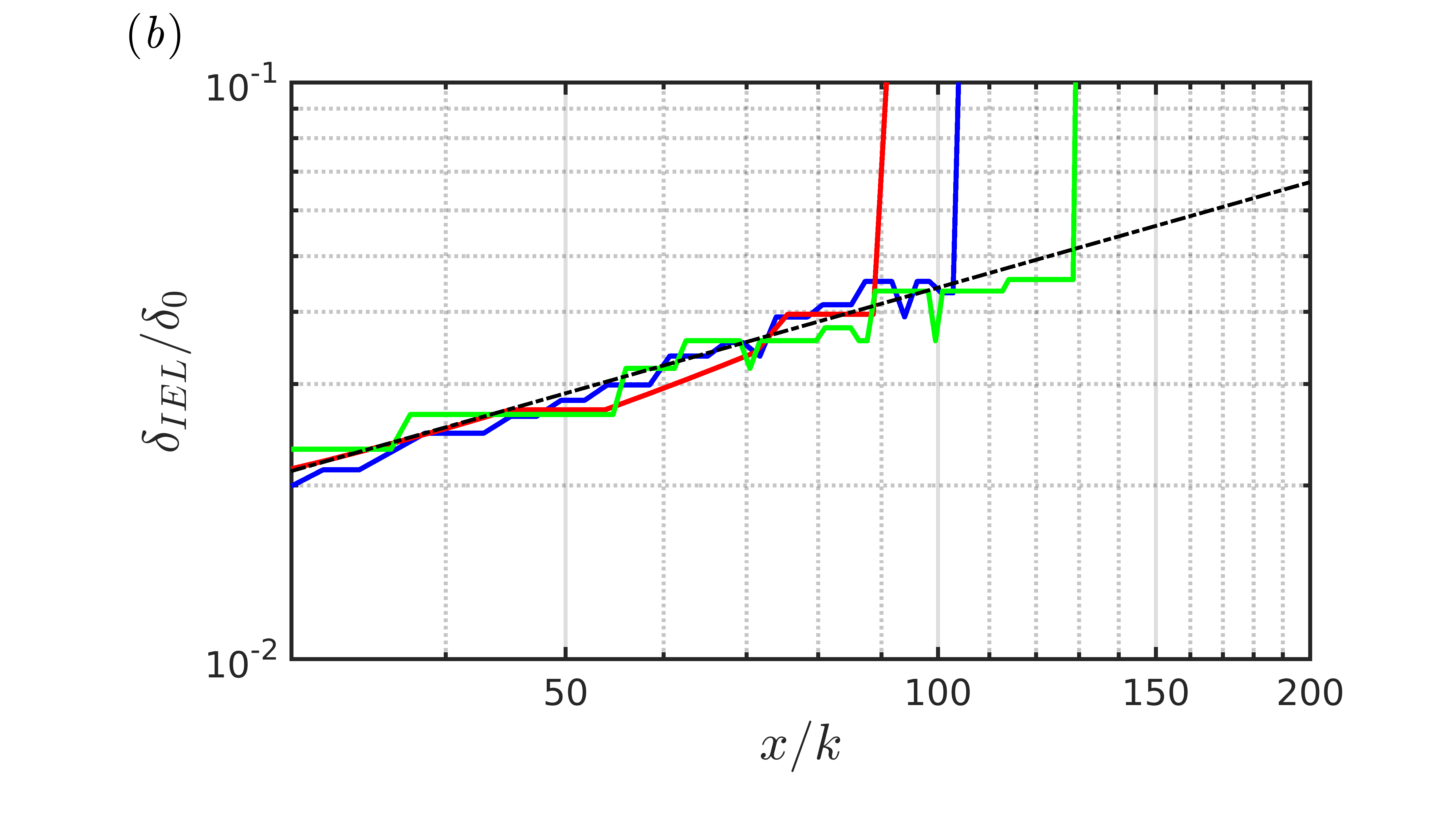}
	\caption{\label{fig:fig14} Streamwise growth of IEL thickness, $\delta_e$, during Stage I (identified in figure \ref{fig:fig6}) for riblet-to-smooth step change cases. Abscissa is scaled by (\textit{a}) boundary layer thickness at the reference location ($\delta_0$) and (\textit{b}) riblet height, $k$. Dash-dotted black line in (\textit{b}) represents the best fit line, corresponding to $\delta_{IEL}/\delta_0 = a \left( x/k \right)^b$, with $a=0.00268$ and $b=0.608$. Line legends: T950\_SM (blue), T615\_SM (red), T650\_SM (green).}
	\end{center}
\end{figure}
%%--------------------------------------------------%%

%In fact, when plotted against $x/k$ (figure~\ref{fig:fig14}\textit{b})

%The recovery in Stage I is complete \ar{by} $x \simeq 100k$ \ar{downstream of} the step change. A best-fit linear relation of the form,

Finally, we analyze the growth rates of $\delta_\text{IEL}$ during Stage I for all the riblet-to-smooth cases in figure~\ref{fig:fig14}. Similar to rough-to-smooth studies, during Stage I, the growth rate of $\delta_\text{IEL}/\delta_0$ downstream of the riblet-to-smooth step change is linear in log-log space  with $x/\delta_0$ (figure~\ref{fig:fig14}\textit{a}), i.e.\ $\delta_\text{IEL}/\delta_0 = a'_k(x/\delta_0)^b$. Figure~\ref{fig:fig14}(\textit{a}) highlights that the exponent $b$ is invariant to the upstream riblet geometry, but the proportionality factor $a'_k$ depends on the riblet geometry; the larger is $k$, the lower is $a'_k$. In figure~\ref{fig:fig14}(\textit{b}), we plot $\delta_\text{IEL}/\delta_0$ versus $x/k$, and interestingly all curves collapse on each other following a unified log-log fit.
\begin{align}
    \frac{\delta_\text{IEL}}{\delta_0} = a\left( \frac{x}{k} \right)^b, \label{eq:delta_IEL_scaling}
\end{align}
By fitting to the data, we obtain $a = 0.00268$ and $b = 0.608$; thus, the growth-rate exponent for these simulated cases is 0.608. By re-expressing (\ref{eq:delta_IEL_scaling}) in terms of $x/\delta_0$, we obtain

\begin{align}
    \frac{\delta_\text{IEL}}{\delta_0} = a\left( \frac{x}{k} \right)^b \implies \frac{\delta_\text{IEL}}{\delta_0} =\underbrace{a\left( \frac{\delta_0}{k} \right)^b}_{a'_k}\left( \frac{x}{\delta_0} \right)^b, \label{eq:delta_IEL_scaling_2}
    % a = 0.00268, b = 0.608
\end{align}
Equation (\ref{eq:delta_IEL_scaling_2}) presents the power-law scaling for each individual case based on $x/\delta_0$, which is then plotted in figure~\ref{fig:fig14}\textit{a} (dashed lines). We observe an excellent agreement with the individual growth rates. A growth rate of $\approx 0.6$ indicates faster recovery than most rough-to-smooth studies in the literature (reported growth rate of IBL $\sim 0.2-0.6$; see table~\ref{tab:studies_step_change}). The faster recovery in riblets-to-smooth configuration could be attributed to the rapid recovery of pressure fluctuations, since riblet surface impose no pressure drag.

%has $Re_{\theta} \simeq 680$ ($Re_\tau \simeq 283$) right upstream of the step change, and 

\section{ Conclusions \label{sec:conclusion}} %%%%%%%%%%%%%%%%%%
We investigated the recovery of ZPG TBLs downstream of riblet-to-smooth and smooth-to-riblet step changes. We focused on three triangular riblet shapes for riblet-to-smooth calculations: riblets with tip angle, $\alpha=\ang{90}$ and viscous-scaled spacing, $s^+ = 50$ (T950), $\alpha=60, s^+=15$ (T615), and $\alpha=\ang{60}, s^+=50$ (T650). Viscous units are calculated based on the friction velocity $u_{\tau_0}$ at a reference location upstream of the step change, where $Re_{\theta_0} \simeq 680$ ($Re_{\tau_0} \simeq 283$). While T615 is a drag-decreasing riblet, T950 and T650 are drag-increasing. For smooth-to-riblet step change, we considered only T950 riblet. The ZPG TBL grows up to $Re_\theta \simeq 1000$ ($Re_\tau \simeq 400$) by the end of the computational domain downstream of the step change. To evaluate the flow recovery to equilibrium, we conducted additional reference calculations of a ZPG TBL over an entirely smooth wall, as well as a ZPG TBL over an entirely riblet-covered wall by T950 riblets. To make the computations affordable, we followed our recently proposed $\eta-$grid generation approach of \citet{Rouhi2025} for unstructured solvers, where the element size grows proportionate to the local Kolmogorov scale. 

We observed that $C_f$ converged towards its equilibrium value with a faster rate during smooth-to-T950 step change (within $4\delta_0$), than during T950-to-smooth step change (within $7\delta_0$). However, for both step changes, $C_f$ did not completely reach equilibrium, even after $45\delta_0$ downstream of the step change. For the riblet-to-smooth cases, the larger is the riblet height $k^+$ (at matching $s^+$), the greater is the departure of $C_f/C_{f_\text{Smooth}}$ from expected equilibrium value of 1.0. This was verified by simulating T650-to-smooth ($s^+ = 50, h^+=43$) step change case, which resulted in $C_f/C_{f_\text{Smooth}}\sim 0.93$ by a distance of $45\delta_0$ (compared to $C_f/C_{f_\text{Smooth}}\sim 0.97$ in T950-to-smooth; $s^+ = 50, h^+ = 25$). Consistent with the trends in $C_f$, the internal equilibrium layer thickness $\delta_\text{IEL}$ did not reach the TBL thickness for larger riblets, and stayed within $0.4\delta_0$ for T950-to-smooth and T650-to-smooth cases. 

Assessment of the profiles of the mean velocity derivative $dU^+/d\ln(y^+)$, streamwise velocity $U^+$, and streamwise turbulent stress $\overline{u'^2}^+$ revealed that the recovery in near wall region ($y^+\lesssim 30$) is fast. We identify this recovery as Stage I recovery. The Stage I growth rate of $\delta_\text{IEL}$ follows a unified scaling of $ \delta_\text{IEL}/\delta_0 = 0.00268\left( x/k \right)^{0.608}$ for all  riblet-to-smooth configurations simulated in this study, where $x$ is the streamwise distance from the step change; stage I is complete by $x\sim 100\,k$. Stage II recovery is associated with the recovery in the wake region ($y^+\gtrsim 70$); this recovery is not completed for T950-to-smooth and T650-to-smooth cases, even up to $x \simeq 50\delta_0$. Consistently, $C_f/C_{f_\text{Smooth}}$ does not reach $1.0$, and $\delta_\text{IEL}$ does not reach the TBL thickness for these cases. The incomplete recovery is related to the `frozen' wake region, that is advected from upstream of the step change; this observation is supported by tracking the evolution of $dU^+/d\ln(y^+)$ profiles, as well as pre-multiplied energy spectrograms of streamwise velocity fluctuations.
%The wake region ($y^+\gtrsim 70$), on the other hand, does not fully recover (Stage II recovery) for larger riblets even by $x \simeq 50\delta_0$, leading to the incomplete recovery in $C_f$ and $\delta_\text{IEL}$. Furthermore, the wake region seems `frozen', when compared to wake profile from upstream (to step change) location. This phenomenon of frozen wake, due to step change, can be attributed to the incomplete recovery of intermediate range scales ($\lambda_z^+ \simeq 100-400$) on the downstream smooth surface, as evidenced by pre-multiplied energy spectra of streamwise fluctuations.

%\ar{Further, the growth rate for riblet-to-smooth adheres to a higher exponent ($\delta_\text{IEL} \propto (x/\delta_0)^{0.608}$) than rough-to-smooth transitions ($\delta_\text{IBL} \propto (x/\delta_0)^{0.2-0.6}$).}

%%% Insert here acknowledgments if necessary %%%

\section*{ Acknowledgments }

VK acknowledges his AI4S fellowship within the Generaci\'on D initiative by Red.es, Ministerio para la Transformaci\'on Digital y de la Funci\'on P\'ublica, for talent attraction (C005/24-ED CV1), funded by NextGenerationEU through PRTR. AR acknowledges the support from the Air Force Office of Scientific Research (AFOSR) under award number FA8655-24-1-7008, monitored by Dr.\ Douglas Smith and Dr.\ Barrett Flake. WW acknowledges the support from AFOSR Grant No.\ FA9550-25-1-0033, monitored by Dr.\ Gregg Abate. 
OL has been partially supported by a Ramon y Cajal postdoctoral contract (Ref: RYC2018- 025949-I). The authors acknowledge the support given by the Departament de Recerca i Universitats de la Generalitat de Catalunya to the Large-Scale Computational Fluid Dynamics Research Group (Code: 2021 SGR 00902).
We thank EPSRC for the computational time made available on ARCHER2 via the UK Turbulence Consortium (EP/X035484/1), and the UKRI access to the HPC call 2024. We also acknowledge the computational resources provided by Barcelona Supercomputing Center and Red Espa\~nola de Supercomputaci\'on (RES) on MareNostrum V (Nos.\ IM-2025-3-0053, IM-2026-1-0036).
% We also acknowledge the Barcelona Supercomputing Center for awarding us access to the MareNostrum V machine based in Barcelona, Spain.

\begin{comment}
\end{comment}

% \bibliographystyle{plainnat}
% \bibliographystyle{abbrvnat}
\bibliographystyle{unsrtnat}
\bibliography{./references_vishal_apr26.bib}

\end{document}